\documentclass[aps,prl,twocolumn,longbibliography,superscriptaddress,10pt,floatfix]{revtex4-2}
\usepackage[colorlinks,bookmarks=true,citecolor=blue,linkcolor=red,urlcolor=blue]{hyperref}

\usepackage{amsmath}
\usepackage{amssymb}
\usepackage{dsfont}
\usepackage{bbold}
\usepackage{tikz}
\usepackage{graphicx}
\usepackage{svg}
\usepackage{mathtools}
\usepackage{mathrsfs}
\usepackage{times}
\usepackage{placeins}
\usepackage{ulem}
\usepackage[utf8]{inputenc}

\newcommand{\ket}[1]{\left| #1 \right\rangle}
\newcommand{\bra}[1]{\left\langle #1 \right|}

\begin{document}

\title{Chiral Edge Excitations of $\nu=1/2$ Fractional Chern Insulators in the Bosonic Hofstadter Model}

\author{Xiao-Han Yang}
\affiliation{Hefei National Laboratory, University of Science and Technology of China, Hefei 230088, China}
\affiliation{Hefei National Research Center for Physical Sciences at the Microscale and School of Physical Sciences, University of Science and Technology of China, Hefei 230026, China}

\author{Ji-Yao Chen}
\email{chenjiy3@mail.sysu.edu.cn}
\affiliation{Center for Neutron Science and Technology, Guangdong Provincial Key Laboratory of Magnetoelectric Physics and Devices, School of Physics, Sun Yat-sen University, Guangzhou 510275, China}

\author{Xiao-Yu Dong}
\email{dongxyphys@ustc.edu.cn}
\affiliation{Hefei National Laboratory, University of Science and Technology of China, Hefei 230088, China}

\date{\today}

\begin{abstract}
Edge excitations are the defining signature of chiral topologically ordered systems. In continuum fractional quantum Hall (FQH) states, these excitations are described by the chiral Luttinger liquid ($\chi$LL) theory. Whether these field theory predictions can be precisely identified in discrete lattice systems of finite width, however, remains a longstanding question. Here we numerically demonstrate that the charge-one edge spectral function of a $\nu=1/2$ FCI on an infinitely long strip with width $L_y=10$ quantitatively follows the predictions of $\chi$LL theory. The edge spectrum is gapless, chiral, and linear, with spectral weight increasing linearly with both momentum and energy. We further analyze the influence of lattice size, particle number, trapping potential, and charge sector of excitations on the edge properties. Our results establish a clear correspondence between lattice FCIs and continuum FQH systems and provide guidance for future experimental detection of chiral edge modes.
\end{abstract} 

\maketitle

\textit{Introduction.---}
Fractional Chern insulators (FCIs)~\cite{regnault2011} are the lattice analogues of fractional quantum Hall (FQH) states, exhibiting features such as quantized Hall conductivity, anyonic excitations, and chiral edge modes. Their potential as a controllable platform for topological quantum computation~\cite{Nayak2008nonAbelian} makes the identification of universal signatures of FCIs both a central theoretical pursuit and an ongoing experimental challenge. Proposed realizations span the systems of optical lattices~\cite{popp2004adiabatic, sorensen2005, hafezi2007, cooper2013reaching, yao2013realizing, hudomal2019bosonic, leonard2023}, optical tweezers~\cite{lunt2024realization}, twisted bilayer MoTe$_2$~\cite{cai2023, park2023, zeng2023, xu2023, he2025, wang2024c}, and interacting photons~\cite{wang2024a}.

One of the simplest and most prominent lattice models hosting FCIs is the Harper-Hofstadter-Hubbard model of strongly interacting bosons. In this model, FCIs emerge as ground states at specific filling factors $\nu = n_b/n_{\phi}$, where $n_b$ denotes the particle density and $n_{\phi}$ (in units of $2\pi$) is the magnetic flux per unit cell~\cite{moller2009composite,he2017realizing}. Notably, the lattice analogue of $\nu=1/2$ Laughlin state arises for $n_b=1/8$ and $n_{\phi}=1/4$ on a square lattice~\cite{hafezi2007,cincio2013characterizing,gerster2017fractional,motruk2017,dong2018}. Numerical studies, primarily using tensor networks and exact diagonalization methods, have revealed a range of diagnostic features of this model. For instance, quantized Hall conductivity can be characterized by fractionalized charge pumping~\cite{dong2018,motruk2017,wang2022measurable}, Str\v{e}da’s formula~\cite{peralta2023,repellin2020fractional,wu2025optimal,wang2024b}, or center of mass Hall drift~\cite{repellin2020fractional,Motruk2020Detecting}, while fractional charge excitations can be captured by local pinning potentials~\cite{Raciunas2018creating,wang2022measurable}. 

The $\nu=1/2$ FCI as a ground state of the Harper-Hofstadter-Hubbard model has been experimentally realized on a $4\times 4$ square lattice using ultracold atoms in an optical lattice with synthetic artificial gauge fields~\cite{leonard2023}, and independently with interacting photons in a two-dimensional circuit quantum electrodynamics system~\cite{wang2024a}. Local density measurements in these systems revealed key signatures of FCI physics, including nearly quantized Hall conductivity and vortex structure of correlations. Although current realizations are limited to two strongly interacting bosons on a small lattice, these results provide compelling evidence for the existence of FCIs. Scaling to larger system sizes is essential for accessing more physical phenomena. A variety of theoretical proposals have explored routes toward this goal using ultracold atoms~\cite{sorensen2005,motruk2017,he2017realizing,hudomal2019bosonic,michen2023adiabatic,wang2024b,wu2025optimal,palm2024}. In particular, Wu et al.~\cite{wu2025optimal} have proposed optimal-control protocols to accelerate state preparation, and Palm et al.~\cite{palm2024} introduced a patchwork preparation scheme that assembles multiple $4 \times 4$ blocks into larger systems. These advances lay the groundwork for future experiments aimed at probing anyonic statistics of excitations and, crucially, pave the way for direct observation of chiral edge states in systems with open boundaries, which is the topic of this work. 

Chiral edge excitations are a hallmark of continuum FQH states, and their low-energy behavior is effectively described by the chiral Luttinger liquid ($\chi$LL) theory ~\cite{wen1990a,wen1995,wen1992}. For a $\nu = 1/s$ Laughlin state with $s\in \mathbb{N}^+$, the theory predicts that the spectral function of charge-one edge excitations takes the form
\begin{eqnarray}
\mathcal{A}(k,w)\propto (\omega+v k)^{s-1}\delta(\omega-vk),
\label{Eq:Akw}
\end{eqnarray}
where $\omega$, $k$, and $v$ denote the energy, momentum, and velocity of the edge excitations, respectively. A central open question is whether the edge excitations of FCIs, realized on discrete lattices with finite width and open boundaries, exhibit the same spectral characteristics predicted by $\chi$LL theory for continuum FQH systems. To date, even in numerical simulations, this characteristic spectral function has not been definitively observed as far as we know. 

Two key features of $\mathcal{A}(k,\omega)$ are expected: first, it should exhibit chirality, as indicated by the delta function $\delta(\omega-vk)$; second, its weight $(\omega+v k)^{s-1}$ should increase with $k$ and $\omega$. In particular, for $\nu=1/2$ ($s=2$), a linear increase of the weight is anticipated. One of the authors (Dong et al.~\cite{dong2018}) attempted to compute the edge spectral function by evaluating real-time dynamical correlations on an infinitely long strip with width $L_y = 8$. While the resulting spectral function was indeed chiral, its weight deviated from the theoretical prediction. This discrepancy may arise from several factors, including numerical inaccuracies in the time-evolution simulations due to the entanglement growth or finite-size effects, where insufficient strip width leads to hybridization between opposite edges. Another possible reason, as suggested in Refs.~\cite{binanti2024,vashisht2025,kjall2012,luo2013}, is that the observation of chiral edge excitations may require a suitably smooth trapping potential. On a finite disk, when such a trap is employed, the energy levels of charge-neutral excitations exhibit the counting predicted by the corresponding conformal field theory.

\begin{figure}[hb]
    \includegraphics[width=0.95\linewidth]{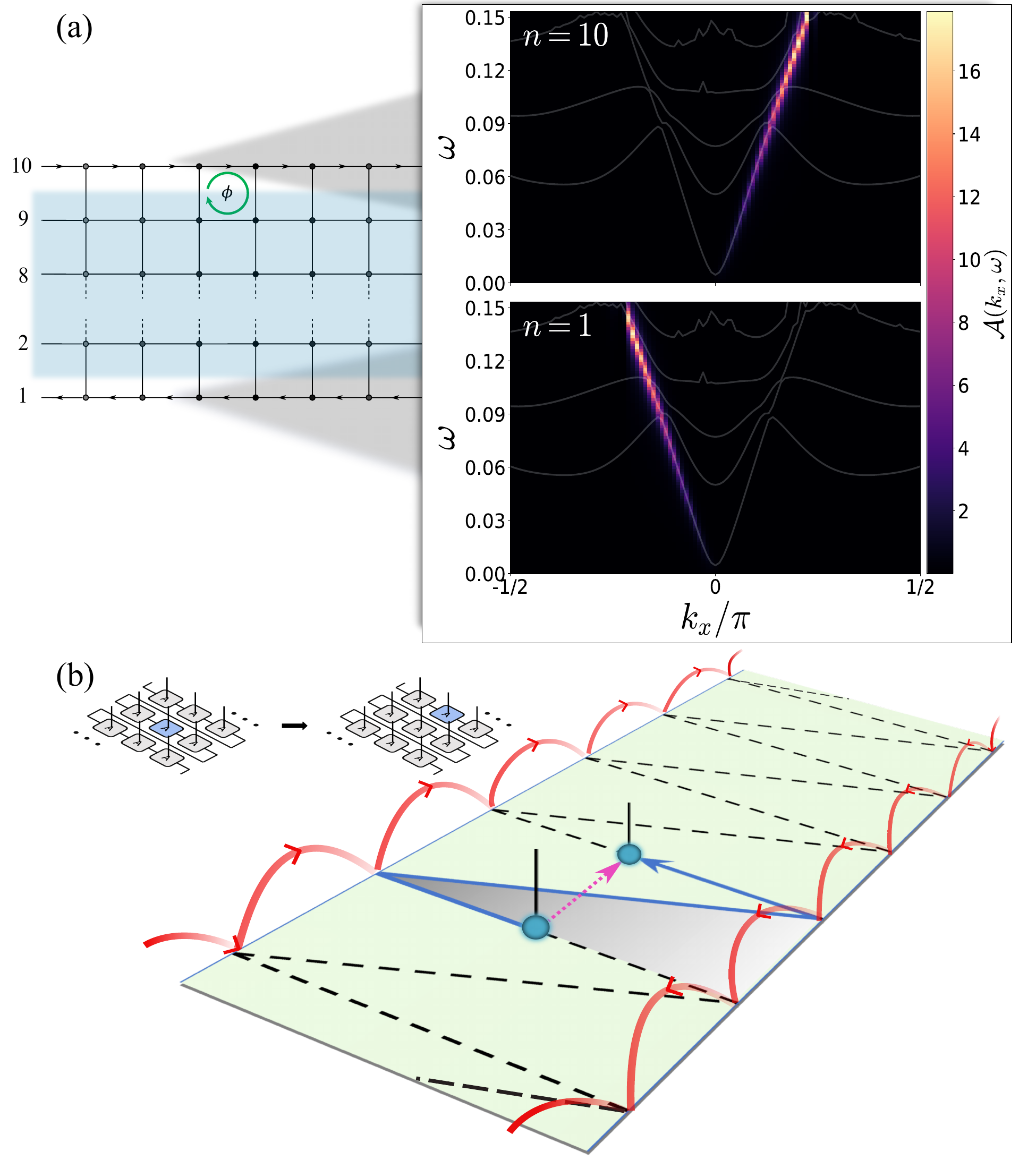}
    \caption{(a) Lattice geometry of a strip with $L_y=10$, in which $\phi$ denotes the magnetic flux per plaquette. The spectral functions $\mathcal{A}_n(k_x,\omega)$ of charge-one edge excitations on the top ($n=10$) and bottom ($n=1$) rows are computed using an iMPS with bond dimension $D=2000$ and a Lorentzian broadening factor $\eta=0.005$. The gray lines show the energies of excited states. (b) Illustration of the path of a local tensor in the iMPS (blue) and its mapped trajectory on the physical lattice (pink). The shaded area marks the region where the Stokes’ theorem is applied to evaluate the momentum shift.}
    \label{fig:Spectrum_Mapping}
\end{figure}

In this Letter, we focus on the chiral edge excitations of $\nu=1/2$ FCIs. The key result is that we obtain the spectral functions of charge-one edge excitations, which quantitatively agree with the predictions of $\chi$LL theory, on an infinitely long strip with width $L_y = 10$ (see Fig.~\ref{fig:Spectrum_Mapping}). The edge spectrum is linear and chiral, with spectral weight increasing linearly with momentum and energy. We further compare our results with previous studies on strips with width $L_y=8$ and $L_y=11$, and also our results of $L_y=12$, without or with a smooth trapping potential, and discuss the possible origins of the difficulties in observing the theoretically predicted spectral behavior. 

\textit{Ground state and bulk FCI.---}
We study the bosonic Harper-Hofstadter-Hubbard model on a square lattice with the Hamiltonian 
\begin{eqnarray}
    \hat{H}&=&\sum_{m,n}(-t_xe^{i\phi n}\hat{a}^{\dag}_{m+1,n}\hat{a}_{m,n}-t_y \hat{a}^{\dag}_{m,n+1}\hat{a}_{m,n}+h.c.)\nonumber\\
    &&+\frac{U}{2}\sum_{m,n}\hat{n}_{m,n}(\hat{n}_{m,n}-1).
\end{eqnarray}
Here, $(m,n)$ denotes the coordinate of a site on the lattice, with $m=1,2,...,L_x$ and $n=1,2,...,L_y$. We consider the lattice with open boundary conditions in both directions. The width $L_y$ is finite, while the length $L_x$ can be finite or infinite. The operator $\hat{a}^{\dag}_{m,n}(\hat{a}_{m,n})$ is the creation (annihilation) operator of a spinless boson on site $(m,n)$, and $\hat{n}_{m,n}=\hat{a}^{\dag}_{m,n}\hat{a}_{m,n}$ is the corresponding particle density operator. The non-zero Peierls phase factor $e^{i\phi n}$ of the hoppings in $x$-direction leads to a finite magnetic flux $\phi$ in each plaquette. The flux density $n_{\phi}=1/4$ is obtained when $\phi=\pi/2$. We set hopping coefficients $t_x=t_y=1$, and on-site interaction $U\rightarrow \infty$ to achieve the hard-core boson limit.

To evaluate the effects of the smooth trapping potential suggested in~\cite{vashisht2025}, we can also add the term $\hat{H}_{\mathrm{trap}}=V\sum_{m,n}(n-(L_y+1)/2)^2\hat{n}_{m,n}$ into the Hamiltonian, which provides a harmonic trapping of particles in the $y$-direction.

\begin{figure}[t]
    \includegraphics[width=1.0\linewidth]{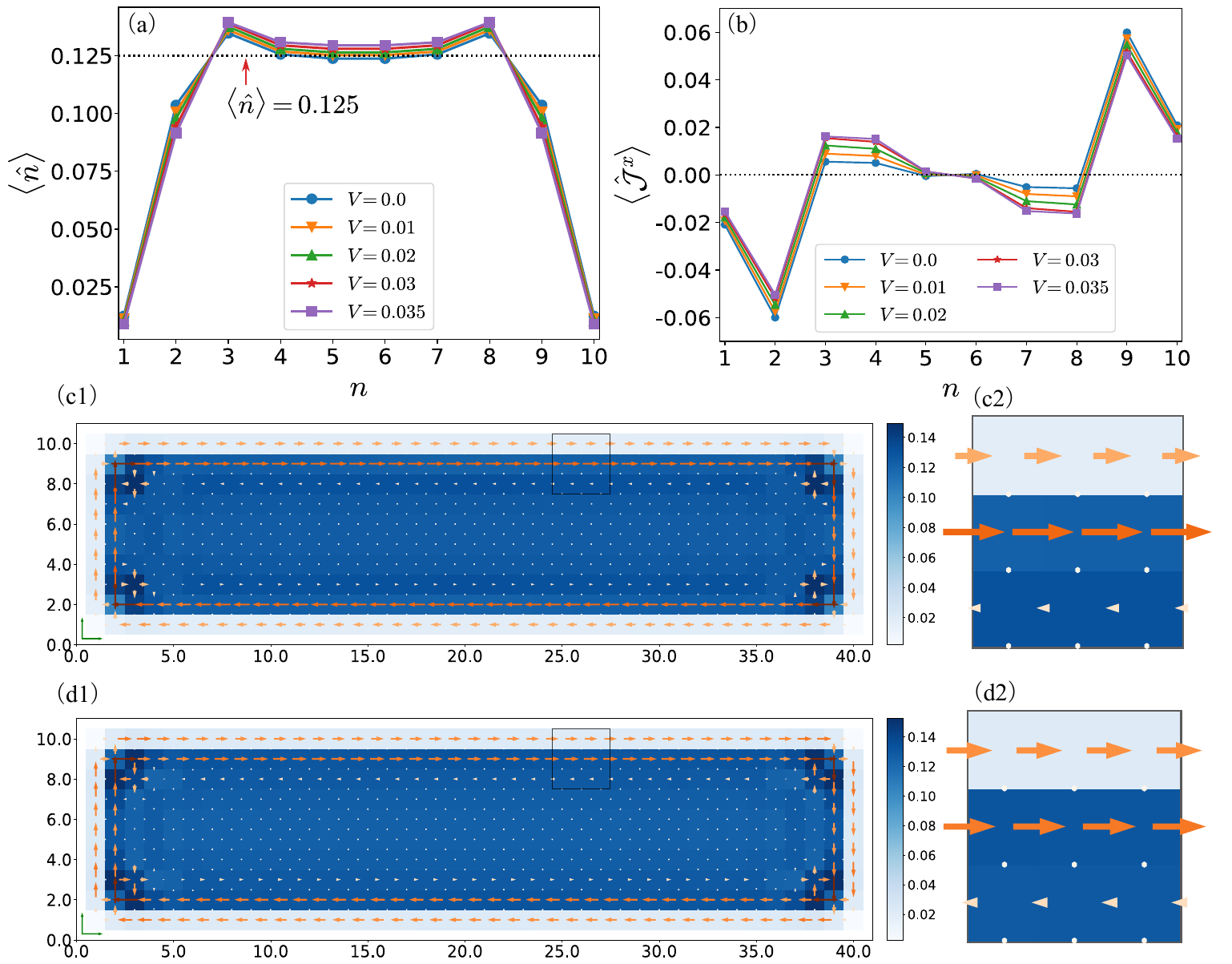}
    \caption{Ground-state particle densities and current distributions of an $\nu=1/2$ FCI on a strip. Panels (a) and (b) show results of iMPS on an $\infty \times 10$ lattice with one particle per column and trapping strength varied from $V=0$ to $V=0.035$. Panels (c1)–(d2) present data on a finite strip with $L_x=40$, $L_y=10$, and $V=0$. The total particle number is $N_{\mathrm{total}}=L_x$ for (c1-c2) and $N_{\mathrm{total}}=L_x+1$ for (d1-d2). Arrow size and color indicate current strength, while square colors denote local particle densities. Panels (c2) and (d2) zoom in on nine sites near the top boundary, revealing enhanced edge currents $\langle \hat{\mathcal{J}}^x \rangle$ after adding one more particle.
    }
    \label{fig:Density_Current_Profile}
\end{figure}

To realize an incompressible $\nu=1/2$ FCI with $n_b=1/8$ and $n_{\phi}=1/4$, the real-space particle density $\langle \hat{n}_{m,n}\rangle$ must be uniform and close to $n_b=1/8$ in the bulk.  For our model on an infinitely long strip ($L_x\rightarrow\infty$), we impose translational symmetry along the $x$-direction and place one particle per column ($N_{\mathrm{total}} = L_x$).  This configuration yields the required bulk density, which, as shown in Fig.~\ref{fig:Density_Current_Profile}(a), remains stable against a weak trapping potential $\hat{H}_{\mathrm{trap}}$. Competing charge-density-wave order is excluded by the uniform density distribution observed along $x$-direction in simulations with finite $L_x = 40$, shown in Fig.~\ref{fig:Density_Current_Profile}(c1). As discussed in Ref.~\cite{wang2022measurable}, incompressibility manifests as the insensitivity of the bulk density to small variations in total particle number. This is demonstrated by comparing Fig.~\ref{fig:Density_Current_Profile}(c) and (d), where one extra particle is added in (d).

An FCI can also be characterized by static local particle currents
%\begin{eqnarray}
$\hat{\mathcal{J}}^x_{(m,m+1);n} = it_x(e^{i\phi n}\hat{a}^\dagger_{m+1,n}\hat{a}_{m,n}-h.c.)$, $ \hat{\mathcal{J}}^y_{m;(n,n+1)} = it_y(\hat{a}^\dagger_{m,n+1}\hat{a}_{m,n}-h.c.)$,
%\end{eqnarray}
as derived from the continuity equation of the local particle density~\cite{wang2022measurable, de1992probability}. These currents are experimentally accessible in ultracold atom systems in optical lattices~\cite{atala2014observation}. In an incompressible FCI, the particle currents exhibit a chiral structure and are predominantly localized near the edges, as shown in Fig.~\ref{fig:Density_Current_Profile}(b). Upon adding an extra boson as in Fig.~\ref{fig:Density_Current_Profile}(d) relative to (c), the boundary currents are enhanced, indicating that the additional particle is expelled to the edge while the incompressible bulk remains unaffected.

\textit{Charge-one edge excitations.---}
We consider an infinitely long strip with width $L_y=10$, and represent the states on it by an infinite MPS (iMPS) with a multi-site unit cell. The ground state is obtained using the VUMPS algorithm~\cite{zauner2018a, laurens2019} and takes the form  
%\begin{eqnarray}
%\label{eq:psi_GS}
$|\psi^{\mathrm{gs}}(\{A\})\rangle=
|\psi^{\mathrm{gs}}(A_1, A_2, \cdots, A_{L_y})\rangle$. 
%\nonumber\\
%&&=\phantom{.}\cdots
%\begin{tikzpicture}[baseline={(A_1.base)}]
%    \node[draw, rounded corners, minimum size=2em, text width=1.5em, text height = 1em, align=center] (A_1) at (0,-1) {$A_1$};
%    \node[draw, rounded corners, minimum size=2em, text width=1.5em, text height = 1em, align=center] (A_2) at (1.2,-1) {$A_2$};
%    \node[rounded corners, minimum size=2em, text width=1.5em, text height = 1em, align=center] (A_cdots) at (2.4,-1) {$\cdots$};
%    \node[draw, rounded corners, minimum size=2em, text width=1.5em, text height = 1em, align=center] (A_Ly) at (3.6,-1) {$A_{L_y}$};
%    \draw[black, thick] (A_1)--(A_2);
%    \draw[black, thick] (A_2)--(A_cdots);
%    \draw[black, thick] (A_cdots)--(A_Ly);
%    \draw[thick] (A_1.west) -- ++(-0.5,0);
%    \draw[thick] (A_1.south) -- ++(0,-0.5);
%    \draw[thick] (A_2.south) -- ++(0,-0.5);
%    \draw[thick] (A_Ly.east) -- ++(+0.5,0);
%    \draw[thick] (A_Ly.south) -- ++(0,-0.5);
%\end{tikzpicture}\cdots.
%\end{eqnarray}
The tensors $\{A_1, A_2, \cdots, A_{L_y}\}$ form a multi-site unit cell of the iMPS, where the subscript $n$ of $A_n$ denotes the $n$th site within the unit cell. The iMPS is formed by repeating this unit cell to the infinite left and right, where the translational symmetry along the strip ($x$-direction) is imposed by construction.  The virtual bond dimension is denoted as $D$, and the physical bond dimension is $d=2$ due to the hard-core boson constraint. The lattice indices are mapped to the iMPS indices in the sawtooth order, as illustrated in Fig.~\ref{fig:Spectrum_Mapping}(b).

The excited state $\ket{\mathcal{E}_p,p}$ with momentum $p$ along $x$-direction and energy $\mathcal{E}_p$ can be obtained using the quasiparticle excitation ansatz~\cite{haegeman2012, haegeman2013, zauner2018b, laurens2019, osborne2025EAobservable}, which can be regarded as a generalization of Feynman's single-mode approximation~\cite{feynman1954}. The ansatz has the form:
\begin{eqnarray}   &&|\psi^{\mathrm{ex}}_{\{A\}}(\{B\};p)\rangle=\phantom{.}\sum_{m\in\mathbb{Z}} e^{i(p+p_0)m}\hat{T}_x^m\sum_{n=1}^{L_y}\nonumber\\
    &&\cdots
    \begin{tikzpicture}[baseline={(A_1.base)}]
        \node[draw, rounded corners, minimum size=2em, text width=1.5em, text height = 1em, align=center] (A_1) at (0,-1) {$A_1$};
        \node[draw, rounded corners, minimum size=2em, text width=1.5em, text height = 1em, align=center] (A_2) at (1.0,-1) {$A_2$};
        \node[rounded corners, minimum size=2em, text width=1.5em, text height = 1em, align=center] (A_cdots1) at (2.0,-1) {$\cdots$};
        \node[draw, rounded corners, minimum size=2em, text width=1.5em, text height = 1em, align=center] (B_n) at (3.0,-1) {$B_n$};
        \node[rounded corners, minimum size=2em, text width=1.5em, text height = 1em, align=center] (A_cdots2) at (4.0,-1) {$\cdots$};
        \node[draw, rounded corners, minimum size=2em, text width=1.5em, text height = 1em, align=center] (A_Ly) at (5.0,-1) {$A_{L_y}$};
        \draw[thick] (A_1.west) -- ++(-0.3,0);
        \draw[thick] (A_1.south) -- ++(0,-0.3);
        \draw[thick] (A_1.east) -- (A_2.west);
        \draw[thick] (A_2.south) -- ++(0,-0.3);
        \draw[thick] (A_2.east) -- (A_cdots1.west);
        \draw[thick] (B_n.west) -- (A_cdots1.east);
        \draw[thick] (B_n.east) -- (A_cdots2.west);
        \draw[thick] (B_n.south) -- ++(0,-0.3);
        %%\draw[thick] (B_n.north east) -- ++(0.3, 0.3);
        \draw[thick] (A_Ly.west) -- (A_cdots2.east);
        \draw[thick] (A_Ly.east) -- ++(+0.3,0);
        \draw[thick] (A_Ly.south) -- ++(0,-0.3);
    \end{tikzpicture}\cdots%\nonumber\\
    %&&\equiv\phantom{.}\sum_{m\in\mathbb{Z}} e^{i(p+p_0)m}\hat{T}_x^m\sum_{n=1}^{L_y}|\phi_{\{A\}}(B_n)\rangle,
    \label{eq:Modified_Ansatz}
\end{eqnarray}
where $m$ indexes the unit cells and $n$ labels the $n$th site within each unit cell, %. The $|\phi_{\{A\}}(B_n)\rangle$ is obtained by replacing 
and in each summand the tensor $A_n$ in a single unit cell of the ground state $\ket{\psi^{\mathrm{gs}}(\{A\})}$ is replaced with $B_n$. The translation operator $\hat{T}_x$ shifts the system by one unit cell, equivalent to a single-site translation along the $x$-direction of the strip. Orthogonality to the ground state is ensured by imposing gauge-fixing conditions on  $B_n$. The relative $\mathrm{U}(1)$ charge $q$ of $B_n$ with respect to $A_n$ labels the charge sector of the excitation relative to that of the ground state.

When the charge sector $q$ of an excited state is nonzero, it induces an energy offset $\mathcal{E}_0$ and a momentum shift $p_0$, such that $\hat{H}\ket{\mathcal{E}_p,p}=(\mathcal{E}_p+\mathcal{E}_0)\ket{\mathcal{E}_p,p}$ and $\hat{T}_x\ket{\mathcal{E}_p,p}=e^{-i(p+p_0)}\ket{\mathcal{E}_p,p}$. Here, $\mathcal{E}_0$ ($p_0$) is defined as the difference in ground-state energy (momentum) between the system with total particle number $N_{\mathrm{total}}+q$ and $N_{\mathrm{total}}$. In the $q=1$ sector of interest, the energy shift $\mathcal{E}_0$ corresponds to the chemical potential $\mu$. Since the iMPS formalism cannot represent globally charged ground states, we extract $\mu$ via finite-size scaling, using $\mu(D, L_x)=\mu_{0}+\mu_D/D+\mu'_D/D^2+\mu_{L_x}/L_x+\mathcal{O}(1/D^3)+\mathcal{O}(1/L_x^2)$, and find $\mu_0\approx -2.65323$ for system with $L_y=10$ and $V=0$. The momentum shift $p_0$ is determined from the real-space path associated with the action of $\hat{T}_x$ on the iMPS.
In the presence of a gauge field, the momentum shift corresponds to the Aharonov-Bohm phase acquired by the added charge along the path,     as illustrated in Fig.~\ref{fig:Spectrum_Mapping}(b). 
Applying Stokes' theorem, we obtain $p_0=q\pi(L_y-3)/4$. This shift is explicitly separated in the ansatz as Eq.~(\ref{eq:Modified_Ansatz}), such that the lowest energy excitation appears at $p=0$. Further details and numerical verification of these shifts are provided in the Supplemental Material~\cite{SM}.

Employing the ansatz $|\psi^{\mathrm{ex}}_{\{A\}}(\{B\};p)\rangle$ for excited states, we obtain the eigenstate $\ket{\mathcal{E}_p,p}$ and its corresponding energy $\mathcal{E}_p$ by variationally minimizing the energy expectation value at fixed momentum $p$, and the variation problem can be converted to a generalized eigenvalue problem~\cite{SM}. %$\mathcal{E}^{\mathrm{ex}}(p)=\langle \psi^{\mathrm{ex}}_{\{A\}}(\{B\};p)|(\hat{H}-\mathcal{E}^{\mathrm{gs}})|\psi^{\mathrm{ex}}_{\{A\}}(\{B\};p)\rangle/\langle \psi^{\mathrm{ex}}_{\{A\}}(\{B\};p)|\psi^{\mathrm{ex}}_{\{A\}}(\{B\};p)\rangle$, where $\mathcal{E}^{\mathrm{gs}}$ denotes the ground state energy, 

%\begin{equation}
%\mathcal{E}^{\mathrm{ex}}(p)=\frac{\langle \psi^{\mathrm{ex}}_{\{A\}}(\{B\};p)|(\hat{H}-\mathcal{E}^{\mathrm{gs}})|\psi^{\mathrm{ex}}_{\{A\}}(\{B\};p)\rangle}{\langle \psi^{\mathrm{ex}}_{\{A\}}(\{B\};p)|\psi^{\mathrm{ex}}_{\{A\}}(\{B\};p)\rangle}
%\end{equation} 

To assess the applicability of the $\chi$LL theory—originally developed for continuum FQH systems—to the case of FCIs on a discrete lattice, we compute the spectral function $\mathcal{A}(k_x, \omega)$. Since the system is simulated on a strip with finite width, the spectral function can be resolved on each row $n = 1, \dots, L_y$. %\textcolor{green}{The row-resolved spectral function $\mathcal{A}_n(k_x,\omega)$ of charge-one excitations for row $n$ is obtained as follows:
%\begin{eqnarray}
%\mathcal{A}_n(k_x,\omega) = &&-\frac{1}{\pi}\mathbf{Im}[\sum_m\int \frac{dt}{2\pi} e^{-i((k_x+p_0) m-(\omega+\mu) t)}\nonumber\\
%&&\quad\cdot\mathcal{G}_n^R(m,t)],
%\end{eqnarray}
%where $\mathcal{G}_n^R(m,t)=-i\theta(t)\bra{\psi^{\mathrm{gs}}}\hat{a}_{m,n}(t)\hat{a}_{0,n}^\dagger(0)\ket{\psi^{\mathrm{gs}}}$.} 
In Lehmann's spectral representation, the spectral function can be evaluated as 
$\mathcal{A}_n(k_x,\omega)=\sum_{\mathcal{E}_p,p}\mathcal{I}_n(\mathcal{E}_p,p)\delta(\omega-\mathcal{E}_p)\delta(k_x-p)$, 
with $\mathcal{I}_n(\mathcal{E}_p,p)=|\bra{\mathcal{E}_p,p}\hat{a}^\dagger_{0,n}(0)\ket{\psi^{\mathrm{gs}}}|^2$, where $\ket{\mathcal{E}_p,p}$ here is an excited state carrying one more $\mathbf{U}(1)$ charge than $\ket{\psi^{\mathrm{gs}}}$.

\begin{figure}[t]
    \includegraphics[width=1.0\linewidth]{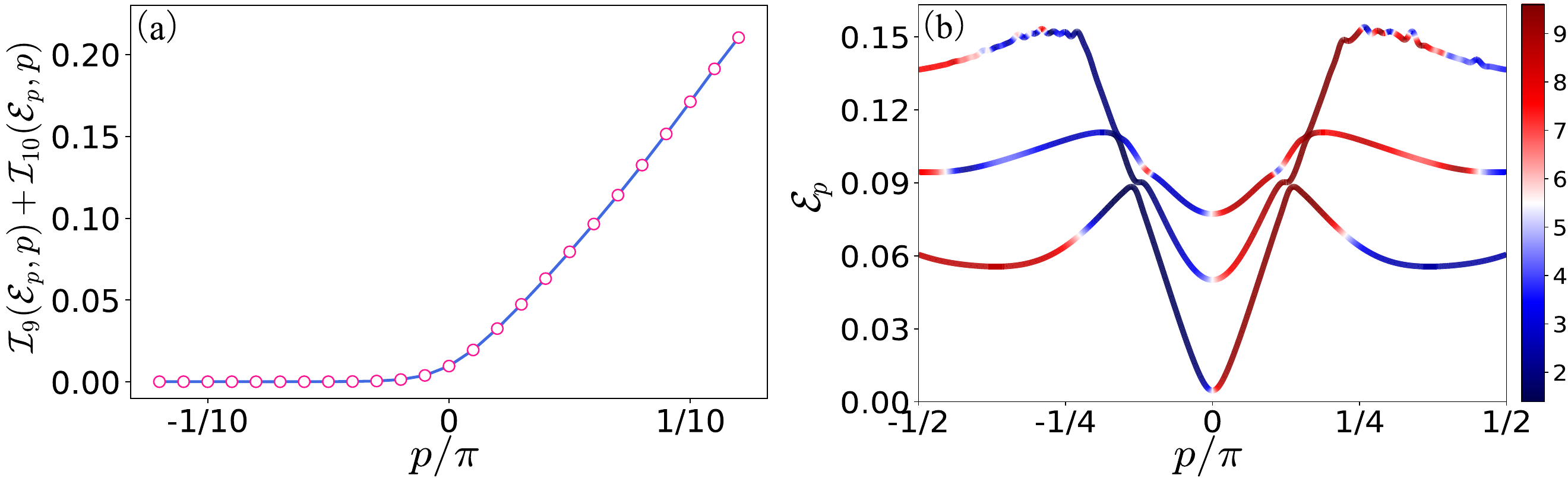}
    \caption{Spectral weight and average positions of charge-one excited states on an infinite long strip with $L_y=10$ and $V=0$. (a) Spectral weight $\mathcal{I}_9(\mathcal{E}_p,p)+\mathcal{I}_{10}(\mathcal{E}_p,p)$ of the lowest chiral excitations near $p=0$. (b) Average positions $\overline{n}$ of the three lowest energy levels of the excited states.}
    \label{fig:average_position}
\end{figure}

The edge spectral functions $\mathcal{A}_1(k_x,\omega)$ and $\mathcal{A}_{10}(k_x,\omega)$ are shown in Fig.~\ref{fig:Spectrum_Mapping}(a). They display the characteristic features predicted by $\chi$LL theory for the $\nu=1/2$ FQH state. First, the edge excitations are gapless, linear, and chiral. The tiny gap observed at $k_x=0$ originates from the finite bond dimension used in the simulations and residual numerical errors in the finite-size scaling of the chemical potential. By increasing the bond dimension $D$, we confirm that the lowest excitation energy decreases, indicating convergence toward a gapless spectrum. Second, the spectral weight increases linearly with both $k_x$ and $\omega$. The weight of $\mathcal{A}_n(k_x,\omega)$ can be extracted from $\mathcal{I}_n(\mathcal{E}_p, p)$. On the lattice, physical edge modes are not confined to a single row; in this case, the low-energy weight $\mathcal{I}_n$ is mainly distributed over rows $n=9,10$ (and symmetrically over $n=1,2$ for the opposite edge)~\cite{SM}. We therefore use $\mathcal{I}_9+\mathcal{I}_{10}$ as a quantitative measure of the edge spectral weight. As shown in Fig.~\ref{fig:average_position}(a), this measure increases linearly with momentum.

The spatial distribution of the excited states can also be characterized by their average position $\overline{n}= \sum_n [n\mathcal{I}_n(\mathcal{E}_p,p) ]/[\sum_n \mathcal{I}_n(\mathcal{E}_p,p)]$.
%\end{equation}
Figure~\ref{fig:average_position}(b) shows the average positions of the three lowest energy levels. The two chiral branches are sharply localized at opposite edges, while the gapped modes display broader and more variable spatial profiles.

\textit{Other system settings.---}
On discrete lattices with finite width, obtaining well-behaved spectral functions, such as those in Fig.~\ref{fig:Spectrum_Mapping} (a), that agree with theoretical predictions originally proposed for continuous systems is challenging. To elucidate the origin of this difficulty, we perform additional calculations for $L_y=8,10,12$ and compare our results with those for $L_y=8$ reported in Dong et al.~\cite{dong2018} and $L_y=11$ in Vashisht et al.~\cite{vashisht2025}.

\begin{figure}[b]    \includegraphics[width=0.9\linewidth]{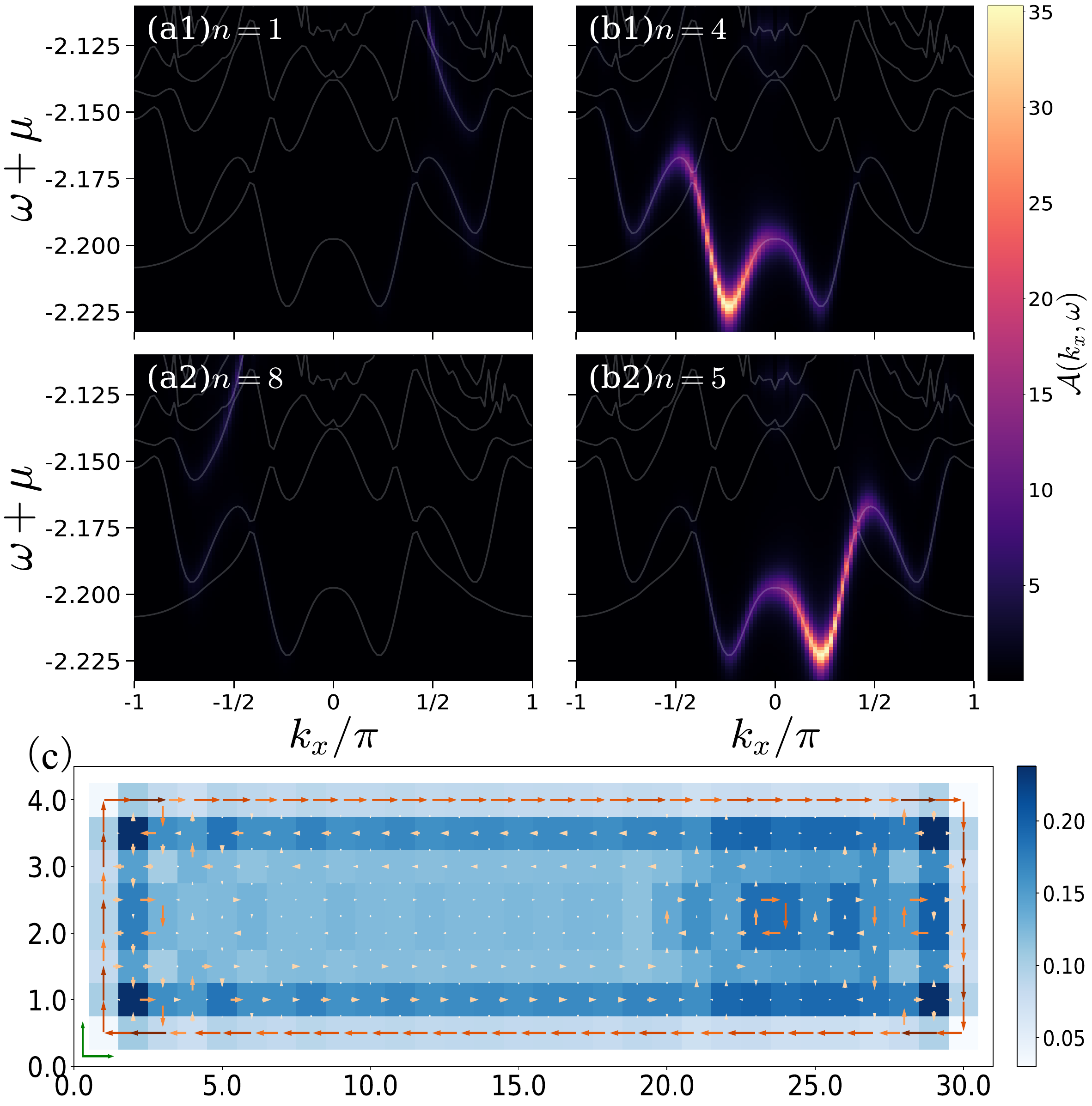}
    \caption{(a,b) Spectral functions $\mathcal{A}_n(k_x,\omega)$ on an $\infty\times 8$ lattice with iMPS bond dimension $D=1500$, for $n=1,8,4,5$ in (a1,a2,b1,b2), respectively. (c) Particle density and current distributions of the ground state with $L_x+1$ particles on the finite strip with $L_x=30$ and $L_y=8$.}
    \label{fig:Ly8}
\end{figure} 

To obtain the correct charge-one edge spectral functions, the edge must provide sufficient space to accommodate an extra boson without disturbing the bulk. In the ground state with one boson per column, a strip with width $L_y=8$ is too narrow. As shown in Fig.~\ref{fig:Ly8}(c), adding an extra boson produces a nonuniform bulk density and induces current vortices in regions of enhanced occupation. In contrast, for $L_y=10$ the extra boson can propagate along the edge without perturbing the bulk [see Fig.~\ref{fig:Density_Current_Profile}(d)]. For charge-one excitations on an infinitely long strip shown in Fig.~\ref{fig:Ly8}(a,b), the spectral functions $\mathcal{A}_4(k_x,\omega)$ and $\mathcal{A}_5(k_x,\omega)$ carry the dominant weight, indicating that the lowest-energy modes propagate primarily in the bulk along rows $n=4$ and $5$. The spectral functions on the edge rows $n=1$ and $8$ have nearly vanishing spectral weight in the low-energy regime. The incorrect spectral weight reported in Ref.~\cite{dong2018} may originate from this finite-width constraint, which also provides a natural explanation for the particle leakage from the edge to the bulk observed in the dynamical process discussed there.

\begin{figure}[t]
    \includegraphics[width=1.0\linewidth]{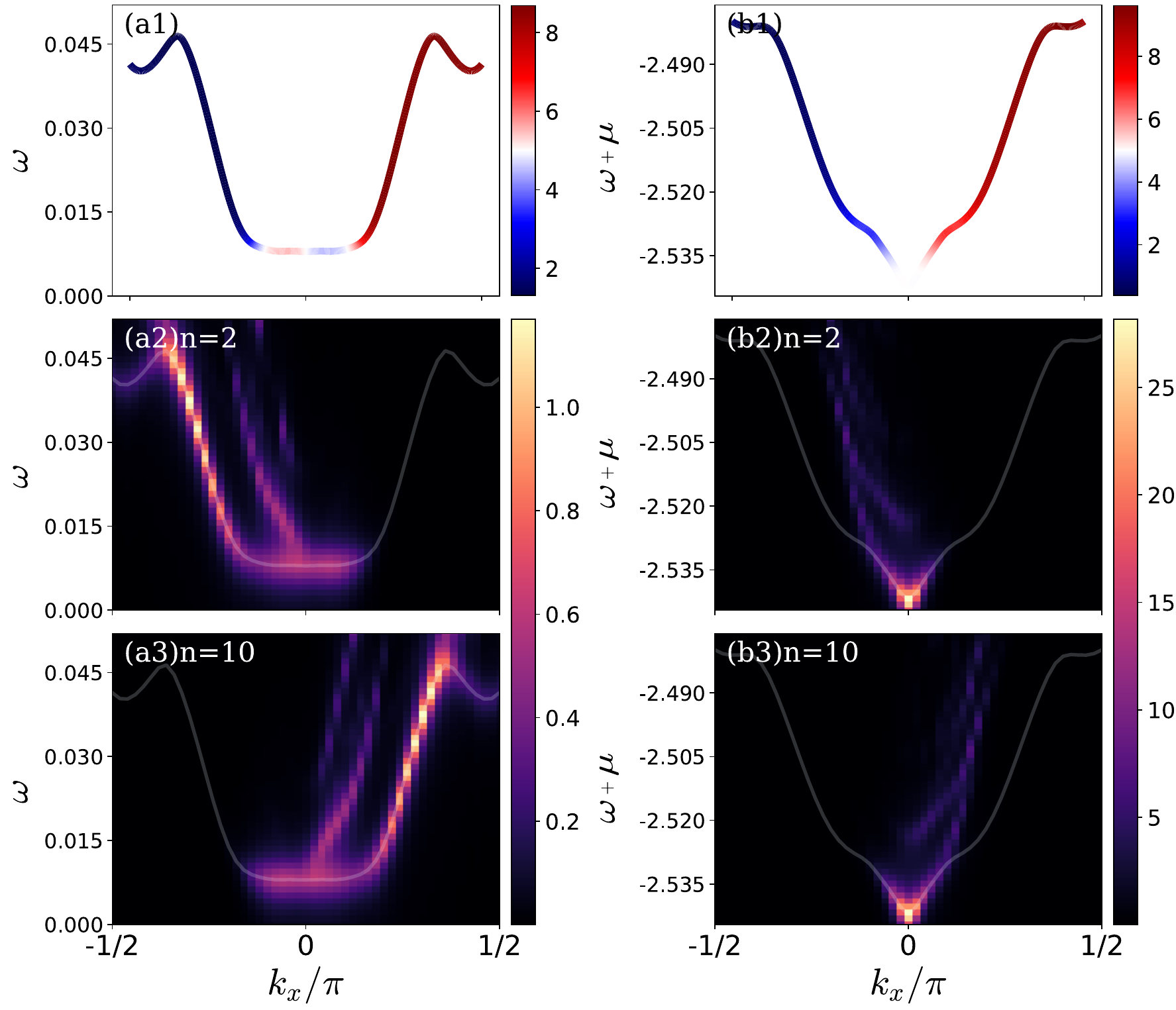}
    \caption{
    Average positions and spectral functions on an infinite strip with width $L_y=11$ and harmonic trapping potential $V=0.01$. (a1) and (b1) show the average positions of charge-zero and charge-one excitations, respectively.
    (a2,a3) display the spectral functions of charge-zero excitations on rows $n=2,10$, while (b2,b3) show the corresponding charge-one spectra. All data are obtained with MPS bond dimension $D=2000$. }
    \label{fig:Compare_Ly_11}
\end{figure}

The harmonic trapping potential has been argued to be essential in certain lattice geometries for obtaining the correct low-energy edge excitations~\cite{binanti2024,vashisht2025,kjall2012,luo2013}. Our results, however, show that on a long strip with finite width—where a natural box-like confinement exists on the edges of the lattice—the harmonic trap is neither necessary nor sufficient once a bulk FCI is established. As shown in the previous section, on a strip with width $L_y=10$, we have obtained the correct charge-one edge spectrum without any harmonic trap. For $L_y=11$, Ref.~\cite{vashisht2025} reported that a harmonic trap is necessary. This necessity, however, arises because a single boson per column fails to stabilize the bulk FCI when the strip is too wide: with $V=0$, the bulk density deviates from $1/8$, signaling the absence of bulk FCI. A finite trap $V$ confines the particle distribution and restores the bulk FCI. Ref.~\cite{vashisht2025} focused on the charge-zero excitations,  defined by the Fourier transform of dynamical density-density correlations, which in the low-energy limit share the same features as charge-one excitations predicted by the field theory of chiral free bosons. They observed that the charge-zero spectrum bends away from linearity and loses edge localization around $k_x=0$. Using the same harmonic potential $V=0.01$ but a larger bond dimension, we compute both charge-zero and charge-one edge spectra in Fig.~\ref{fig:Compare_Ly_11}. We find that, in both cases, the spectral weight is concentrated near the rows $n=2,3$ and $n=9,10$ and exhibits nonchiral behavior at the lowest energies. Notably, the charge-zero spectrum even flattens around $k_x=0$. These results demonstrate that the harmonic trap alone does not guarantee the correct edge spectrum.

To demonstrate the generality of our results and establish the conditions under which a well-behaved spectrum of chiral excitations is observable in discrete lattice systems, we have performed an extensive range of supplemental calculations. These include: (1) for $L_y=10$, charge-one excitations with $V \in \{0.01, 0.02\}$ and charge-zero excitations with $V \in \{0, 0.01\}$; (2) for $L_y=12$, charge-one excitations with $V \in \{0.01, 0.015, 0.02\}$; and (3) for $L_y=12$ with a modified magnetic flux $n_{\phi}=1/5$, where both charge-zero and charge-one sectors are analyzed at $V=0$. As predicted in Ref.~\cite{he2017realizing}, the latter configuration also hosts a $\nu=1/2$ FCI in the bulk. The full numerical results for these cases are provided in the Supplemental Material~\cite{SM}.

Our results demonstrate that the observation of gapless chiral edge states consistent with $\chi$LL theory requires several key conditions: (i) a well-defined bulk FCI state with appropriate magnetic flux and particle density; (ii) sufficient lattice space near the boundary to accommodate the edge excitations; and (iii) a large enough width $L_y$ to suppress hybridization between opposing edges. Additionally, we find that charge-zero edge excitations penetrate further into the bulk than their charge-one counterparts, necessitating a larger system width to recover the gapless limit. We further establish that the specific form of the confining potential is non-essential, as both intrinsic box-like and external harmonic traps support well-behaved edge modes. While a weak harmonic trap may allow for the emergence of extraneous low-energy boundary modes at $k_x\neq 0$, these are efficiently suppressed by increasing the trapping potential $V$, whereas the fundamental gapless chiral modes remain remarkably robust.

\textit{Conclusion and Discussions.---}
In this Letter, we address the fundamental question of whether the spectral functions predicted by chiral Luttinger liquid theory can be observed at the edges of FCIs in discrete lattice systems. Using numerical methods based on tensor-networks on an infinitely long strip of a square lattice with width $L_y=10$, we demonstrate that the charge-one edge excitations of a $\nu=1/2$ FCI exhibit spectral functions in remarkable agreement with theoretical predictions. Specifically, the low-energy modes display clear chiral, linear dispersions, with spectral weights that increase linearly with both momentum and energy. Furthermore, we systematically examine the roles of lattice width, particle density, and confining potentials across different charge sectors, thereby establishing the necessary conditions for the emergence of robust, well-behaved chiral edge excitations in lattice systems with finite width.

Our results provide strong evidence for the universality of the bulk-edge correspondence in fractional Chern insulators, paving the way for future studies of edge physics at other filling factors~\cite{grushin2015, he2017realizing}. Beyond theoretical interest, observing chiral edge states is a central goal in the experimental study of topologically ordered systems. Proposals for detecting chiral edge excitations in optical lattice experiments have been put forward in Refs.~\cite{unal2024,vashisht2025,binanti2024}. Meanwhile, chiral edge dynamics associated with topological order have also been observed on other quantum simulation platforms, including a periodically driven Floquet Kitaev model~\cite{kitaev2006honeycomb} implemented on a superconducting quantum processor~\cite{will2025probing} and ground-state simulations of the Kitaev honeycomb model on a trapped-ion quantum processor~\cite{ali2025robust}. Whether edge states of FCIs from the Harper–Hofstadter–Hubbard model can be realized in these platforms remains an open question. Our results provide useful guidance for designing future experimental explorations.

\textit{Acknowledgments.---}
We thank Botao Wang, Xing-Yu Zhang, Yang Liu, Lei Wang, Hai-Jun Liao, Wei Zheng, Jun-Sen Wang, Qiao-Yi Li, Jia-Lin Chen, Frank Verstraete, Jutho Haegeman, Laurens Vanderstraeten, and Maarten Van Damme for helpful discussions.  XHY and XYD are supported by the Innovation Program for Quantum Science and Technology (Grant No.~2021ZD0301900) and the National Natural Science Foundation of China (Grant No.~12504175).
JYC is supported by National Natural Science Foundation of China (Grants No.~12447107, No.~12304186), and Guangdong Basic and Applied Basic Research Foundation (Grant No.~2024A1515013065).
All numerical codes employed in this study were independently developed by the authors, based on the data structures of TeNPy~\cite{tenpy2024}. The complete implementation is publicly available at \url{https://github.com/xiaohanyangggg/VUMPS_Excitation}. 

\bibliographystyle{apsrev4-2}
\bibliography{reference}

\end{document}

\title{Supplemental Material for \\ ``Chiral Edge Excitations of $\nu=1/2$ Fractional Chern Insulators in the Bosonic Hofstadter Model''}

\author{Xiao-Han Yang}
\affiliation{Hefei National Laboratory, University of Science and Technology of China, Hefei 230088, China}
\affiliation{Hefei National Research Center for Physical Sciences at the Microscale and School of Physical Sciences, University of Science and Technology of China, Hefei 230026, China}

\author{Ji-Yao Chen}
\email{chenjiy3@mail.sysu.edu.cn}
\affiliation{Center for Neutron Science and Technology, Guangdong Provincial Key Laboratory of Magnetoelectric Physics and Devices, School of Physics, Sun Yat-sen University, Guangzhou 510275, China}

\author{Xiao-Yu Dong}
\email{dongxyphys@ustc.edu.cn}
\affiliation{Hefei National Laboratory, University of Science and Technology of China, Hefei 230088, China}

\maketitle

%\section*{Appendix}
\section{Details of Algorithms}
\label{supp:Tensor_Detail}
This section provides some details about the VUMPS algorithm and the method for calculating excited states. 
\subsection{VUMPS and canonical form}
\label{supp:VUMPS}
The VUMPS algorithm uses an iMPS with a multi-site unit cell, as Eq.~(5) in the main text, and imposes translational invariance to construct a variational ansatz for the ground state. The number of sites in a unit cell is $L_y$, and $A_n$ denotes the local tensor at the $n$th site in a unit cell. The mixed canonical form of an iMPS is:  

\begin{equation}
    \cdots
    \begin{tikzpicture}[baseline={(C.base)}]
        \node[draw, rounded corners, minimum height=2.0em, minimum width=2.20em, text width= 2.2em, text height = 1em, inner sep=0pt, align=left] (AL_nm1) at (0,-1) {$A^L_{n-1}$};
        \node[draw, rounded corners, minimum size=2em, text width=1.5em, text height = 1em, align=center] (AL_n) at (1.2,-1) {$A^L_n$};
        \node[draw, rounded corners, minimum size=2em, text width=1.5em, text height = 1em, align=center] (C) at (2.4,-1) {$C_n$};
        \node[draw, rounded corners, minimum height=2.0em, minimum width=2.20em, text width= 2.2em, text height = 1em, inner sep=0pt, align=left] (AR_n_p_1) at (3.6,-1) {$A^R_{n+1}$};
        \node[draw, rounded corners, minimum height=2.0em, minimum width=2.20em, text width= 2.2em, text height = 1em, inner sep=0pt, align=left] (AR_np2) at (4.8,-1) {$A^R_{n+2}$};
        \draw[thick] (AL_nm1.south) -- ++(0,-0.5);
        \draw[thick] (AL_n.south) -- ++(0,-0.5);
        \draw[thick] (AL_nm1.east) -- (AL_n.west);
        \draw[thick] (AL_nm1.west) -- ++(-0.5,0);
        \draw[thick] (AL_n.east) -- (C.west);

        \draw[thick] (C.east) -- (AR_n_p_1.west);
        \draw[thick] (AR_n_p_1.south) -- ++(0,-0.5);
        %%\draw[thick] (B_n.north east) -- ++(0.3, 0.3);
        \draw[thick] (AR_n_p_1.east) -- (AR_np2.west);
        \draw[thick] (AR_np2.east) -- ++(+0.5,0);
        \draw[thick] (AR_np2.south) -- ++(0,-0.5);
    \end{tikzpicture}
    \cdots,
\end{equation}
where $A^L_n$ and $A^R_n$ are left and right isometric tensors at site $n$, respectively, satisfying:
\begin{equation}
    \begin{tikzpicture}[baseline={(mid.base)}]
        \node[draw, rounded corners, minimum size=2em, text width=1.5em, text height = 1em, align=center] (A_L) at (0,0) {$A^L_n$};
        \node[coordinate] (mid) at (0,-0.8) {};
        \node[draw, rounded corners, minimum size=2em, text width=1.5em, text height = 1em, align=center] (A_L_c) at (0,-1.5) {$\bar{A}^{L}_n$};
        \draw[thick] (A_L.west) arc[start angle=90,end angle=270,radius=0.75];
        \draw[thick] (A_L.east) -- ++(0.5,0);
        \draw[thick] (A_L.south) -- (A_L_c.north);
        \draw[thick] (A_L_c.east) -- ++(0.5,0);
    \end{tikzpicture}=\phantom{..}
    \begin{tikzpicture}[baseline={(mid.base)}]
        \node[coordinate] (A_L) at (0,0) {};
        \node[coordinate] (mid) at (0,-0.8) {};
        \node[coordinate] (A_L_c) at (0,-1.5) {};
        \draw[thick] (A_L.west) arc[start angle=90,end angle=270,radius=0.75];
        \draw[thick] (A_L.east) -- ++(0.5,0);
        \draw[thick] (A_L_c.east) -- ++(0.5,0);
    \end{tikzpicture},
\end{equation}
\begin{equation}
    \begin{tikzpicture}[baseline={(mid.base)}]
        \node[draw, rounded corners, minimum size=2em, text width=1.5em, text height = 1em, align=center] (A_R) at (0,0) {$A^R_n$};
        \node[coordinate] (mid) at (0,-0.8) {};
        \node[draw, rounded corners, minimum size=2em, text width=1.5em, text height = 1em, align=center] (A_R_c) at (0,-1.5) {$\bar{A}^{R}_n$};
        \draw[thick] (A_R.east) arc[start angle=90,end angle=-90,radius=0.75];
        \draw[thick] (A_R.west) -- ++(-0.5,0);
        \draw[thick] (A_R.south) -- (A_R_c.north);
        \draw[thick] (A_R_c.west) -- ++(-0.5,0);
    \end{tikzpicture}=\phantom{..}
    \begin{tikzpicture}[baseline={(mid.base)}]
        \node[coordinate] (A_R) at (0,0) {};
        \node[coordinate] (mid) at (0,-0.8) {};
        \node[coordinate] (A_R_c) at (0,-1.5) {};
        \draw[thick] (A_R.west) arc[start angle=90,end angle=-90,radius=0.75];
        \draw[thick] (A_R.west) -- ++(-0.5,0);
        \draw[thick] (A_R_c.west) -- ++(-0.5,0);
    \end{tikzpicture},
\end{equation}
and the bond tensor $C_n$ satisfies:
\begin{equation}
    \begin{tikzpicture}[baseline={(AR_n.base)}]
        \node[draw, rounded corners, minimum size=2em, text width=1.5em, text height = 1em, align=center] (AR_n) at (1.2,-1) {$A^R_n$};
        \node[draw, rounded corners, minimum height=2.0em, minimum width=2.20em, text width= 2.2em, text height = 1em, inner sep=0pt, align=left] (Cn_m_1) at (0,-1) {$C_{n-1}$};
        \draw[thick] (Cn_m_1.east) -- (AR_n.west);
        \draw[thick] (Cn_m_1.west) -- ++(-0.5,0);
        \draw[thick] (AR_n.east) -- ++(0.5,0);
        \draw[thick] (AR_n.south) -- ++(0,-0.5);
    \end{tikzpicture}=
    \begin{tikzpicture}[baseline={(AL_n.base)}]
        \node[draw, rounded corners, minimum size=2em, text width=1.5em, text height = 1em, align=center] (AL_n) at (0,-1) {$A^L_n$};
        \node[draw, rounded corners, minimum size=2em, text width=1.5em, text height = 1em, align=center] (Cn) at (1.2,-1) {$C_n$};
        \draw[thick] (Cn.west) -- (AL_n.east);
        \draw[thick] (Cn.east) -- ++(+0.5,0);
        \draw[thick] (AL_n.west) -- ++(-0.5,0);
        \draw[thick] (AL_n.south) -- ++(0,-0.5);
    \end{tikzpicture}\coloneqq
    \begin{tikzpicture}[baseline={(Ac_n.base)}]
        \node[draw, rounded corners, minimum size=2em, text width=1.5em, text height = 1em, align=center] (Ac_n) at (0,-1) {$A^C_n$};
        \draw[thick] (Ac_n.south) -- ++(0,-0.5);
        \draw[thick] (Ac_n.east) -- ++(0.5,0);
        \draw[thick] (Ac_n.west) -- ++(-0.5,0);
    \end{tikzpicture}.
\end{equation}
The $A^C_n$ defined here is called the center-site tensor. 

The local tensors $\{A_n \in\mathbb{C}^{D_{n-1}\times d_n \times D_{n}}\}$ can be mapped to iMPS $|\psi^{\mathrm{gs}}(\{A\})\rangle\in\mathbb{H}\simeq \otimes_{m,n}\mathbb{C}^{d_{m,n}}$. 
Thus $|\psi^{\mathrm{gs}}(\{A\})\rangle$ can be parameterized with $A$'s entries, and we can view the wave funtion $|\psi^{\mathrm{gs}}(\{A\})\rangle$ as the mapping from the manifold $\mathbf{A}$ of tensor entries to the manifold $\mathbf{M}\subset\mathbb{H}$ spanned by $|\psi^{\mathrm{gs}}\rangle$, $\psi^{\mathrm{gs}}:A\rightarrow M:\{(A_n)_{\alpha\beta i}\}\mapsto|\psi^{\mathrm{gs}}(\{A\})\rangle$.
The tangent vector~\cite{laurens2019, haegeman2013} for the state $|\psi^{\mathrm{gs}}(\{A\})\rangle$ can be written in the canonical form as:

\begin{equation}
    \label{A6}
    |\mathcal{T}(\{G\};\{A\})\rangle=\sum_{m\in{\mathbb{Z}}}\sum_{n=1}^{L_y} \hat{T}_x^m
    \begin{tikzpicture}[baseline={(A_1.base)}]
        \node[draw, rounded corners, minimum size=2em, text width=1.5em, text height = 1em, align=center] (A_1) at (0,-1) {$A^L_1$};
        \draw[thick] (A_1.west) -- ++(-0.5,0);
        \draw[thick] (A_1.south) -- ++(0,-0.5);
        \draw[thick] (A_1.east) -- ++(0.5,0);
    \end{tikzpicture}
    \cdots
    \begin{tikzpicture}[baseline={(A_1.base)}]
        \node[draw, rounded corners, minimum height=2.0em, minimum width=2.20em, text width= 2.2em, text height = 1em, inner sep=0pt, align=left] (AL_nm1) at (1.2,-1) {$A^L_{n-1}$};
        \node[draw, rounded corners, minimum size=2em, text width=1.5em, text height = 1em, align=center] (G_n) at (2.4,-1) {$G_n$};
        \node[draw, rounded corners, minimum height=2.0em, minimum width=2.20em, text width= 2.2em, text height = 1em, inner sep=0pt, align=left] (AR_np1) at (3.6,-1) {$A^R_{n+1}$};
        \draw[thick] (G_n.west) -- (AL_nm1.east);
        \draw[thick] (AL_nm1.south) -- ++(0,-0.5);
        \draw[thick] (AL_nm1.west) -- ++(-0.5,0);
        \draw[thick] (G_n.east) -- (AR_np1.west);
        \draw[thick] (AR_np1.south) -- ++(0,-0.5);
        \draw[thick] (AR_np1.east) -- ++(0.5,0);
        \draw[thick] (G_n.south) -- ++(0,-0.5);
    \end{tikzpicture}
    \cdots
    \begin{tikzpicture}[baseline=(A_1.base)]
        %%\draw[thick] (B_n.north east) -- ++(0.3, 0.3);
        \node[draw, rounded corners, minimum size=2em, text width=1.5em, text height = 1em, align=center] (A_Ly) at (4.8,-1) {$A^R_{L_y}$};
        \draw[thick] (A_Ly.west) -- ++(-0.5,0);
        \draw[thick] (A_Ly.east) -- ++(+0.5,0);
        \draw[thick] (A_Ly.south) -- ++(0,-0.5);
    \end{tikzpicture}.
\end{equation}
The  pair $(\{|\psi^{\mathrm{gs}}\{A\}\rangle\}, \{|\mathcal{T}(\{G\};\{A\})\rangle\})$ forms a tangent bundle $\mathcal{T}\mathbf{M}$, where the tangent space $\mathcal{T}_{|\psi^{\mathrm{gs}}(\{A\})\rangle}\mathbf{M}$ at point $|\psi^{\mathrm{gs}}(\{A\})\rangle$ is also embedded in $\mathbb{H}$.
Although the mapping $\mathbf{A}$ to $\mathbf{M}$ is well-defined now, the tangent vector has a gauge freedom:
\begin{equation}
    \label{A7}
    \begin{aligned}
        |\mathcal{T}(\{G\};\{A\})\rangle&=|\mathcal{T}(\{G+YA-AY\};\{A\})\rangle,
    \end{aligned}
\end{equation}
where $\{G+YA-AY\}$ denotes $G_n+Y_{n-1}A^R_n-A^L_nY_{n}$, $ n=1,2,\cdots,L_y$, and $Y_n$ are arbitrary rank-2 tensors with bond dimension $D_{n-1}\times D_n$. 

We can check that if $G_n$ is left-orthogonal to $A^L_n$, the gauge degrees of freedom are fixed.
So we fix the gauge by requiring that $G_n$ is parameterized with a $D_n(d-1)\times D_n$ tensor $X_n$:
\begin{equation}
    \label{A8}
    \begin{tikzpicture}[baseline={(Gn.base)}]
        \node[draw, rounded corners, minimum size=2em, text width=1.5em, text height = 1em, align=center] (Gn) at (0,-1) {$G_n$};
        \draw[thick] (Gn.east) -- ++( +0.5,0);
        \draw[thick] (Gn.south) -- ++(0,-0.5);
        \draw[thick] (Gn.west) -- ++(-0.5,0);
    \end{tikzpicture}=
    \begin{tikzpicture}[baseline={(NL.base)}]
        \node[draw, rounded corners, minimum size=2em, text width=1.5em, text height = 1em, align=center] (NL) at (0,-1) {$N_n^{L}$};
        \node[draw, rounded corners, minimum size=2em, text width=1.5em, text height = 1em, align=center] (XG) at (1.2,-1) {$X_n$};
        \draw[thick] (XG.west) -- (NL.east);
        \draw[thick] (XG.east) -- ++(+0.5,0);
        \draw[thick] (NL.west) -- ++(-0.5,0);
        \draw[thick] (NL.south) -- ++(0,-0.5);
    \end{tikzpicture},
\end{equation}
where $N_n^L$ is defined by:
\begin{equation}
    \begin{tikzpicture}[baseline={(mid.base)}]
        \node[draw, rounded corners, minimum size=2em, text width=1.5em, text height = 1em, align=center] (AL_n) at (0,0) {$A^L_n$};
        \node[coordinate] (mid) at (0,-0.8) {};
        \node[draw, rounded corners, minimum size=2em, text width=1.5em, text height = 1em, align=center] (NL_n_c) at (0,-1.5) {$\bar{N}^L_n$};
        \draw[thick] (AL_n.west) arc[start angle=90,end angle=270,radius=0.75];
        \draw[thick] (AL_n.east) -- ++(0.5,0);
        \draw[thick] (AL_n.south) -- (NL_n_c.north);
        \draw[thick] (NL_n_c.east) -- ++(0.5,0);
    \end{tikzpicture}=0,
\end{equation}
\begin{equation}
    \begin{tikzpicture}[baseline={(mid.base)}]
        \node[draw, rounded corners, minimum size=2em, text width=1.5em, text height = 1em, align=center] (NL_n) at (0,0) {$N^L_n$};
        \node[coordinate] (mid) at (0,-0.8) {};
        \node[draw, rounded corners, minimum size=2em, text width=1.5em, text height = 1em, align=center] (NL_n_c) at (0,-1.5) {$\bar{N}^L_n$};
        \draw[thick] (NL_n.west) arc[start angle=90,end angle=270,radius=0.75];
        \draw[thick] (NL_n.east) -- ++(0.5,0);
        \draw[thick] (NL_n.south) -- (NL_n_c.north);
        \draw[thick] (NL_n_c.east) -- ++(0.5,0);
    \end{tikzpicture}=\phantom{..}
    \begin{tikzpicture}[baseline={(mid.base)}]
        \node[coordinate] (A_L) at (0,0) {};
        \node[coordinate] (mid) at (0,-0.8) {};
        \node[coordinate] (A_L_c) at (0,-1.5) {};
        \draw[thick] (A_L.west) arc[start angle=90,end angle=270,radius=0.75];
        \draw[thick] (A_L.east) -- ++(0.2,0);
        \draw[thick] (A_L_c.east) -- ++(0.2,0);
    \end{tikzpicture}.
\end{equation}
The definition means if we group $A^L_{n}$ and $N^L_{n}$'s left virtual and physical legs together and write their indices as $((\alpha i),\beta)$, the matrix $N^L_n$'s columns will be $D(d-1)$ orthonormal basis orthogonal to $A^L_n$'s column space.

There is a `gradient vector'~\cite{laurens2019, zauner2018a} in the tangent space with the energy $\langle\psi^{\mathrm{gs}}(\{A\})|\hat{H}|\psi^{\mathrm{gs}}(\{A\})\rangle$ being the variational optimum. We can first vary the energy and then expand it to the first order of $\delta(|\psi^{\mathrm{gs}}(\{A\})\rangle)$ to get its cotangent vector correspondence. 
Or we can directly write it in a form with explicit physical meaning:
\begin{equation}
    \label{A9}
    |\mathcal{T}(\{\tilde{G}\};\{A\}\})\rangle=\mathcal{P}_{\{A\}}(\hat{H}-\mathcal{E}^{gs})|\psi^{\mathrm{gs}}(\{A\})\rangle.
\end{equation}
The projector $\mathcal{P}_{\{A\}}$ projects an arbitrary state to the tangent space $\mathcal{T}_{\psi^{\mathrm{gs}}(\{A\})\rangle}\mathbf{M}$, which means within the iMPS formalism, the gradient vector is obtained by projecting an usual energy gradient vector to the iMPS tangent space. 
Eq.~($\ref{A9}$) turns out to be the tangent vector with :
\begin{equation}
    \tilde{G}_n=H_{A^C,n}(A_n^C)-A^L_n H_{C,n}(C_n).
\end{equation}

We write the Hamiltonian as an infinite matrix product operator (iMPO)~\cite{crosswhite2008, michel2010, cirac2021}, where the local tensors $O_n$ are rank-4 tensors with bond dimension $d^2\times\chi^2$. Here we adopt the convention that for fixed physical indices $i,i'$, the matrix $(O_n)_{i, i';l,m}$ is an upper-triangular matrix. $H_{A_n^C}$ and $H_{C_n}$ are defined by:
\begin{equation}
    H_{A^C,n}(A_n^C)=
    \begin{tikzpicture}[baseline={(O_n.base)}]
        \node[draw, rounded corners, minimum size=2em, text width=1.5em, text height = 1em, align=center] (AC_n) at (0,0) {$A^C_n$};
        \node[draw, rounded corners, minimum size=2em, text width=1.5em, text height = 1em, align=center] (O_n) at (0,-1.0) {$O_n$};
        \node[draw, rounded corners, minimum width=2.2em, minimum height=7.7em, text width=2.2em, text height = 1em, align=center] (E_L) at (-1.5, -1.0) {$E_{L,n}$};
        \node[draw, rounded corners, minimum width=2.2em, minimum height=7.7em, text width=2.2em, text height = 1em, align=center] (E_R) at (1.5, -1.0) {$E_{R,n}$};
        \node[coordinate] (newnodeuL) at ([xshift=0.0cm, yshift=+1.0cm]E_L.east) {};
        \node[coordinate] (newnodeuR) at ([xshift=0.0cm, yshift=+1.0cm]E_R.west) {};
        \draw[thick] (AC_n.south) -- (O_n.north);
        \draw[thick] (AC_n.west) -- (newnodeuL);
        \draw[thick] (AC_n.east) -- (newnodeuR);
        \draw[thick] (O_n.west) -- (E_L.east);
        \draw[thick] (O_n.east) -- (E_R.west);
        \draw[thick] (O_n.south) -- ++(0,-1.5);
        \node[coordinate] (newnodeL) at ([xshift=0.0em, yshift=-1.0cm]E_L.east) {};
        \draw[thick] (newnodeL) -- ++(+0.5,0) -- ++(0,-0.6) -- ++(-0.5,0);
        \node[coordinate] (newnodeR) at ([xshift=0.0em, yshift=-1.0cm]E_R.west) {};
        \draw[thick] (newnodeR) -- ++(-0.5,0) -- ++(0,-0.6) -- ++(+0.5,0);
    \end{tikzpicture}\coloneqq 
    \begin{tikzpicture}[baseline={(O_n.base)}]
        \node[draw, rounded corners, minimum size=2em, text width=1.5em, text height = 1em, align=center] (AC_n) at (0,-1.0) {$A'^C_n$};
        \draw[thick] (AC_n.south) -- ++(0,-0.5);
        \draw[thick] (AC_n.west) -- ++(-0.43,0);
        \draw[thick] (AC_n.east) -- ++(0.43,0);
    \end{tikzpicture},
\end{equation}
and
\begin{equation}
    \label{A14}
    H_{C,n}(C_n)=
    \begin{tikzpicture}[baseline={(AL_c.base)}]
        \node[draw, rounded corners, minimum size=2em, text width=1.5em, text height = 1em, align=center] (AC_n) at (0,0) {$A'^C_n$};
        \node[draw, rounded corners, minimum size=2em, text width=1.5em, text height = 1em, align=center] (AL_c) at (0,-1.5) {$\bar{A}^L_n$};
        \draw[thick] (AC_n.west) arc[start angle=90,end angle=270,radius=0.75];
        \draw[thick] (AC_n.east) -- ++(0.8,0) -- ++(0,-2.0) -- ++(0.6,0);
        \draw[thick] (AC_n.south) -- (AL_c.north);
        \draw[thick] (AL_c.east) -- ++(0.4,0) -- ++(0,-0.5) -- ++(-1.6,0);
    \end{tikzpicture}\coloneqq
    \begin{tikzpicture}[baseline={(C_n.base)}]
        \node[draw, rounded corners, minimum size=2em, text width=1.5em, text height = 1em, align=center] (C_n) at (0,-1.0) {$C'_n$};
        \draw[thick] (C_n.west) -- ++(-0.43,0);
        \draw[thick] (C_n.east) -- ++(0.43,0);
    \end{tikzpicture},
\end{equation}
where $E_{L,n}$ and $E_{R,n}$ are rank-3 MPO environments obtained by contraction in the unit cell to the $n$th site:
\begin{equation}
    \begin{tikzpicture}[baseline={(E_L.base)}]
        \node[draw, rounded corners, minimum width=2.2em, minimum height=7.7em, text width=2.2em, text height = 1em, align=center] (E_L) at (0.0, -1.0) {$E_{L,n}$};
        \node[coordinate] (newnodeuL) at ([xshift=0.0cm, yshift=+1.0cm]E_L.east) {};
        \node[coordinate] (newnodedL) at ([xshift=0.0cm, yshift=-1.0cm]E_L.east) {};
        \draw[thick] (E_L.east) -- ++(+0.5,0);
        \draw[thick] (newnodeuL) -- ++(+0.5,0);
        \draw[thick] (newnodedL) -- ++(+0.5,0);
    \end{tikzpicture}=
    \begin{tikzpicture}[baseline={(E_L1.base)}]
        \node[draw, rounded corners, minimum width=2.2em, minimum height=7.7em, text width=2.2em, text height = 1em, align=center] (E_L1) at (0.0, -1.0) {$E_{L,1}$};
        \node[coordinate] (newnodeuL) at ([xshift=0.0cm, yshift=+1.0cm]E_L1.east) {};
        \node[coordinate] (newnodedL) at ([xshift=0.0cm, yshift=-1.0cm]E_L1.east) {};
        \node[draw, rounded corners, minimum size=2em, text width=1.5em, text height = 1em, align=center] (A_L1) at (1.2, 0.0) {$A^L_1$};
        \node[draw, rounded corners, minimum size=2em, text width=1.5em, text height = 1em, align=center] (O_1) at (1.2,-1.0) {$O_1$};
        \node[draw, rounded corners, minimum size=2em, text width=1.5em, text height = 1em, align=center] (A_L1_c) at (1.2,-2.0) {$\bar{A}^L_1$};
        \draw[thick] (newnodeuL) -- (A_L1.west);
        \draw[thick] (E_L1.east) -- (O_1.west);
        \draw[thick] (newnodedL) -- (A_L1_c.west);
        \draw[thick] (A_L1.south) -- (O_1.north);
        \draw[thick] (O_1.south) -- (A_L1_c.north);
        \draw[thick] (A_L1.east) -- ++(0.25,0);
        \draw[thick] (A_L1_c.east) -- ++(0.25,0);
        \draw[thick] (O_1.east) -- ++(0.25,0);
    \end{tikzpicture}\cdots
    \begin{tikzpicture}[baseline={(E_L1.base)}]
        \node[draw, rounded corners, minimum height=2em, minimum width=2.2em, text width=2.2em, text height = 1em, inner sep=0pt, align=left] (AL_n_m_1) at (0,0) {$A^L_{n-1}$};
        \node[draw, rounded corners, minimum height=2em, minimum width=2.2em, text width=2.2em, text height = 1em, inner sep=0pt, align=left] (O_n_m_1) at (0,-1.0) {$O_{n-1}$};
        \node[draw, rounded corners, minimum height=2em, minimum width=2.2em, text width=2.2em, text height = 1em, inner sep=0pt, align=left] (AL_n_m_1_c) at (0,-2.0) {$\bar{A}^L_{n-1}$};
        \draw[thick] (AL_n_m_1.south) -- (O_n_m_1.north);
        \draw[thick] (O_n_m_1.south) -- (AL_n_m_1_c.north);
        \draw[thick] (AL_n_m_1.west) -- ++(-0.25,0);
        \draw[thick] (AL_n_m_1.east) -- ++(+0.25,0);
        \draw[thick] (O_n_m_1.west) -- ++(-0.25,0);
        \draw[thick] (O_n_m_1.east) -- ++(+0.25,0);
        \draw[thick] (AL_n_m_1_c.west) -- ++(-0.25,0);
        \draw[thick] (AL_n_m_1_c.east) -- ++(+0.25,0);
    \end{tikzpicture},
\end{equation}
and 
\begin{equation}
    \begin{tikzpicture}[baseline={(E_R.base)}]
        \node[draw, rounded corners, minimum width=2.2em, minimum height=7.7em, text width=2.2em, text height = 1em, align=center] (E_R) at (0.0, -1.0) {$E_{R,n}$};
        \node[coordinate] (newnodeuL) at ([xshift=0.0cm, yshift=+1.0cm]E_R.west) {};
        \node[coordinate] (newnodedL) at ([xshift=0.0cm, yshift=-1.0cm]E_R.west) {};
        \draw[thick] (E_R.west) -- ++(-0.5,0);
        \draw[thick] (newnodeuL) -- ++(-0.5,0);
        \draw[thick] (newnodedL) -- ++(-0.5,0);
    \end{tikzpicture}=
    \begin{tikzpicture}[baseline={(O_n_p_1.base)}]
        \node[draw, rounded corners, minimum height=2em, minimum width=2.2em, text width=2.2em, text height = 1em, inner sep=0pt, align=left] (AR_n_p_1) at (0.0, 0.0) {$A^R_{n+1}$};
        \node[draw, rounded corners, minimum height=2em, minimum width=2.2em, text width=2.2em, text height = 1em, inner sep=0pt, align=left] (O_n_p_1) at (0.0,-1.0) {$O_{n+1}$};
        \node[draw, rounded corners, minimum height=2em, minimum width=2.2em, text width=2.2em, text height = 1em, inner sep=0pt, align=left] (AR_n_p_1_c) at (0.0,-2.0) {$\bar{A}^R_{n+1}$};
        \draw[thick] (AR_n_p_1.south) -- (O_n_p_1.north);
        \draw[thick] (O_n_p_1.south) -- (AR_n_p_1_c.north);
        \draw[thick] (AR_n_p_1.west) -- ++(-0.25,0);
        \draw[thick] (AR_n_p_1.east) -- ++(+0.25,0);
        \draw[thick] (O_n_p_1.west) -- ++(-0.25,0);
        \draw[thick] (O_n_p_1.east) -- ++(+0.25,0);
        \draw[thick] (AR_n_p_1_c.west) -- ++(-0.25,0);
        \draw[thick] (AR_n_p_1_c.east) -- ++(+0.25,0);
    \end{tikzpicture}\cdots
    \begin{tikzpicture}[baseline={(ER_Ly.base)}]
        \node[draw, rounded corners, minimum width=2.8em, minimum height=7.7em, text width=2.6em, text height = 1em, align=center] (ER_Ly) at (1.2, -1.0) {$E_{R,L_y}$};
        \node[coordinate] (newnodeuR) at ([xshift=0.0cm, yshift=+1.0cm]ER_Ly.west) {};
        \node[coordinate] (newnodedR) at ([xshift=0.0cm, yshift=-1.0cm]ER_Ly.west) {};
        \node[draw, rounded corners, minimum height=2em, minimum width=2.2em, text width=2.2em, text height = 1em, inner sep=0pt, align=center] (AR_Ly) at (0.0,0) {$A^R_{L_y}$};
        \node[draw, rounded corners, minimum height=2em, minimum width=2.2em, text width=2.2em, text height = 1em, inner sep=0pt, align=center] (O_Ly) at (0.0,-1.0) {$O_{L_y}$};
        \node[draw, rounded corners, minimum height=2em, minimum width=2.2em, text width=2.2em, text height = 1em, inner sep=0pt, align=center] (AR_Ly_c) at (0.0,-2.0) {$\bar{A}^R_{L_y}$};
        \draw[thick] (AR_Ly.south) -- (O_Ly.north);
        \draw[thick] (O_Ly.south) -- (AR_Ly_c.north);
        \draw[thick] (AR_Ly.east) -- (newnodeuR);
        \draw[thick] (AR_Ly_c.east) -- (newnodedR);
        \draw[thick] (O_Ly.west) -- ++(-0.25,0);
        \draw[thick] (O_Ly.east) -- (ER_Ly.west);
        \draw[thick] (AR_Ly_c.west) -- ++(-0.25,0);
        \draw[thick] (AR_Ly.west) -- ++(-0.25,0);
    \end{tikzpicture}.
\end{equation}
The environments $E_{L,1}$ and $E_{R,L_y}$ can be obtained by applying MPO transfer matrices to the left and right arbitrary initial environments. 
However, the contraction will lead to divergence in the evaluation of energy since the system is extensive. So we need to modify the action of MPO transfer matrices:
\begin{equation}
    \begin{aligned}
        \mathbb{T}^{L}_{L}\rightarrow&\tilde{\mathbb{T}}^{L}_{L}=\mathbb{T}^{L}_{L}-\mathbb{P}_L\\
        \coloneqq&
        \begin{tikzpicture}[baseline={(O_1.base)}]
            \node[draw, rounded corners, minimum size=2em, text width=1.5em, text height = 1em, align=center] (A_L1) at (1.2, 0.0) {$A^L_1$};
            \node[draw, rounded corners, minimum size=2em, text width=1.5em, text height = 1em, align=center] (O_1) at (1.2,-1.0) {$O_1$};
            \node[draw, rounded corners, minimum size=2em, text width=1.5em, text height = 1em, align=center] (A_L1_c) at (1.2,-2.0) {$\bar{A}^L_1$};
            \draw[thick] (A_L1.south) -- (O_1.north);
            \draw[thick] (O_1.south) -- (A_L1_c.north);
            \draw[thick] (A_L1.east) -- ++(0.25,0);
            \draw[thick] (A_L1_c.east) -- ++(0.25,0);
            \draw[thick] (O_1.east) -- ++(0.25,0);
            \draw[thick] (A_L1.west) -- ++(-0.25,0);
            \draw[thick] (A_L1_c.west) -- ++(-0.25,0);
            \draw[thick] (O_1.west) -- ++(-0.25,0);
        \end{tikzpicture}\cdots
        \begin{tikzpicture}[baseline={(O_1.base)}]
            \node[draw, rounded corners, minimum height=2em, minimum width=2.2em, text width=2.2em, text height = 1em, inner sep=0pt, align=center] (AL_Ly) at (0.0,0) {$A^L_{L_y}$};
            \node[draw, rounded corners, minimum height=2em, minimum width=2.2em, text width=2.2em, text height = 1em, inner sep=0pt, align=center] (O_Ly) at (0.0,-1.0) {$O_{L_y}$};
            \node[draw, rounded corners, minimum height=2em, minimum width=2.2em, text width=2.2em, text height = 1em, inner sep=0pt, align=center] (AL_Ly_c) at (0.0,-2.0) {$\bar{A}^L_{L_y}$};
            \draw[thick] (AR_Ly.south) -- (O_Ly.north);
            \draw[thick] (O_Ly.south) -- (AL_Ly_c.north);
            \draw[thick] (AL_Ly.east) -- ++(0.25,0);
            \draw[thick] (AL_Ly_c.east) -- ++(0.25,0);
            \draw[thick] (O_Ly.east) -- ++(0.25,0);
            \draw[thick] (AL_Ly.west) -- ++(-0.25,0);
            \draw[thick] (AL_Ly_c.west) -- ++(-0.25,0);
            \draw[thick] (O_Ly.west) -- ++(-0.25,0);
        \end{tikzpicture}-
        \begin{tikzpicture}[baseline={(O_1.base)}]
            \node[circle, draw, inner sep=0pt, minimum size=0.6cm] (rho) at (0,-1.0) {$\rho_R$};
            \node[circle, draw, minimum size=0.5cm] (Id) at (0.75,-1.0) {$\mathbf{I}$};
            \draw (-0.45,-1.0) circle (0.08cm);
            \draw (1.2, -1.0) circle (0.08cm);
            \draw[thick] (-0.525, -1.0) -- (-0.95, -1.0);
            \draw[thick] (1.275, -1.0) -- (1.7, -1.0);
            \draw[thick] (rho.north) arc[start angle=0,end angle=90,radius=0.8];
            \draw[thick] (rho.south) arc[start angle=0,end angle=-90,radius=0.8];
            \draw[thick] (Id.north) arc[start angle=180,end angle=90,radius=0.8];
            \draw[thick] (Id.south) arc[start angle=180,end angle=270,radius=0.8];
        \end{tikzpicture},
    \end{aligned}
\end{equation}
\begin{equation}
    \begin{aligned}
        \mathbb{T}^{R}_{R}\rightarrow&\tilde{\mathbb{T}}^{R}_{R}=\mathbb{T}^{R}_{R}-\mathbb{P}_R\\
        \coloneqq&
        \begin{tikzpicture}[baseline={(O_1.base)}]
            \node[draw, rounded corners, minimum size=2em, text width=1.5em, text height = 1em, align=center] (A_R1) at (1.2, 0.0) {$A^R_1$};
            \node[draw, rounded corners, minimum size=2em, text width=1.5em, text height = 1em, align=center] (O_1) at (1.2,-1.0) {$O_1$};
            \node[draw, rounded corners, minimum size=2em, text width=1.5em, text height = 1em, align=center] (A_R1_c) at (1.2,-2.0) {$\bar{A}^R_1$};
            \draw[thick] (A_R1.south) -- (O_1.north);
            \draw[thick] (O_1.south) -- (A_R1_c.north);
            \draw[thick] (A_R1.east) -- ++(0.25,0);
            \draw[thick] (A_R1_c.east) -- ++(0.25,0);
            \draw[thick] (O_1.east) -- ++(0.25,0);
            \draw[thick] (A_R1.west) -- ++(-0.25,0);
            \draw[thick] (A_R1_c.west) -- ++(-0.25,0);
            \draw[thick] (O_1.west) -- ++(-0.25,0);
        \end{tikzpicture}\cdots
        \begin{tikzpicture}[baseline={(O_1.base)}]
            \node[draw, rounded corners, minimum height=2em, minimum width=2.2em, text width=2.2em, text height = 1em, inner sep=0pt, align=center] (AL_Ly) at (0.0,0) {$A^R_{L_y}$};
            \node[draw, rounded corners, minimum height=2em, minimum width=2.2em, text width=2.2em, text height = 1em, inner sep=0pt, align=center] (O_Ly) at (0.0,-1.0) {$O_{L_y}$};
            \node[draw, rounded corners, minimum height=2em, minimum width=2.2em, text width=2.2em, text height = 1em, inner sep=0pt, align=center] (AL_Ly_c) at (0.0,-2.0) {$\bar{A}^R_{L_y}$};
            \draw[thick] (AR_Ly.south) -- (O_Ly.north);
            \draw[thick] (O_Ly.south) -- (AL_Ly_c.north);
            \draw[thick] (AL_Ly.east) -- ++(0.25,0);
            \draw[thick] (AL_Ly_c.east) -- ++(0.25,0);
            \draw[thick] (O_Ly.east) -- ++(0.25,0);
            \draw[thick] (AL_Ly.west) -- ++(-0.25,0);
            \draw[thick] (AL_Ly_c.west) -- ++(-0.25,0);
            \draw[thick] (O_Ly.west) -- ++(-0.25,0);
        \end{tikzpicture}-
        \begin{tikzpicture}[baseline={(O_1.base)}]
            \node[circle, draw, minimum size=0.5cm] (Id) at (0,-1.0) {$\mathbf{I}$};
            \node[circle, draw, inner sep=0pt, minimum size=0.6cm] (rho) at (0.75,-1.0) {$\rho_L$};
            \fill (-0.45,-1.0) circle (0.08cm);
            \fill (1.2, -1.0) circle (0.08cm);
            \draw[thick] (-0.525, -1.0) -- (-0.95, -1.0);
            \draw[thick] (1.275, -1.0) -- (1.7, -1.0);
            \draw[thick] (Id.north) arc[start angle=0,end angle=90,radius=0.8];
            \draw[thick] (Id.south) arc[start angle=0,end angle=-90,radius=0.8];
            \draw[thick] (rho.north) arc[start angle=180,end angle=90,radius=0.8];
            \draw[thick] (rho.south) arc[start angle=180,end angle=270,radius=0.8];
        \end{tikzpicture},
    \end{aligned}
\end{equation}
where \tikz[baseline=-0.5ex]{\draw[line width=1pt] (0,0) -- (0.395,0);\draw (0.5,0) circle (0.1cm);} and \tikz[baseline=-0.5ex]{\draw[line width=1pt] (0,0) -- (0.395,0);\fill (0.5,0) circle (0.1cm);} are rank-1 tensors. Their compotents are
    $(\tikz[baseline=-0.5ex]{
    \draw[line width=1pt] (0,0) -- (0.395,0);
    \draw (0.5,0) circle (0.1cm);})_i=\delta_{i,\chi}$ and
    $(\tikz[baseline=-0.5ex]{
    \draw[line width=1pt] (0,0) -- (0.395,0);
    \fill (0.5,0) circle (0.1cm);})_i=\delta_{i,1}$
    with $\chi$ being the largest index of the MPO virtual bond.
The rank-2 tensors $\rho_L$ and $\rho_R$ are left and right density matrices: $\rho_L=C^\dagger C$ and $\rho_R=CC^\dagger$, and their traces are 1.
We can check that the product of two $\tilde{\mathbb{T}}_{L}^{L}$ is:
\begin{equation}
    \begin{aligned}
        (\tilde{\mathbb{T}}_{L}^{L})^2&=(\mathbb{T}^{L}_{L}-\mathbb{P}_L)\cdot(\mathbb{T}^{L}_{L}-\mathbb{P}_L)
        \\
        &=(\mathbb{T}^{L}_{L})^2-\mathbb{P}_L\mathbb{T}_L^L-\mathbb{T}_L^L\mathbb{P}_L+\mathbb{P}_L^2
        \\
        &=(\mathbb{T}^{L}_{L})^2-\mathbb{T}_L^L\mathbb{P}_L
        \\
        &=(\mathbb{T}^{L}_{L})\tilde{\mathbb{T}}^{L}_{L}.
    \end{aligned}
\end{equation}
For $\tilde{\mathbb{T}}^R_R$, similar relation holds. The third line follows from the fact that $(\mathbf{I}\otimes\tikz[baseline=-0.5ex]{\draw[line width=1pt] (0.105,0) -- (0.5,0);\draw (0.0,0) circle (0.1cm);})\cdot\mathbb{T}^{L}_{L}=\mathbf{I}$.
By definition, applying $(\tikz[baseline=-0.5ex]{\draw[line width=1pt] (0,0) -- (0.395,0);\draw (0.5,0) circle (0.1cm);}\otimes\rho)$ to an arbitrary enviroment ${E}^0_{L}$ will gives the `internel energy' left to the unit cell we consider. So if we denote the unit cell's transfer-matrix as ${\mathbb{T}}_{C}$, we conclude that the energy we calculate with the environments $E_{L,1}$ and $E_{R,L_y}$ is:
\begin{equation}
    \begin{aligned}
        &E_{L,1}\cdot\mathbb{T}_C\cdot E_{R,L_y}
        \\
        =&\lim_{n_L\rightarrow\infty}\lim_{n_R\rightarrow\infty}E^0_L\cdot(\tilde{\mathbb{T}}^L_L)^{n_L}\mathbb{T}_C(\tilde{\mathbb{T}}^R_R)^{n_R}\cdot E^0_R
        \\
        =&\lim_{n_L\rightarrow\infty}\lim_{n_R\rightarrow\infty}E^0_L\cdot(\mathbb{T}^L_L)^{n_L-1}(\tilde{\mathbb{T}}^L_L)\mathbb{T}_C\tilde{\mathbb{T}}^R_R(\mathbb{T}^R_R)^{n_R-1}\cdot E^0_R
        \\
        =&\phantom{.}\mathcal{E}_{\infty}-\mathcal{E}_L-\mathcal{E}_R\coloneqq \mathcal{E}_{UC}.
    \end{aligned}
\end{equation}
The result can be checked by expanding $\tilde{\mathbb{T}}_{L}^{L}$ and $\tilde{\mathbb{T}}_{R}^{R}$ and do the contraction explicitly.
\\
\indent The convergence conditions of the VUMPS algorithm are the following equations:
\begin{equation}
    \label{A15}
    \begin{aligned}
        H_{A^C,n}(A^C_n) &= \mathcal{E}_{UC}A^C_n,\\
        H_{C,n}(C_n) &= \mathcal{E}_{UC}C_n,\\
        H_{A^C,n}(A^C_n)&=A^L_n H_{C,n}(C_n).
    \end{aligned}
\end{equation}
The eigenvalue $\mathcal{E}_{UC}$ is the internal interaction energy in an iMPS unit cell plus its interaction energy with neighboring unit cells in the ground state if we subtract the environments' energy correctly during calculation.
So the VUMPS algorithm's iteration can be organized as: 
(\romannumeral 1) Input the tensors $A^L_n$, $A^R_n$ and $C_n$ from the last step to construct $\{H_{A^C}(\cdot)\}$ and $\{H_{C}(\cdot)\}$ of all sites in a unit cell; (\romannumeral 2) solve the first and second eigenvalue equations in (\ref{A15}) and accept the lowest eigenvalue and eigenvector as $E_{UC}$ and $A'^C_n$ and $C'_n$; (\romannumeral 3) calculate the new $A'^L_n$'s and $A'^R_n$'s by solve $\mathbf{min}_{A'^L_n}|A'^C_n-A'^L_nC'_n|$ and $\mathbf{min}_{A'^R_n}|A'^C_n-C'_nA'^R_n|$; (\romannumeral 4) calculate the module of the gradient vector by $|H_{A'^C,n}(A'^C_n)-A'^L_n H_{C',n}(C'_n)|$ and check if it is smaller than the error tolerance. If not, go back to (\romannumeral 1).

\subsection{Excited state ansatz}\label{App:tangent_space}
In VUMPS we have assumed that the ground state is the 0-eigenvector of the momentum operator, thus invariant under the action of translation. 
However, the excitation states can be in different momentum sectors, and they should be viewed as living in a larger tangent space $\mathcal{T}\mathbf{M'}$, where $\mathbf{M'}$ is the manifold spanned by the states with different $A$ tensors at different sites. 
This change makes it possible for $B$'s at different unit cells to be different, while the $A$ tensors are restricted to the original ground state manifold with $A$'s at different unit cells being the same. 
Therefore, we can adopt the \textbf{quasiparticle excitation ansatz} proposed in~\cite{laurens2019} to describe the excitation states by taking the $n$th $B$ in the $m$th unit cell as $e^{imk_x}B_n$, the ansatz's form is defined in Eq.~(6) in the main text, where the translation operator $\hat{T}_x$ acts on the iMPS as:
\begin{equation}
    \begin{aligned}
        &\hat{T}_x|\phi_{\{A\}}(B_n)\rangle\\
        =&\hat{T}_x\left(\cdots
        \begin{tikzpicture}[baseline={(A_1.base)}]
            \node[draw, rounded corners, minimum size=2em, text width=1.5em, text height = 1em, align=center] (A_1) at (0,-1) {$A_1$};
            \node[draw, rounded corners, minimum size=2em, text width=1.5em, text height = 1em, align=center] (A_2) at (1.2,-1) {$A_2$};
            \draw[black, thick] (A_1.east)--(A_2.west);
            \draw[black, thick] (A_2.east) -- ++(0.5,0);
            \draw[thick] (A_1.west) -- ++(-0.5,0);
            \draw[thick] (A_1.south) -- ++(0,-0.5);
            \draw[thick] (A_2.south) -- ++(0,-0.5);
        \end{tikzpicture}\cdots
        \begin{tikzpicture}[baseline={(A_1.base)}]
            \node[draw, rounded corners, minimum size=2em, text width=1.5em, text height = 1em, align=center] (A_Ly) at (3.6,-1) {$A_{L_y}$};
            \draw[thick] (A_Ly.west) -- ++(-0.5,0);
            \draw[thick] (A_Ly.east) -- ++(+0.5,0);
            \draw[thick] (A_Ly.south) -- ++(0,-0.5);
        \end{tikzpicture}\cdots\right)\\
        =&\phantom{..}\begin{tikzpicture}[baseline={(A_1.base)}]
            \node[draw, rounded corners, minimum size=2em, text width=1.5em, text height = 1em, align=center] (A_1) at (0,-1) {$A_1$};
            \node[draw, rounded corners, minimum size=2em, text width=1.5em, text height = 1em, align=center] (A_2) at (1.2,-1) {$A_2$};
            \draw[thick] (-0.8,-1.9) .. controls (-0.5,-1.9) and (0.1,-1.9) .. (0.1,-2.4);
            \draw[black, thick] (A_1.east)--(A_2.west);
            \draw[thick] (A_1.west) -- ++(-0.5,0);
            \draw[thick] (A_1.south) .. controls  ++(0,-0.8) .. (4.2,-2.5);
            \draw[thick] (A_2.south) .. controls  ++(0,-0.8) .. (4.3,-2.4);
            \draw[thick] (A_2.east) -- ++(0.5,0);
            \node[rounded corners, minimum size=2em, text width=1.5em, text height = 1em, align=center] (cdots) at (2.4,-1) {$\cdots$};
            \node[draw, rounded corners, minimum size=2em, text width=1.5em, text height = 1em, align=center] (A_Ly) at (3.6,-1) {$A_{L_y}$};
            \draw[thick] (A_Ly.east) -- ++(+0.5,0);
            \draw[thick] (A_Ly.west) -- ++(-0.5,0);
            \draw[thick] (A_Ly.south) .. controls  ++(0,-0.8) .. (4.5,-2.2);
        \end{tikzpicture}\cdots.
    \end{aligned}
\end{equation} 

Once the ansatz has been written, we can calculate the excited states. This problem can be solved by (\romannumeral 1) variationally find $|\psi_0^{\mathrm{ex}}(k_x)\rangle$ that makes $\langle\psi^{\mathrm{ex}}(k_x)|\hat{H}|\psi^{\mathrm{gs}}(k_x)\rangle$ minimum with the constriction that $|\psi^{\mathrm{ex}}(k_x)\rangle$ is $\delta$-normalized: $\langle\psi^{\mathrm{ex}}(k_x)|\psi^{\mathrm{ex}}(k_x)\rangle=2\pi\delta(0)$; (\romannumeral 2) then we project out $|\psi_0^{\mathrm{ex}}\rangle$ from $\hat{H}$ and (\romannumeral 3) we do (\romannumeral 1) again to get $|\psi^{\mathrm{ex}}_1(k_x)\rangle$. We repeat this process to the level of excited state we want.
If we also parameterize $B_n$ as $N_n^L X_n$, the process is equivalent to solving the generalized eigenvalue problem~\cite{laurens2019}:
\begin{equation}
    \label{A16}
    H_{\mathrm{eff}}(k_x)\vec{X} = \omega(k_x) N_{\mathrm{eff}}(k_x)\vec{X}.
\end{equation}
Here $\vec{X}=\bigoplus_{n=1}^{L_y}\vec{X}_n$ is the vectorized form of $X_n$'s, and $H_{\mathrm{eff}}$ and $N_{\mathrm{eff}}$ are the effective Hamiltonian and norm matrix acting on $\vec{X}$. 
$H_{\mathrm{eff}}(k_x)$ and $N_{\mathrm{eff}}(k_x)$ are defined by partial derivatives $\partial_{\vec{\bar{X}}}\langle\psi^{\mathrm{ex}}(\{X\};\{A\};k_x)|\hat{H}|\psi^{\mathrm{ex}}(\{X\}; \{A\};k_x)\rangle$ and $\partial_{\vec{\bar{X}}}\langle\psi^{\mathrm{ex}}(\{X\};\{A\};k_x)|\psi^{\mathrm{ex}}(\{X\}; \{A\};k_x)\rangle$, divided by the factor $2\pi\delta(0)$. With all the orthogonal conditions in hand, $N_{\mathrm{eff}}(k_x)$ acts trivially on $\vec{X}$. The calculation of $X'_n$ after the action of $H_{\mathrm{eff}}$ is equivalent to cutting off the $\bar{X}_n$ from the tensor contraction diagram of $\langle\psi^{\mathrm{ex}}(\{X\};\{A\};k_x)|\hat{H}|\psi^{\mathrm{ex}}(\{X\}; \{A\};k_x)\rangle$ and leave a slot with two virtual legs and one physical leg open. The left-hand side of (\ref{A16}) is finally converted to the summation over three tensor diagrams:
\begin{itemize}
    \item The term in which $X$'s are located in the unit cells left to the unit cell where the new $X'_n$ lives: 
    \begin{equation}
        \label{A17}
        \begin{tikzpicture}
            \node[draw, rounded corners, minimum size=2em, text width=1.5em, text height = 1em, align=center] (AR_n) at (0,0) {$A^R_n$};
            \node[draw, rounded corners, minimum size=2em, text width=1.5em, text height = 1em, align=center] (O_n) at (0,-1.0) {$O_n$};
            \node[draw, rounded corners, minimum width=2.2em, minimum height=7.7em, text width=2.2em, text height = 1em, align=center] (EX_L) at (-1.5, -1.0) {$E^X_{L,n}$};
            \node[draw, rounded corners, minimum width=2.2em, minimum height=7.7em, text width=2.2em, text height = 1em, align=center] (E_R) at (1.5, -1.0) {$E_{R,n}$};
            \node[coordinate] (newnodeuL) at ([xshift=0.0cm, yshift=+1.0cm]EX_L.east) {};
            \node[coordinate] (newnodeuR) at ([xshift=0.0cm, yshift=+1.0cm]E_R.west) {};
            \draw[thick] (AR_n.south) -- (O_n.north);
            \draw[thick] (AR_n.west) -- (newnodeuL);
            \draw[thick] (AR_n.east) -- (newnodeuR);
            \draw[thick] (O_n.west) -- (EX_L.east);
            \draw[thick] (O_n.east) -- (E_R.west);
            \draw[thick] (O_n.south) -- ++(0,-1.5);
            \node[coordinate] (newnodeL) at ([xshift=0.0em, yshift=-1.0cm]EX_L.east) {};
            \draw[thick] (newnodeL) -- ++(+0.5,0) -- ++(0,-0.6) -- ++(-0.5,0);
            \node[coordinate] (newnodeR) at ([xshift=0.0em, yshift=-1.0cm]E_R.west) {};
            \draw[thick] (newnodeR) -- ++(-0.5,0) -- ++(0,-0.6) -- ++(+0.5,0);
        \end{tikzpicture},
    \end{equation}
    where $E_{R,n}$ is the same environment as described in the VUMPS algorithm.
    We define the new kind of MPO transfer-matrix $\mathbb{T}_{L,n}^X$ as:
    \begin{equation}
        \label{A18}
        \begin{aligned}        
            \mathbb{T}_{L,n}^X&=
            \begin{tikzpicture}[baseline={(O_1.base)}]
                \node[draw, rounded corners, minimum size=2em, text width=1.5em, text height = 1em, align=center] (AL_1) at (0,0) {$A^L_1$};
                \node[draw, rounded corners, minimum size=2em, text width=1.5em, text height = 1em, align=center] (O_1) at (0,-1.0) {$O_1$};
                \node[draw, rounded corners, minimum size=2em, text width=1.5em, text height = 1em, align=center] (AL_1_c) at (0,-2.0) {$\bar{A}^L_1$};
                \draw[thick] (AL_1.south) -- (O_1.north);
                \draw[thick] (O_1.south) -- (AL_1_c.north);
                \draw[thick] (AL_1.west) -- ++(-0.25,0);
                \draw[thick] (AL_1.east) -- ++(+0.25,0);
                \draw[thick] (O_1.west) -- ++(-0.25,0);
                \draw[thick] (O_1.east) -- ++(+0.25,0);
                \draw[thick] (AL_1_c.west) -- ++(-0.25,0);
                \draw[thick] (AL_1_c.east) -- ++(+0.25,0);
            \end{tikzpicture}
            \cdots
            \begin{tikzpicture}[baseline={(O_n.base)}]
                \node[draw, rounded corners, minimum size=2em, text width=1.5em, text height = 1em, align=center] (B_n) at (0,0) {$B_n$};
                \node[draw, rounded corners, minimum height=2em, minimum width=2.2em, text width=2.2em, text height = 1em, inner sep=0pt, align=left] (AL_n_m_1) at (-1.2,0) {$A^L_{n-1}$};
                \node[draw, rounded corners, minimum height=2em, minimum width=2.2em, text width=2.2em, text height = 1em, inner sep=0pt, align=left] (AR_n_p_1) at (1.2,0) {$A^R_{n+1}$};
                \node[draw, rounded corners, minimum size=2em, text width=1.5em, text height = 1em, align=center] (O_n) at (0,-1.0) {$O_n$};
                \node[draw, rounded corners, minimum height=2em, minimum width=2.2em, text width=2.2em, text height = 1em, inner sep=0pt, align=left] (O_n_m_1) at (-1.2,-1.0) {$O_{n-1}$};
                \node[draw, rounded corners, minimum height=2em, minimum width=2.2em, text width=2.2em, text height = 1em, inner sep=0pt, align=left] (O_n_p_1) at (+1.2,-1.0) {$O_{n+1}$};
                \node[draw, rounded corners, minimum size=2em, text width=1.5em, text height = 1em, align=center] (AL_n_c) at (0,-2.0) {$\bar{A}^L_n$};
                \node[draw, rounded corners, minimum height=2em, minimum width=2.2em, text width=2.2em, text height = 1em, inner sep=0pt, align=left] (AL_n_c_m_1) at (-1.2,-2.0) {$\bar{A}^L_{n-1}$};
                \node[draw, rounded corners, minimum height=2em, minimum width=2.2em, text width=2.2em, text height = 1em, inner sep=0pt, align=left] (AL_n_c_p_1) at (1.2,-2.0) {$\bar{A}^L_{n+1}$};
                \draw[thick] (B_n.south) -- (O_n.north);
                \draw[thick] (B_n.west) -- (AL_n_m_1.east);
                \draw[thick] (B_n.east) -- (AR_n_p_1.west);
                \draw[thick] (O_n.south) -- (AL_n_c.north);
                \draw[thick] (AL_n_c.west) -- (AL_n_c_m_1.east);
                \draw[thick] (AL_n_c.east) -- (AL_n_c_p_1.west);
                \draw[thick] (O_n.west) -- (O_n_m_1.east);
                \draw[thick] (O_n.east) -- (O_n_p_1.west);
                \draw[thick] (AL_n_m_1.south) -- (O_n_m_1.north);
                \draw[thick] (AR_n_p_1.south) -- (O_n_p_1.north);
                \draw[thick] (O_n_m_1.south) -- (AL_n_c_m_1.north);
                \draw[thick] (O_n_p_1.south) -- (AL_n_c_p_1.north);
                \draw[thick] (AL_n_m_1.west) -- ++(-0.25,0);
                \draw[thick] (AR_n_p_1.east) -- ++(+0.25,0);
                \draw[thick] (O_n_m_1.west) -- ++(-0.25,0);
                \draw[thick] (O_n_p_1.east) -- ++(+0.25,0);
                \draw[thick] (AL_n_c_m_1.west) -- ++(-0.25,0);
                \draw[thick] (AL_n_c_p_1.east) -- ++(+0.25,0);
            \end{tikzpicture}\cdots
            \begin{tikzpicture}[baseline={(O_n.base)}]
                \node[draw, rounded corners, minimum height=2em, minimum width=2.2em, text width=2.2em, text height = 1em, inner sep=0pt, align=center] (AR_Ly) at (0,0) {$A^R_{L_y}$};
                \node[draw, rounded corners, minimum height=2em, minimum width=2.2em, text width=2.2em, text height = 1em, inner sep=0pt, align=center] (O_Ly) at (0,-1.0) {$O_{L_y}$};
                \node[draw, rounded corners, minimum height=2em, minimum width=2.2em, text width=2.2em, text height = 1em, inner sep=0pt, align=center] (AL_Ly_c) at (0,-2.0) {$\bar{A}^L_{L_y}$};
                \node[coordinate] (mid) at (0, -1.145) {}; 
                \draw[thick] (AR_Ly.south) -- (O_Ly.north);
                \draw[thick] (O_Ly.south) -- (AL_Ly_c.north);
                \draw[thick] (AR_Ly.west) -- ++(-0.25,0);
                \draw[thick] (AR_Ly.east) -- ++(+0.25,0);
                \draw[thick] (O_Ly.west) -- ++(-0.25,0);
                \draw[thick] (O_Ly.east) -- ++(+0.25,0);
                \draw[thick] (AL_Ly_c.west) -- ++(-0.25,0);
                \draw[thick] (AL_Ly_c.east) -- ++(+0.25,0);
            \end{tikzpicture}
            \\
            &=
            \begin{tikzpicture}[baseline={(O_1.base)}]
                \node[draw, rounded corners, minimum size=2em, text width=1.5em, text height = 1em, align=center] (AL_1) at (0,0) {$A^L_1$};
                \node[draw, rounded corners, minimum size=2em, text width=1.5em, text height = 1em, align=center] (O_1) at (0,-1.0) {$O_1$};
                \node[draw, rounded corners, minimum size=2em, text width=1.5em, text height = 1em, align=center] (AL_1_c) at (0,-2.0) {$\bar{A}^L_1$};
                \draw[thick] (AL_1.south) -- (O_1.north);
                \draw[thick] (O_1.south) -- (AL_1_c.north);
                \draw[thick] (AL_1.west) -- ++(-0.25,0);
                \draw[thick] (AL_1.east) -- ++(+0.25,0);
                \draw[thick] (O_1.west) -- ++(-0.25,0);
                \draw[thick] (O_1.east) -- ++(+0.25,0);
                \draw[thick] (AL_1_c.west) -- ++(-0.25,0);
                \draw[thick] (AL_1_c.east) -- ++(+0.25,0);
            \end{tikzpicture}
            \cdots
            \begin{tikzpicture}[baseline={(O_n.base)}]
                \node[draw, rounded corners, minimum size=2em, text width=1.5em, text height = 1em, align=center] (NL_n) at (0,0) {$N^L_n$};
                \node[draw, rounded corners, minimum size=2em, text width=1.5em, text height = 1em, align=center] (X_n) at (0.9,0) {$X_n$};
                \node[draw, rounded corners, minimum height=2em, minimum width=2.2em, text width=2.2em, text height = 1em, inner sep=0pt, align=left] (AL_n_m_1) at (-1.2,0) {$A^L_{n-1}$};
                \node[draw, rounded corners, minimum height=2em, minimum width=2.2em, text width=2.2em, text height = 1em, inner sep=0pt, align=left] (AR_n_p_1) at (+1.8,0) {$A^R_{n+1}$};
                \node[draw, rounded corners, minimum size=2em, text width=1.5em, text height = 1em, align=center] (O_n) at (0,-1.0) {$O_n$};
                \node[draw, rounded corners, minimum height=2em, minimum width=2.2em, text width=2.2em, text height = 1em, inner sep=0pt, align=left] (O_n_m_1) at (-1.2,-1.0) {$O_{n-1}$};
                \node[draw, rounded corners, minimum height=2em, minimum width=2.2em, text width=2.2em, text height = 1em, inner sep=0pt, align=left] (O_n_p_1) at (+1.8,-1.0) {$O_{n+1}$};
                \node[draw, rounded corners, minimum size=2em, text width=1.5em, text height = 1em, align=center] (AL_n_c) at (0,-2.0) {$\bar{A}^L_n$};
                \node[draw, rounded corners, minimum height=2em, minimum width=2.2em, text width=2.2em, text height = 1em, inner sep=0pt, align=left] (AL_n_c_m_1) at (-1.2,-2.0) {$\bar{A}^L_{n-1}$};
                \node[draw, rounded corners, minimum height=2em, minimum width=2.2em, text width=2.2em, text height = 1em, inner sep=0pt, align=left] (AL_n_c_p_1) at (1.8,-2.0) {$\bar{A}^L_{n+1}$};
                \draw[thick] (NL_n.south) -- (O_n.north);
                \draw[thick] (NL_n.west) -- (AL_n_m_1.east);
                \draw[thick] (NL_n.east) -- (X_n.west);
                \draw[thick] (X_n.east) -- (AR_n_p_1.west);
                \draw[thick] (O_n.south) -- (AL_n_c.north);
                \draw[thick] (AL_n_c.east) -- (AL_n_c_p_1.west);
                \draw[thick] (AL_n_c.west) -- (AL_n_c_m_1.east);
                \draw[thick] (O_n.west) --(O_n_m_1.east);
                \draw[thick] (O_n.east) -- (O_n_p_1.west);
                \draw[thick] (AL_n_m_1.south) -- (O_n_m_1.north);
                \draw[thick] (AR_n_p_1.south) -- (O_n_p_1.north);
                \draw[thick] (O_n_m_1.south) -- (AL_n_c_m_1.north);
                \draw[thick] (O_n_p_1.south) -- (AL_n_c_p_1.north);
                \draw[thick] (AL_n_m_1.west) -- ++(-0.25,0);
                \draw[thick] (AR_n_p_1.east) -- ++(+0.25,0);
                \draw[thick] (O_n_m_1.west) -- ++(-0.25,0);
                \draw[thick] (O_n_p_1.east) -- ++(+0.25,0);
                \draw[thick] (AL_n_c_m_1.west) -- ++(-0.25,0);
                \draw[thick] (AL_n_c_p_1.east) -- ++(+0.25,0);
            \end{tikzpicture}\cdots
            \begin{tikzpicture}[baseline={(O_n.base)}]
                \node[draw, rounded corners, minimum height=2em, minimum width=2.2em, text width=2.2em, text height = 1em, inner sep=0pt, align=center] (AR_Ly) at (0,0) {$A^R_{L_y}$};
                \node[draw, rounded corners, minimum height=2em, minimum width=2.2em, text width=2.2em, text height = 1em, inner sep=0pt, align=center] (O_Ly) at (0,-1.0) {$O_{L_y}$};
                \node[draw, rounded corners, minimum height=2em, minimum width=2.2em, text width=2.2em, text height = 1em, inner sep=0pt, align=center] (AL_Ly_c) at (0,-2.0) {$\bar{A}^L_{L_y}$};
                \node[coordinate] (mid) at (0, -1.145) {}; 
                \draw[thick] (AR_Ly.south) -- (O_Ly.north);
                \draw[thick] (O_Ly.south) -- (AL_Ly_c.north);
                \draw[thick] (AR_Ly.west) -- ++(-0.25,0);
                \draw[thick] (AR_Ly.east) -- ++(+0.25,0);
                \draw[thick] (O_Ly.west) -- ++(-0.25,0);
                \draw[thick] (O_Ly.east) -- ++(+0.25,0);
                \draw[thick] (AL_Ly_c.west) -- ++(-0.25,0);
                \draw[thick] (AL_Ly_c.east) -- ++(+0.25,0);
            \end{tikzpicture}.
        \end{aligned}
    \end{equation} 
    Then $E^X_{L,n}$ is obtained by solving the equation to get $E^X_{L,1}$ first:
    \begin{equation}
        \label{A19}
        E^X_{L,1}\cdot (\mathbb{1}-e^{-ik_x}\mathbb{T}^R_L)=E_{L,1}\cdot (e^{-ik_x}\sum_{n'}\mathbb{T}^X_{L,n'}),
    \end{equation}
    where the dot means contracting of the three right legs of $E_{L,1}$ and the three left legs of the transfer-matrix. And $\mathbb{T}^R_L$ is the MPO transfer-matrix of a unit cell with all the $A_n$'s in right canonical form and $\bar{A}_n$'s in left canonical form.
    After getting $E^X_{L,1}$ we can contract in the unit cell to the left of site $n$ and obtain $E^X_{L,n}$.
    \item The term in which $X$'s are located in the same unit cell as $X'_n$ we consider, which is just a finite summation over all the positions $B_{n'}$ that can appear:
    \begin{equation}
        \label{20}
        \sum_{n'=1}^{L_y}
        \begin{tikzpicture}[baseline={(EL_1.base)}]
            \node[draw, rounded corners, minimum width=2.8em, minimum height=7.7em, text width=2.6em, text height = 1em, align=center] (EL_1) at (-1.5, -1.0) {$E_{L,1}$};
            \node[coordinate] (newnodedL) at ([xshift=0.0cm, yshift=-1.0cm]EL_1.east) {};
            \node[coordinate] (newnodeuL) at ([xshift=0.0cm, yshift=+1.0cm]EL_1.east) {};
            \draw[thick] (newnodedL) -- ++ (+0.25,0);
            \draw[thick] (newnodeuL) -- ++ (+0.25,0);
            \draw[thick] (EL_1.east) -- ++ (+0.25,0);
        \end{tikzpicture}
        \cdots
        \begin{tikzpicture}[baseline={(O_np.base)}]
            \node[draw, rounded corners, minimum size=2em, text width=1.5em, text height = 1em, align=center] (B_np) at (0,0) {$B_{n'}$};
            \node[draw, rounded corners, minimum size=2em, text width=1.5em, text height = 1em, align=center] (O_np) at (0,-1.0) {$O_{n'}$};
            \node[draw, rounded corners, minimum size=2em, text width=1.5em, text height = 1em, align=center] (AL_np_c) at (0,-2.0) {$\bar{A}^L_{n'}$};
            \draw[thick] (B_np.south) -- (O_np.north);
            \draw[thick] (B_np.west) -- ++(-0.25,0);
            \draw[thick] (B_np.east) -- ++(+0.25,0);
            \draw[thick] (O_np.west) -- ++(-0.25,0);
            \draw[thick] (O_np.east) -- ++(+0.25,0);
            \draw[thick] (O_np.south) -- (AL_np_c.north);
            \draw[thick] (AL_np_c.west) -- ++(-0.25,0);
            \draw[thick] (AL_np_c.east) -- ++(+0.25,0);
        \end{tikzpicture}
        \cdots
        \begin{tikzpicture}[baseline={(O_n.base)}]
            \node[draw, rounded corners, minimum size=2em, text width=1.5em, text height = 1em, align=center] (AR_n) at (0,0) {$A_n^R$};
            \node[draw, rounded corners, minimum size=2em, text width=1.5em, text height = 1em, align=center] (O_n) at (0,-1.0) {$O_n$};
            \node[coordinate] (newnodeCL) at ([xshift=-0.25cm, yshift=-1.0cm]O_n.west) {};
            \node[coordinate] (newnodeCR) at ([xshift=+0.25cm, yshift=-1.0cm]O_n.east) {};
            \draw[thick] (AR_n.south) -- (O_n.north);
            \draw[thick] (AR_n.west) -- ++(-0.25,0);
            \draw[thick] (AR_n.east) -- ++(+0.25,0);
            \draw[thick] (O_n.west) -- ++(-0.25,0);
            \draw[thick] (O_n.east) -- ++(+0.25,0);
            \draw[thick] (O_n.south) -- ++(0,-1.2);
            \draw[thick] (newnodeCL) -- ++(+0.5,0) -- ++(0,-0.25) -- ++(-0.5,0);
            \draw[thick] (newnodeCR) -- ++(-0.5,0) -- ++(0,-0.25) -- ++(+0.5,0);
        \end{tikzpicture}
        \cdots
        \begin{tikzpicture}[baseline={(ER_Ly.base)}]
            \node[draw, rounded corners, minimum width=2.8em, minimum height=7.7em, text width=2.6em, text height = 1em, align=left] (ER_Ly) at (0.0, -1.0) {$E_{R,L_y}$};
            \node[coordinate] (newnodedR) at ([xshift=0.0cm, yshift=-1.0cm]ER_Ly.west) {};
            \node[coordinate] (newnodeuR) at ([xshift=0.0cm, yshift=+1.0cm]ER_Ly.west) {};
            \draw[thick] (newnodedR) -- ++ (-0.25,0);
            \draw[thick] (newnodeuR) -- ++ (-0.25,0);
            \draw[thick] (ER_Ly.west) -- ++ (-0.25,0);
        \end{tikzpicture}.
    \end{equation}
    \item The term in which $X$'s are located in the unit cells right to the unit cell where the new $X'_n$ lives:
    \begin{equation}
        \label{A21}
        \begin{tikzpicture}
            \node[draw, rounded corners, minimum size=2em, text width=1.5em, text height = 1em, align=center] (AL_n) at (0,0) {$A^L_n$};
            \node[draw, rounded corners, minimum size=2em, text width=1.5em, text height = 1em, align=center] (O_n) at (0,-1.0) {$O_n$};
            \node[draw, rounded corners, minimum width=2.2em, minimum height=7.7em, text width=2.2em, text height = 1em, align=center] (EX_R) at (1.5, -1.0) {$E^X_{R,n}$};
            \node[draw, rounded corners, minimum width=2.2em, minimum height=7.7em, text width=2.2em, text height = 1em, align=center] (E_L) at (-1.5, -1.0) {$E_{L,n}$};
            \node[coordinate] (newnodeuL) at ([xshift=0.0cm, yshift=+1.0cm]E_L.east) {};
            \node[coordinate] (newnodeuR) at ([xshift=0.0cm, yshift=+1.0cm]EX_R.west) {};
            \draw[thick] (AL_n.south) -- (O_n.north);
            \draw[thick] (AL_n.west) -- (newnodeuL);
            \draw[thick] (AL_n.east) -- (newnodeuR);
            \draw[thick] (O_n.west) -- (E_L.east);
            \draw[thick] (O_n.east) -- (EX_R.west);
            \draw[thick] (O_n.south) -- ++(0,-1.5);
            \node[coordinate] (newnodeL) at ([xshift=0.0em, yshift=-1.0cm]E_L.east) {};
            \draw[thick] (newnodeL) -- ++(+0.5,0) -- ++(0,-0.6) -- ++(-0.5,0);
            \node[coordinate] (newnodeR) at ([xshift=0.0em, yshift=-1.0cm]EX_R.west) {};
            \draw[thick] (newnodeR) -- ++(-0.5,0) -- ++(0,-0.6) -- ++(+0.5,0);
        \end{tikzpicture},
    \end{equation}
    where $E^X_{R,n}$ is obtained by solving the equation to get $E^X_{R,L_y}$ first:
    \begin{equation}
        \label{A22}
        (\mathbb{1}-e^{ik_x}\mathbb{T}^R_{L})\cdot E^X_{R,L_y}=(e^{ik_x}\sum_{n'}\mathbb{T}^X_{R,n'})\cdot E_{R,L_y},
    \end{equation}
    and then contracting to the site we want. Here $\mathbb{T^X_{R,n}}$ is defined as (\ref{A18}) with $\bar{A}^R$ replaced by $\bar{A}^L$ and vice versa. The linear equations can be solved by the GMRES method~\cite{saad1986}.
\end{itemize}
The last thing we should notice is that the shift of the ground state energy $\mathcal{E}_{GS}$ from $\hat{H}$ will lead to the subtraction of $\mathcal{E}_{UC}B_n$ from the new $B_n$' after applying $H_{\mathrm{eff}}$. This is easy to understand from the way we get environments, as in~\ref{supp:VUMPS}, and the fact that $B_n$'s are left-orthogonal to $A_n$'s.

\section{Chiral Luttinger liquid theory}
\label{App:ChiLL}
In this section, we give a brief review of the properties of chiral Luttinger liquid theory. In \cite{wen1990a, wen1995}, X.-G. Wen proposed the hydrodynamic approach to FQH systems with boundaries.
It was suggested that for Laughlin state with filling factor $1/m$, $m\in\mathbb{N}^+$, the low energy excitation is described by the effective theory for the chiral boson field $\phi$~\cite{floreanini1988}:
\begin{equation}
    \mathcal{S}=\frac{m}{4\pi}\int dtdx\partial_x\phi(\partial_t\phi-v\partial_x\phi).
\end{equation}
The effective action for the edge mode can be obtained by introducing constrictions to the gauge transformation of the bulk Chern-Simons theory~\cite{wen2007}, and the gauge degrees of freedom become dynamical. Following the routine of canonical quantization, we find that the Fourier modes $\rho_k$ of $\rho\coloneqq\frac{1}{2\pi}\partial_x\phi$ form the $\mathbf{U}(1)$ Kac-Moody algebra, and the Hamiltonian is a bilinear form of $\rho_k$'s.
\begin{equation}
    \begin{aligned}
        &[\rho_k,\rho_{k'}]=\frac{k}{mL}\delta_{k+k',0},
        \\
        &\hat{H}=2\pi mv\sum_{k>0}\rho_k\rho_{-k},
    \end{aligned}
\end{equation}
where $L$ is the size of the edge. The field operator for the system's original particle on the edge is defined by the vertex operator $\hat{\Psi}=\mathopen{:}\exp(-im\phi)\mathclose{:}$ which satisfies $\hat{\Psi}(x,t)\hat{\Psi}(x',t')=(-1)^{m}\hat{\Psi}(x',t')\hat{\Psi}(x,t)$. The commutator means that the physical degree of freedom is bosonic if $m$ is even and fermionic if $m$ is odd. $\Psi$'s retarded Green's function is:
\begin{equation}
    \begin{aligned}
        \mathcal{G}^{R}(x,t)&=-i\theta(t)\langle GS|\hat{\Psi}(x,t)\hat{\Psi}^\dagger(0,0)|GS\rangle
        \\
        &=-i\theta(t)\langle GS |\mathopen{:}\exp(-im\phi(x,t))\mathclose{:}\mathopen{:}\exp(im\phi(0,0))\mathclose{:}| GS\rangle
        \\
        &=-i\theta(t)\exp(m^2\langle GS |\phi(x,t)\phi(0,0)| GS \rangle).
    \end{aligned}
\end{equation}
Since we already have:
\begin{equation}
    \begin{aligned}
        &\langle GS |\phi(x,t)\phi(0,0)| GS \rangle\\
        \sim &\int^x dx_1\int^0 dx_2 \langle GS |\rho(x_1,t)\rho(x_2,0)| GS \rangle\\
        \sim &\frac{1}{m}\ln(x-vt).
    \end{aligned}
\end{equation}
We get the Green's function:
\begin{equation}
    \mathcal{G}^{R}(x,t)\sim -i\theta(t)(x-vt)^{-m}.
\end{equation}
Then we do the Fourier transformation $\mathfrak{F[\cdot]}$ to $\mathcal{G}^{R}$ to calculate the spectral function.
\begin{equation}
    \begin{aligned}
        \mathcal{A}(k,\omega)&=-\frac{1}{\pi}\mathbf{Im}\mathfrak{F}[\mathcal{G}^{R}(x,t)]\\
                   &\sim \mathbf{Im}\frac{(\omega+vk)^{m-1}}{\omega-vk+i0^+\mathrm{sgn}(\omega)}\\
                   &\sim \mathrm{sgn}(\omega)\delta(\omega-vk)(\omega+vk)^{m-1}.
    \end{aligned}
\end{equation}
This result illustrates that for an ideal isolated charged edge excitation, its spectral function should show linearity and chirality, and the spectral weight should increase with $|k|$ and $|\omega|$.

\section{Particle density and current distributions}
We present the particle density and current distributions of ground states on infinite stripes with widths $L_y=9$ and $11$ under varying trapping potentials $V$. All simulations are performed within the parameter range where DMRG converges reliably. Notably, for $L_y=11$ at $V=0$, the bulk density exhibits no plateau at $n = 1/8$, demonstrating that a finite harmonic trapping potential is required to stabilize the FCI in this geometry.
    
\begin{figure}[h]
    \includegraphics[width=0.9\linewidth]{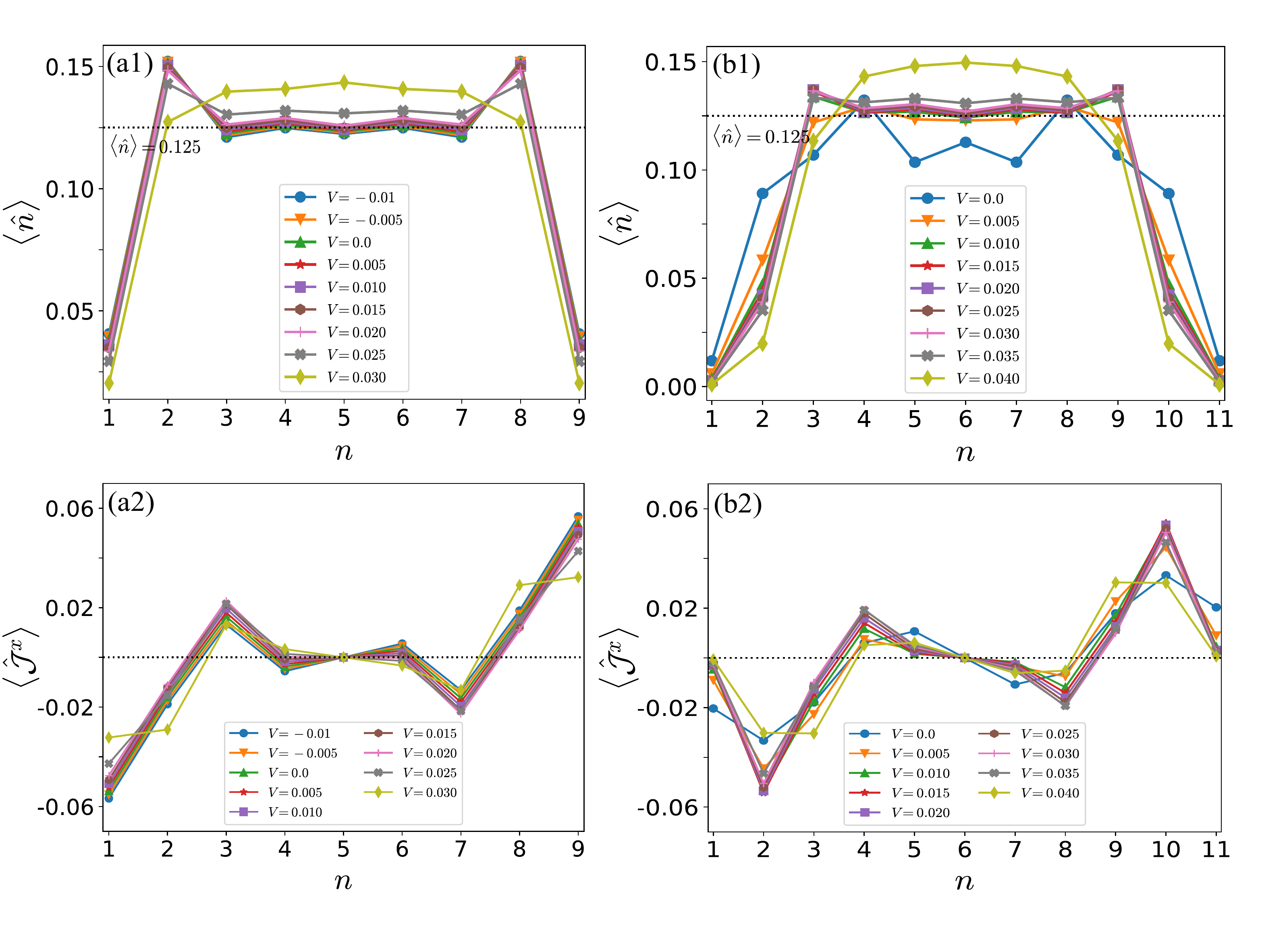}
    \caption{The particle densities and currents of ground states on infinite strips with $L_y=9$, in (a1,a2) and $L_y=11$ in (b1,b2), respectively. The bond dimension is $D=600$.}
    \label{fig:Ly9_11_n_Jx}
\end{figure}
\begin{figure}[h]
\includegraphics[width=0.9\linewidth]{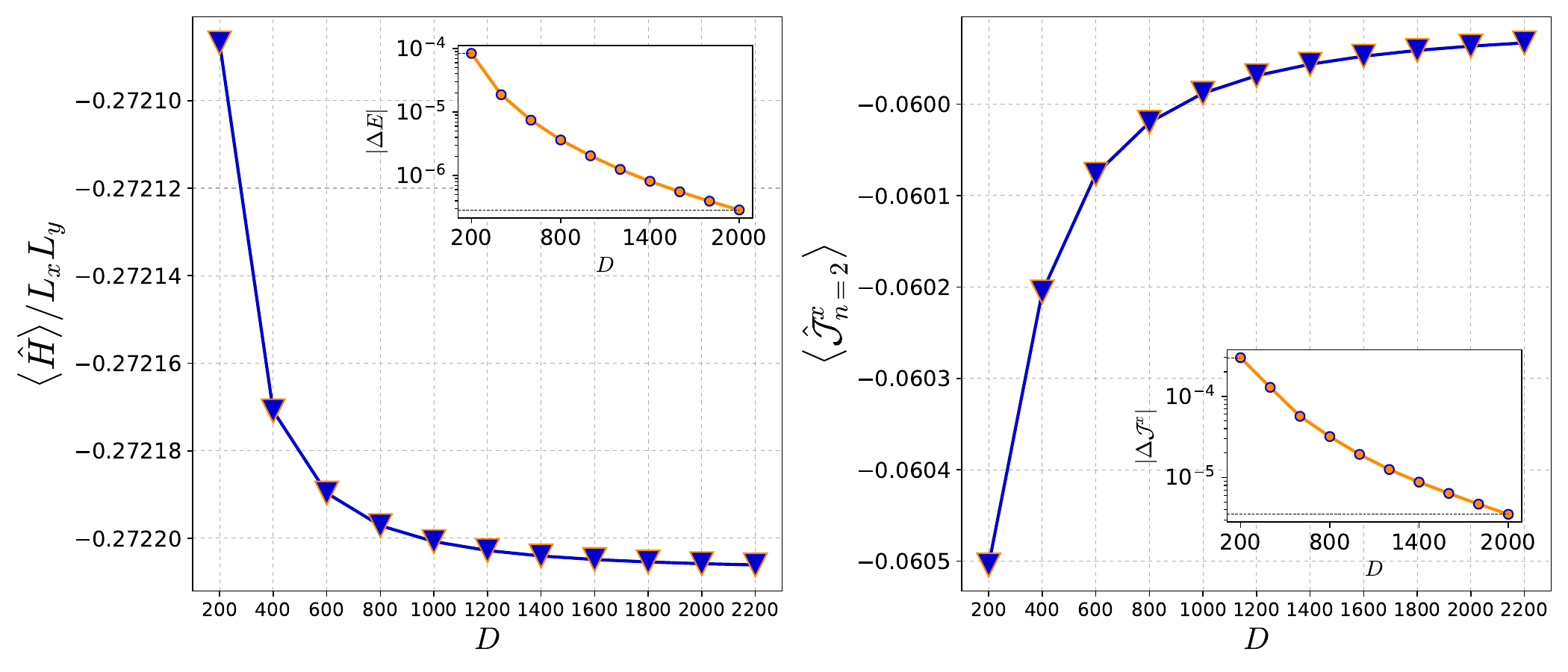}
\caption{Convergence of ground-state observables for $L_y=10,V=0$. (Left) Energy density and (right) current density along the $x$-direction at the second row ($n=2$) as a function of bond dimension $D$. The insets depict the difference in these values between successive bond dimensions, $\Delta D = 200$, demonstrating numerical stability and convergence of the VUMPS results.}
    \label{fig:E_Jx_converg}
\end{figure}

To verify the convergence of the ground state obtained via the VUMPS algorithm at a bond dimension of $D=2000$ (for $L_y=10$ and $V=0$), we plot the energy density and the edge current at row $n=2$ as a function of bond dimension $D$ in Fig.~\ref{fig:E_Jx_converg}. The truncation error (i.e., the sum of squares of the discarded Schmidt values) at $D=2000$ is on the order of $10^{-8}$, confirming the numerical stability of our results.

\section{Determination of chemical potential and the momentum shift}
As noted in the main text, the calculation of excited states in a nontrivial charge sector necessarily introduces shifts in both energy and momentum. When we calculate spectral functions, the expectation value $\bra{\psi^{\mathrm{gs}}}\hat{a}(x,t)\hat{a}^\dagger(0,0)\ket{\psi^{\mathrm{gs}}}$ should be evaluated by inserting $\sum_{\omega,k_x}\ket{\omega,k_x}\bra{\omega,k_x}$, where $\ket{\omega,k_x}$ is charge-one excited states:
\begin{equation}
    \begin{aligned}
            &\bra{\psi^{\mathrm{gs}}}\hat{a}_{m,n}(t)\hat{a}_{0,n}^\dagger(0,0)\ket{\psi^{\mathrm{gs}}}\\
            =&\sum_{\omega,k_x}\bra{\psi^{\mathrm{gs}}}\hat{a}_{m,n}(t)\ket{\omega,k_x}\bra{\omega,k_x}\hat{a}_{0,n}^\dagger(0)\ket{\psi^{\mathrm{gs}}}\\
            =&\sum_{\omega,k_x}\bra{\psi^{\mathrm{gs}}}e^{iHt-P_x m}\hat{a}_{0,n}(0)e^{-iHt+P_x m}\ket{\omega,k_x}\bra{\omega,k_x}\hat{a}_{0,n}^\dagger(0)\ket{\psi^{\mathrm{gs}}}\\
            =&\sum_{\omega,k_x}\exp(-i(\omega-\mathcal{E}^{N_{\mathrm{tot}}}_{GS})t+i(k_x-k^{N_{\mathrm{tot}}}_{GS})m)|\bra{\omega,k_x}\hat{a}_{0,n}^\dagger\ket{\psi^{\mathrm{gs}}}|^2\\
            =&\sum_{\omega,k_x}\exp[-i((\omega-\mathcal{E}^{N_{\mathrm{tot}}+1}_{GS})+(\mathcal{E}^{N_{\mathrm{tot}}+1}_{GS}-\mathcal{E}^{N_{\mathrm{tot}}}_{GS}))t\\&+i((k_x-k^{N_{\mathrm{tot}}+1}_{GS})+(k^{N_{\mathrm{tot}}+1}_{GS}-k^{N_{\mathrm{tot}}}_{GS}))m]|\bra{\omega,k_x}\hat{a}_{0,n}^\dagger\ket{\psi^{\mathrm{gs}}}|^2\\
            \equiv& \sum_{\omega,k_x}\exp[-i((\omega-\mathcal{E}^{N_{\mathrm{tot}}+1}_{GS})+\mu)t+i((k_x-k^{N_{\mathrm{tot}}+1}_{GS})+p_0)m]|\bra{\omega,k_x}\hat{a}_{0,n}^\dagger\ket{\psi^{\mathrm{gs}}}|^2.
    \end{aligned}
\end{equation}
Here, the superscripts in $\mathcal{E}_{GS}$ and $k_{GS}$ label the charge sectors of ground states. Nonzero chemical potential $\mu$ and momentum shift $p_0$ appear because we calculate excited states with $N_{\mathrm{tot}}+1$ charges based on the ground state with $N_{\mathrm{tot}}$ charges, as we have mentioned in the main text.

\begin{figure}[htb]
    \includegraphics[width=0.9\linewidth]{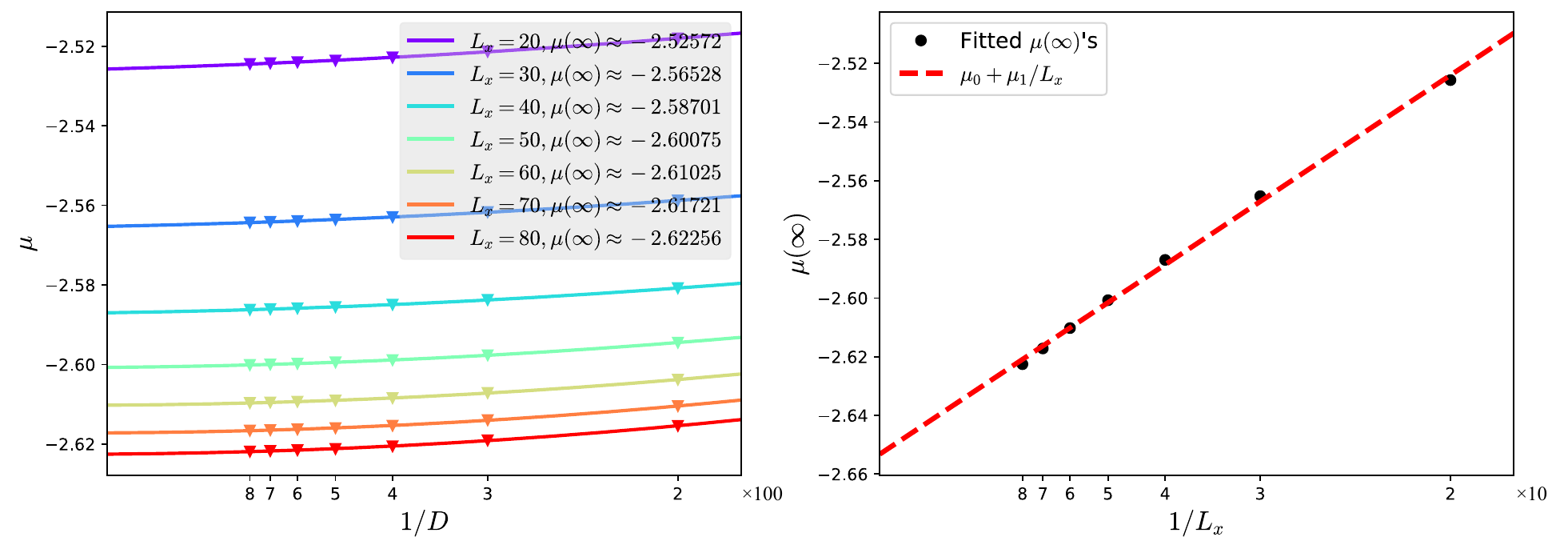}
    \caption{Determination of the chemical potential $\mu$ for $L_y=10$, defined by $\mathcal{E}_{GS}^{N_{\mathrm{tot}}+1}-\mathcal{E}_{GS}^{N_{\mathrm{tot}}}$, where the boson number $N_{\mathrm{tot}}$ in our setup is $L_x$. We first fix system sizes $L_x\times 10$ and fit with respect to $D$ to get $\mu(D=\infty, L_x)$ for different $L_x$'s, as shown in the left figure. Then we fit with respect to $L_x$ and obtain that $\mu(D=\infty, L_x=\infty)=-2.65323$ as the final estimated result for the chemical potential for infinite strip geometry, consistent with the lowest energy of the charge-one excitation spectrum. The fitting function is chosen intuitively, and we have required that the fitted functions are monotonic for both $D$ and $L_x$.}
    \label{fig:deter_mu}
\end{figure}

\begin{figure}[htb]
    \centering
    \includegraphics[width=0.05\linewidth]{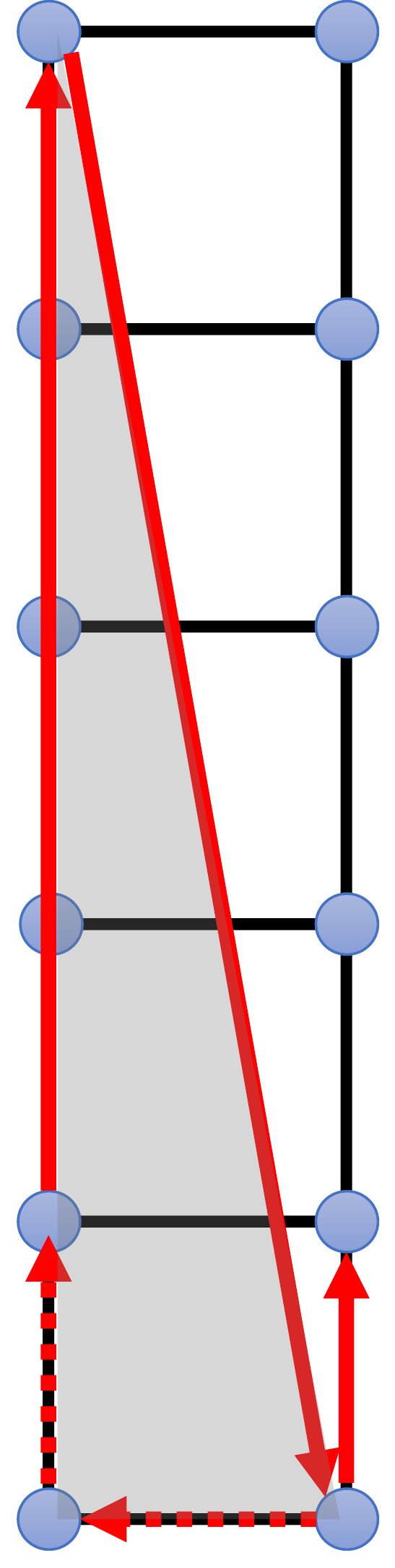}
    \caption{The path induced by the MPS-lattice mapping. The shaded area is the region around which we close the loop and use Stokes' theorem.}
    \label{fig:mpstolatt}
\end{figure}

\begin{figure}[htb]
    \centering
    \includegraphics[width=1.0\linewidth]{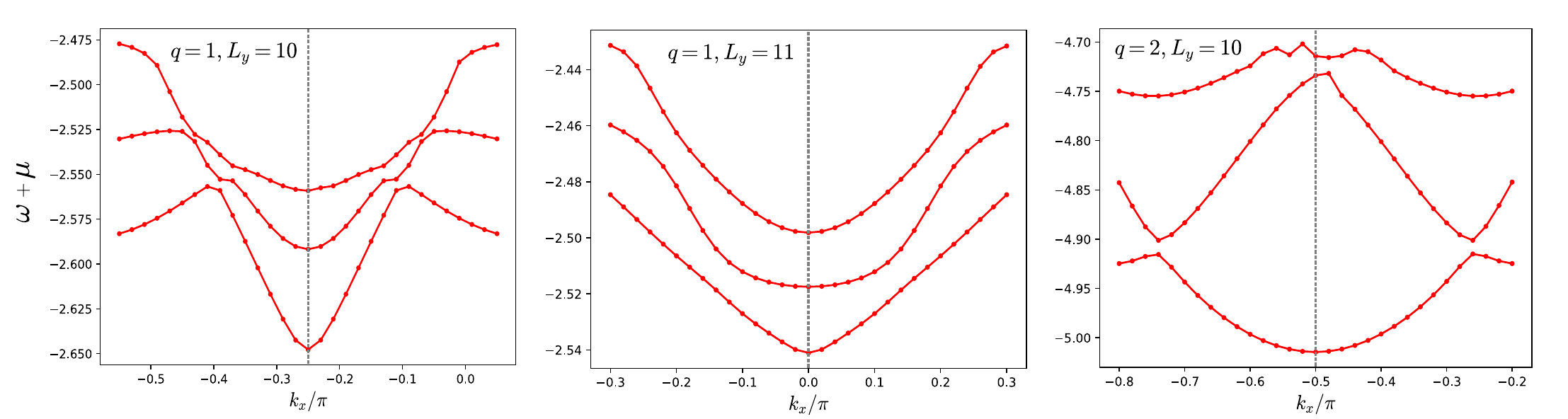}
    \caption{The phase shift $p_0$ is verified by computing the energy spectra for different charge sectors $q$ and system widths $L_y$, with $n_\phi=1/4$. For the illustrated cases, we obtain $p_0=-\pi/4$, $0$, and $-\pi/2$, which are consistent with our theoretical predictions.}
    \label{fig:k_shift}
\end{figure}

We determine $\mu$ by calculating the energy difference of systems with finite sizes $L_x\times L_y$, and using the scaling function $\mu(D, L_x)=\mu_{0}+\mu_D/D+\mu'_D/D^2+\mu_{L_x}/L_x+\mathcal{O}(1/D^3)+\mathcal{O}(1/L_x^2)$ to fit it. Although the fitting function is heuristic and the accuracy is not particularly high, the result is sufficient to qualitatively support the discussion in the main text regarding the chemical potential of the charge-one excitation, especially for our main result with $L_y=10$(Fig.~\ref{fig:deter_mu}).
\\
\indent In terms of the momentum shift $p_0$, we view it as the phase factor introduced by the magnetic translation symmetry of the ground state with $(N_{\mathrm{tot}}+q)$ bosons, which means that the ground state is invariant under the action of the operator $\tilde{T}_x\coloneq\exp[i(P_x-q\int d\mathbf{l}\cdot \mathbf{A}_{\mathrm{gauge}})]$. Here $q\int d\mathbf{l}\cdot \mathbf{A}_{\mathrm{gauge}}$ formally represents the Aharonov–Bohm phase the added charges accumulate under the translation. In MPS language, this phase is intuitively calculated over the path induced by the mapping between MPS indices and lattice sites(Fig.~\ref{fig:mpstolatt}). We can complete the loop and use Stokes' theorem to calculate this phase factor, which in our calculation with $n_\phi=1/4$ is $q\pi(L_y-3)/4$. We check this relation for different charges $q$ and widths $L_y$(Fig.~\ref{fig:k_shift}) and find that they are consistent with the formula.

\section{Spectral functions and average positions for $L_y=8,10,11$}\label{App:Spectrum_Data}
\begin{figure}[htb]
        \includegraphics[width=0.9\linewidth]{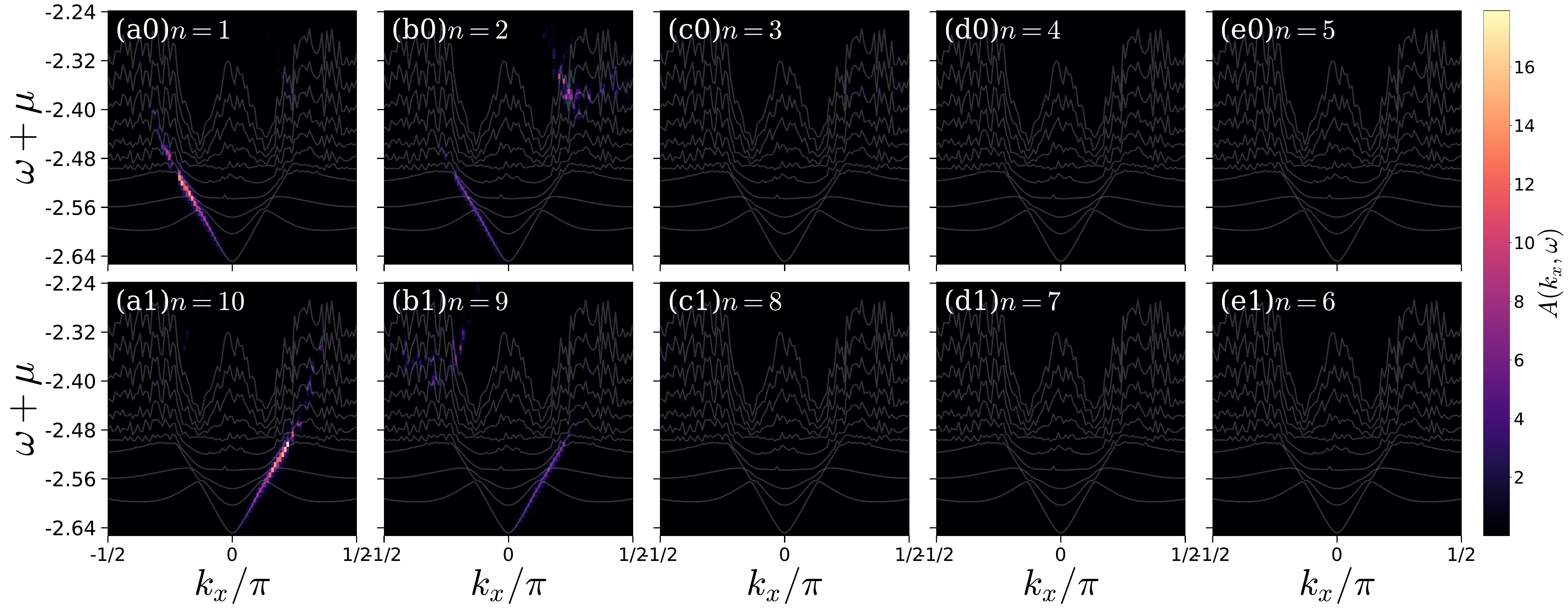}
        \caption{The spectral functions of charge-one edge excitations on an infinitely long strip with width $L_y=10$, $n_\phi=1/4$, trapping potential $V=0.0$, Lorentzian broadening factor $\eta=0.005$, and bond dimension $D=2000$.}
        \label{Ly10_Charge1_Vtrap00_Full}
    \end{figure}

\begin{figure}[htb]
        \includegraphics[width=0.9\linewidth]{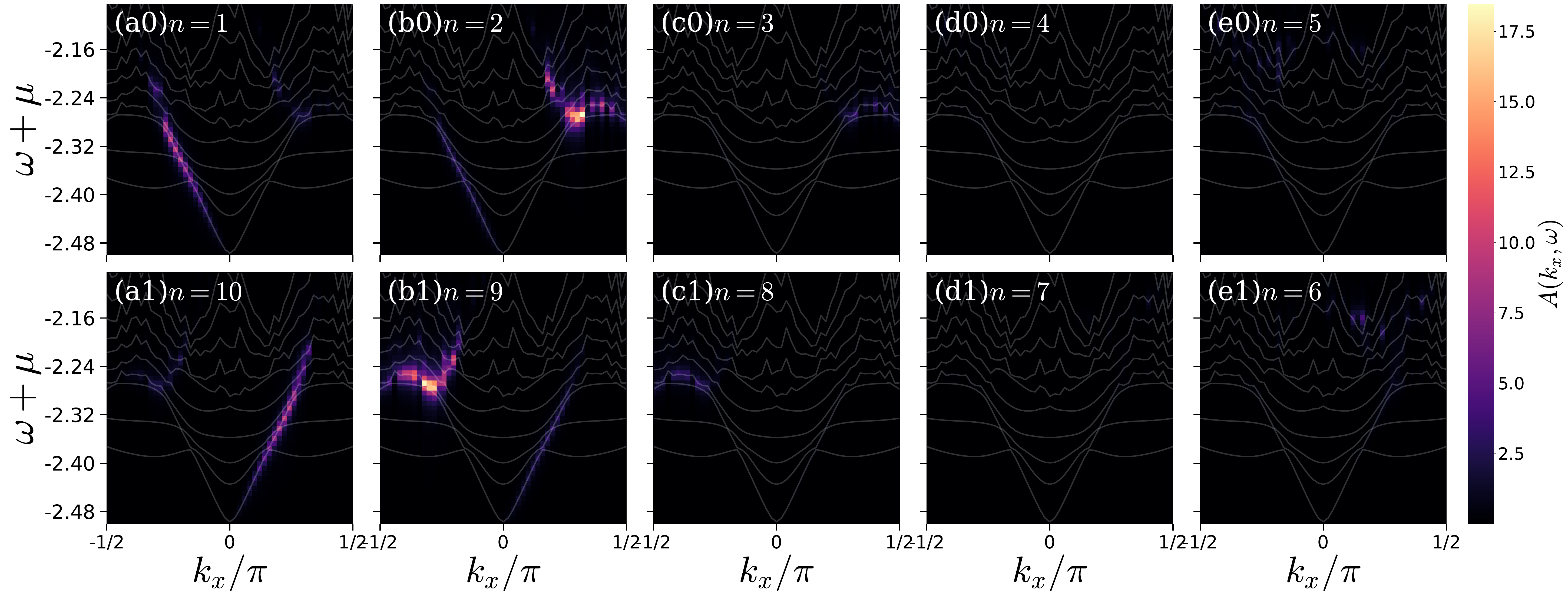}
        \caption{The spectral functions of charge-one edge excitations on an infinitely long strip with width $L_y=10$, $n_\phi=1/4$, trapping potential $V=0.01$, Lorentzian broadening factor $\eta=0.01$, and bond dimension $D=1500$.}
        \label{Ly10_Charge1_Vtrap001_Full}
    \end{figure}

\begin{figure}[htb]
        \includegraphics[width=0.9\linewidth]{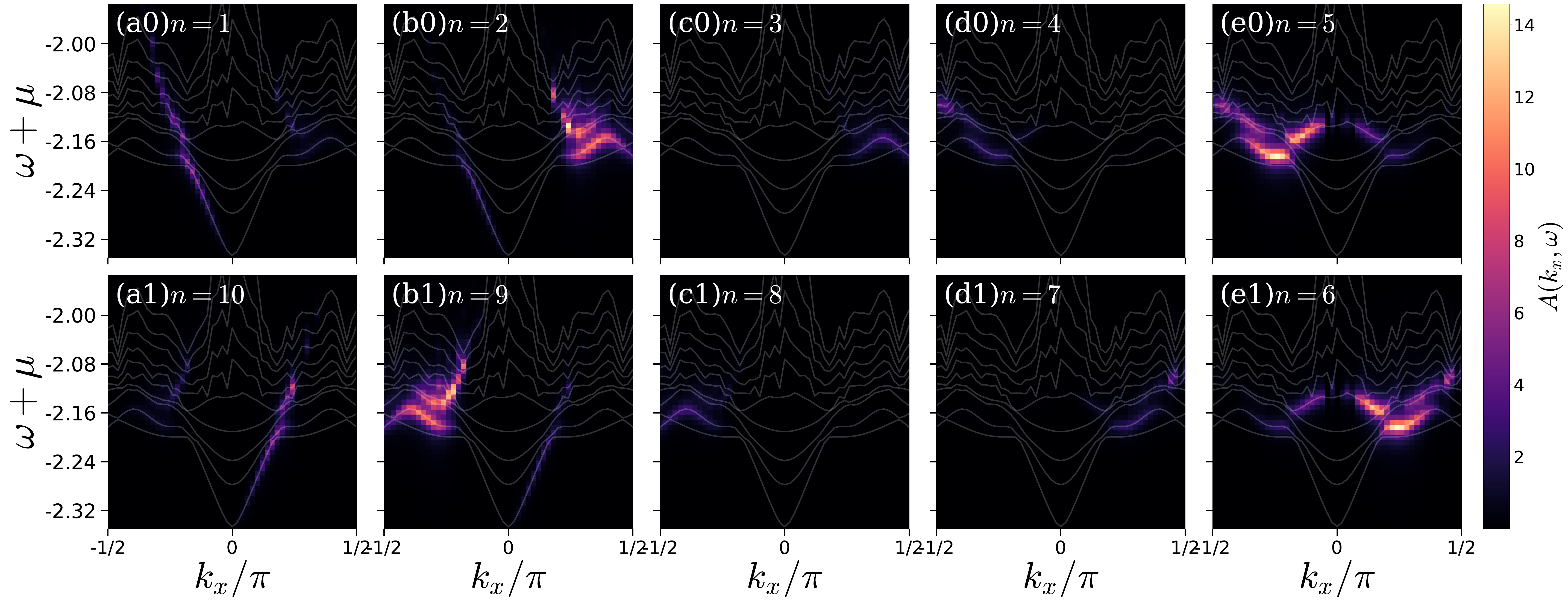}
        \caption{The spectral functions of charge-one edge excitations on an infinitely long strip with width $L_y=10$, $n_\phi=1/4$, trapping potential $V=0.02$, Lorentzian broadening factor $\eta=0.01$, and bond dimension $D=1500$.}
        \label{Ly10_Charge1_Vtrap002_Full}
    \end{figure}

\begin{figure}[htb]
        \includegraphics[width=0.9\linewidth]{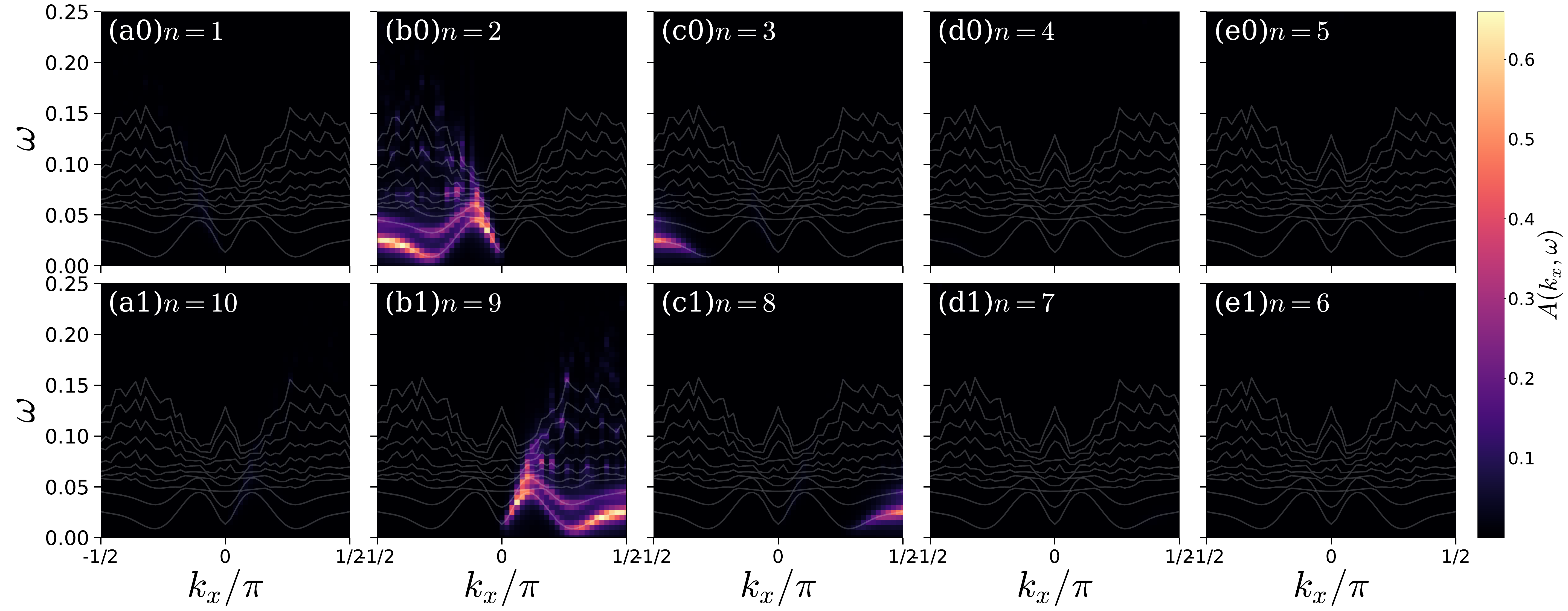}
        \caption{The spectral functions of charge-zero edge excitations on an infinitely long strip with width $L_y=10$, $n_\phi=1/4$, trapping potential $V=0.0$, Lorentzian broadening factor $\eta=0.005$, and bond dimension $D=2000$.}
        \label{Ly10_Charge0_Vtrap00_Full}
    \end{figure}

\begin{figure}[htb]
        \includegraphics[width=0.9\linewidth]{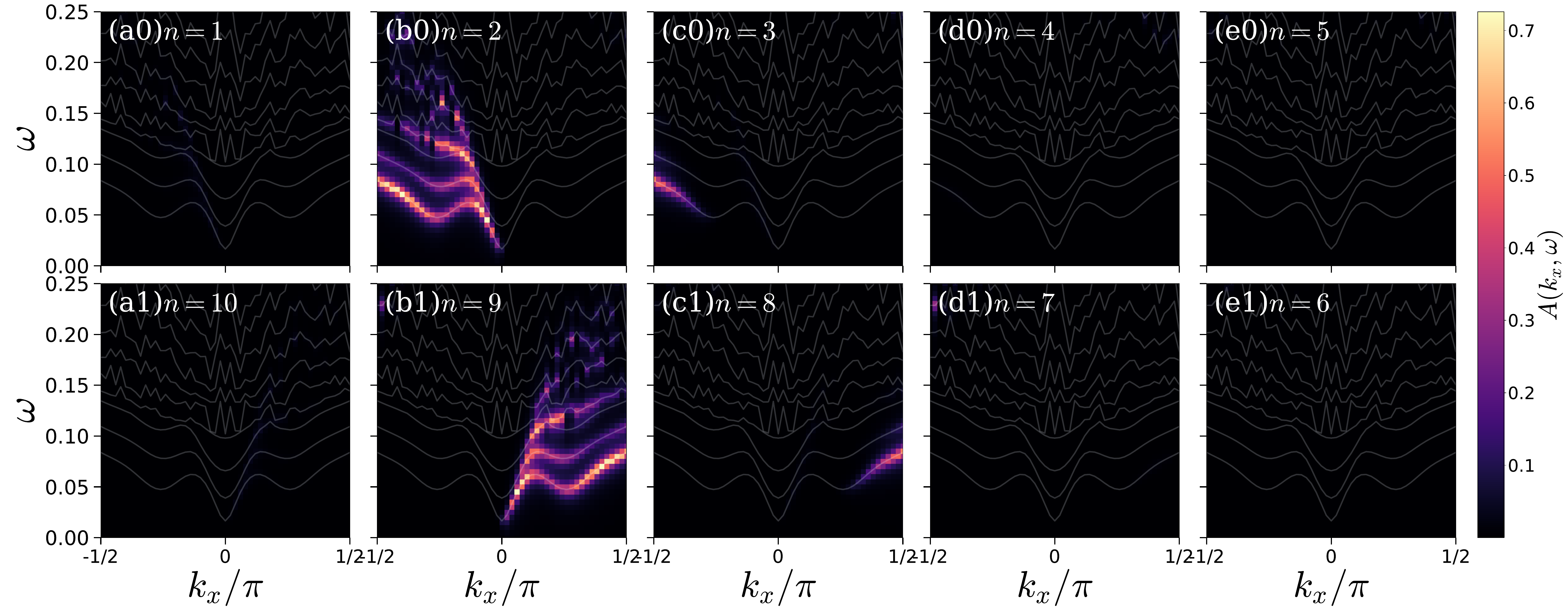}
        \caption{The spectral functions of charge-zero edge excitations on an infinitely long strip with width $L_y=10$, $n_\phi=1/4$, trapping potential $V=0.01$, Lorentzian broadening factor $\eta=0.005$, and bond dimension $D=1500$.}
        \label{Ly10_Charge0_Vtrap001_Full}
    \end{figure}

\begin{figure}[htb]
        \includegraphics[width=0.9\linewidth]{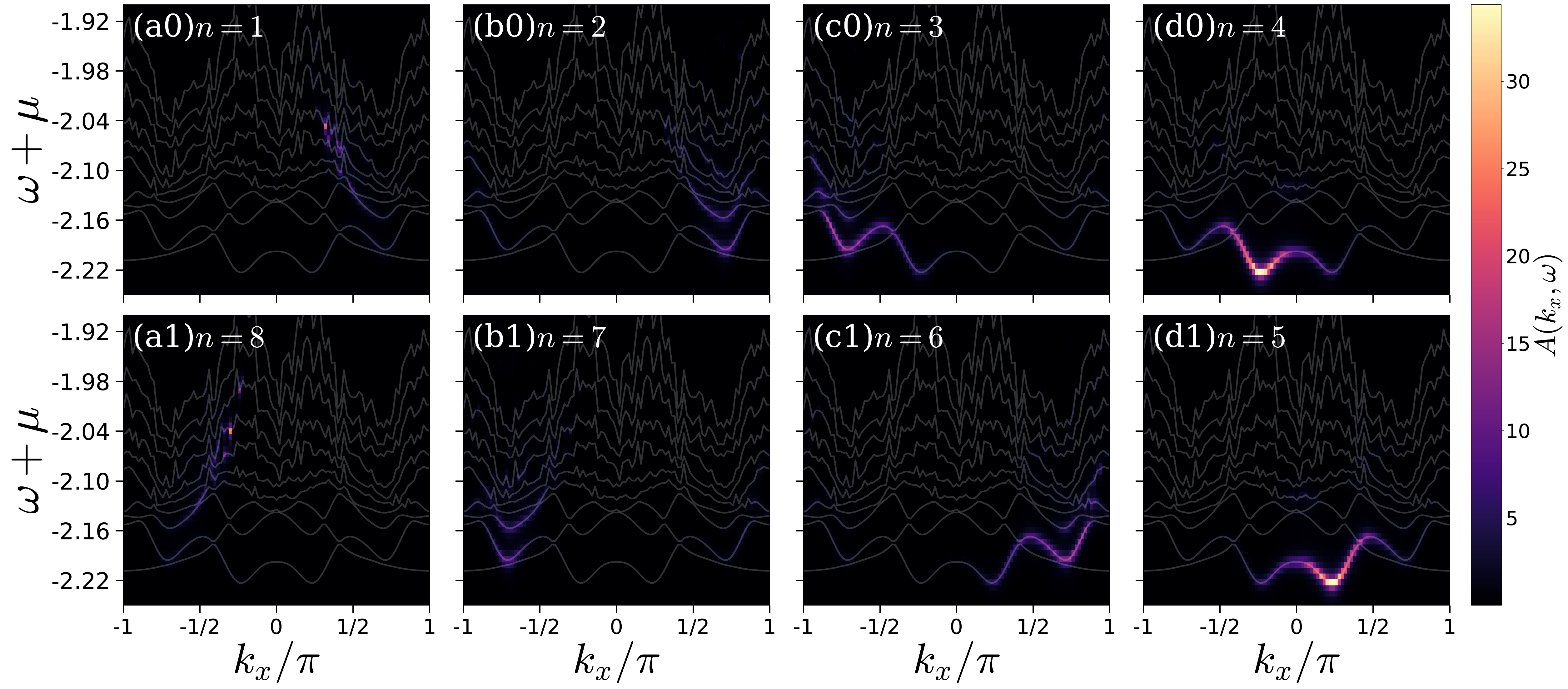}
        \caption{The spectral functions of charge-one edge excitations on an infinitely long strip with width $L_y=8$, $n_\phi=1/4$, trapping potential $V=0.0$, Lorentzian broadening factor $\eta=0.005$, and bond dimension $D=1500$.}
        \label{Ly8_Charge1_Vtrap00_Full}
    \end{figure}

\begin{figure}[htb]
        \includegraphics[width=0.9\linewidth]{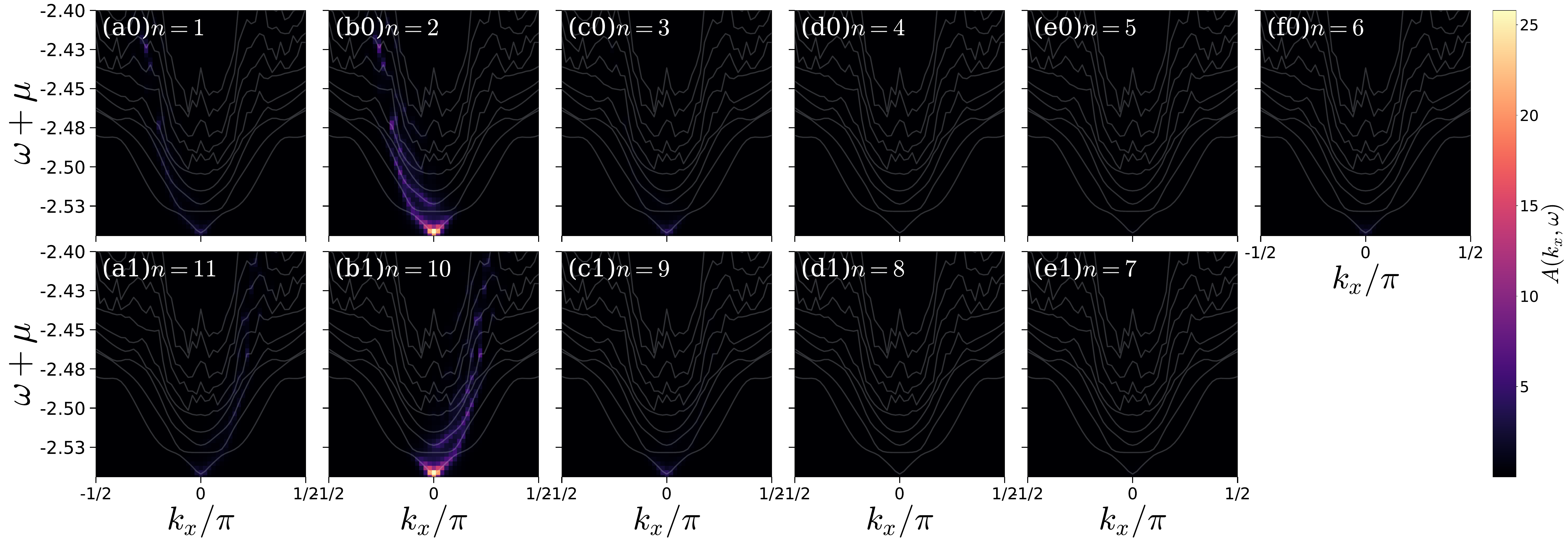}
        \caption{The spectral functions of charge-one edge excitations on an infinitely long strip with width $L_y=11$, $n_\phi=1/4$, trapping potential $V=0.01$, Lorentzian broadening factor $\eta=0.003$, and bond dimension $D=2000$.}
        \label{Ly11_Charge1_Vtrap001_Full}
    \end{figure}

\begin{figure}[htb]
        \includegraphics[width=0.9\linewidth]{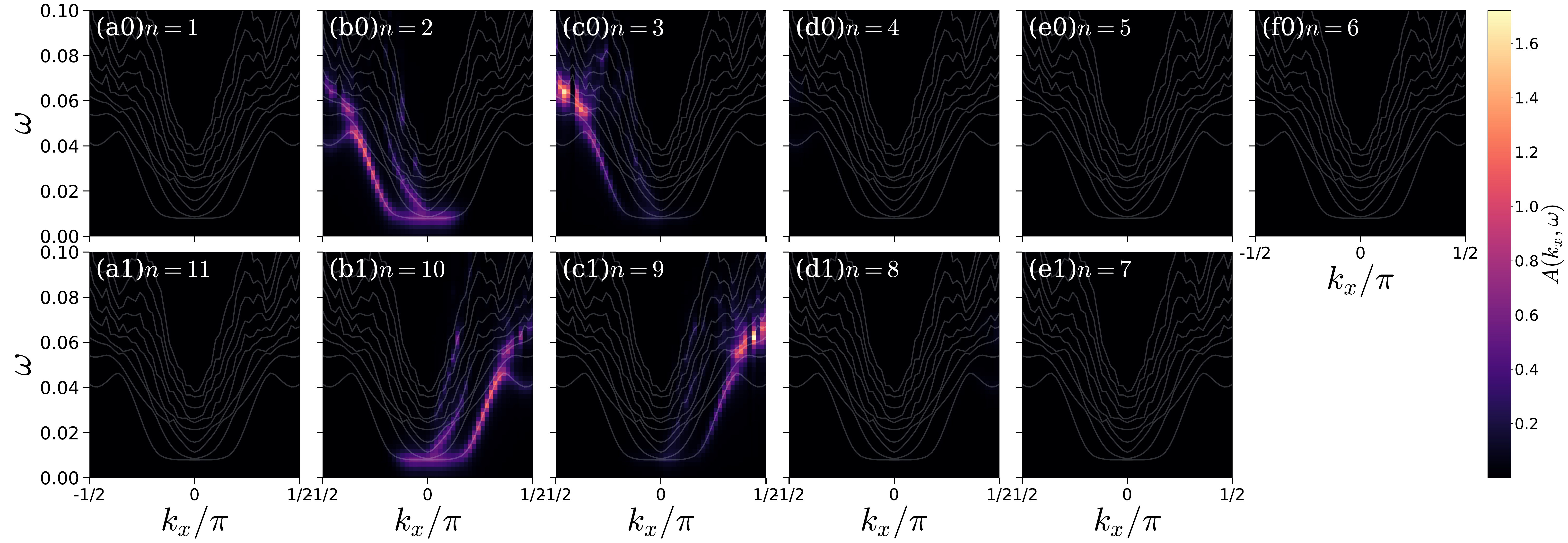}
        \caption{The spectral functions of charge-zero edge excitations on an infinitely long strip with width $L_y=11$, $n_\phi=1/4$, trapping potential $V=0.01$, Lorentzian broadening factor $\eta=0.003$, and bond dimension $D=2000$.}
        \label{Ly11_Charge0_Vtrap001_Full}
    \end{figure}

The following figures present the row-resolved spectral functions $\mathcal{A}_n(k_x,\omega)$ of charge-one or charge-zero edge excitations for all rows $n=1,\dots, L_y$ of infinitely long strips with widths $L_y=8,10,11$, under various trapping potentials. The results for $L_y=10$ are shown in Figs.~\ref{Ly10_Charge1_Vtrap00_Full}–\ref{Ly10_Charge0_Vtrap001_Full}, for $L_y=8$ in Fig.~\ref{Ly8_Charge1_Vtrap00_Full}, and for $L_y=11$ in Figs.~\ref{Ly11_Charge1_Vtrap001_Full} and \ref{Ly11_Charge0_Vtrap001_Full}.

    \FloatBarrier
    \FloatBarrier
    \FloatBarrier
    \FloatBarrier
    \FloatBarrier
     \FloatBarrier
\FloatBarrier
\FloatBarrier

\begin{figure}[htb]
    \centering
    \includegraphics[width=0.9\linewidth]{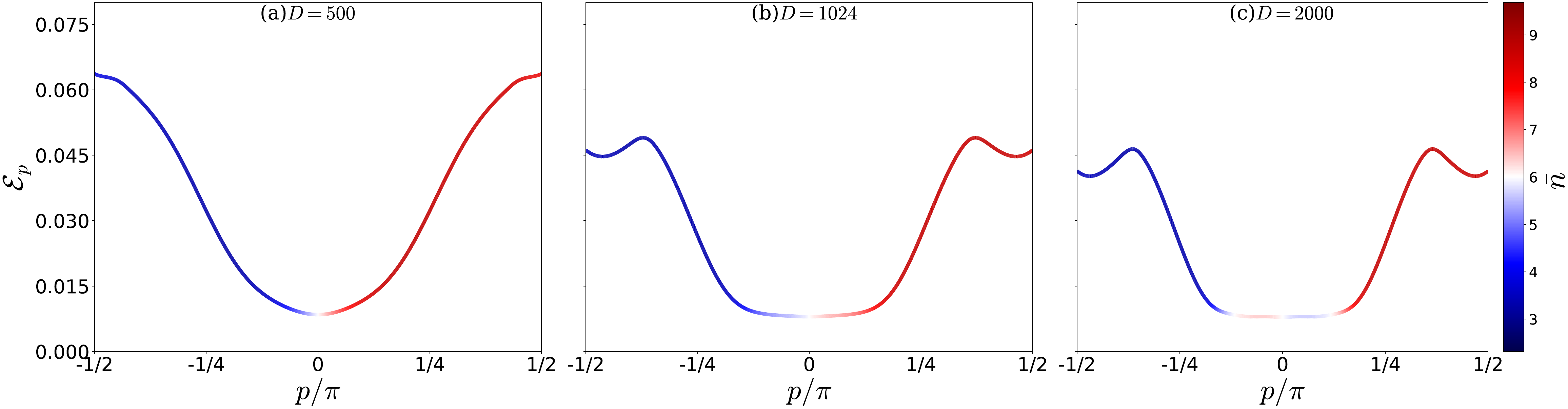}
    \caption{Average position $\overline{n}$ of charge-zero excitations for $L_y=11, V=0.01$, $n_\phi=1/4$, with bond dimensions 500 in (a), 1024 in (b), 2000 in (c).}
    \label{fig:Ly11_diff_bd}
\end{figure}

To compare with the previous results for $L_y=11$ in Ref.~\cite{vashisht2025}, we compute the average position $\overline{n}$ of charge-zero excitations using different bond dimensions of iMPS, as shown in Fig.~\ref{fig:Ly11_diff_bd}. For a small bond dimension $D=500$, the results resemble those in Ref.~\cite{vashisht2025}, where the lowest energy level exhibits a finite energy gap and the two branches appear chiral. However, upon increasing the bond dimension, the energy gap remains open and a plateau develops near $p=0$, indicating that the numerically converged low-energy excitations are no longer chiral.
    
\begin{figure}[htb]
\includegraphics[width=0.9\textwidth]{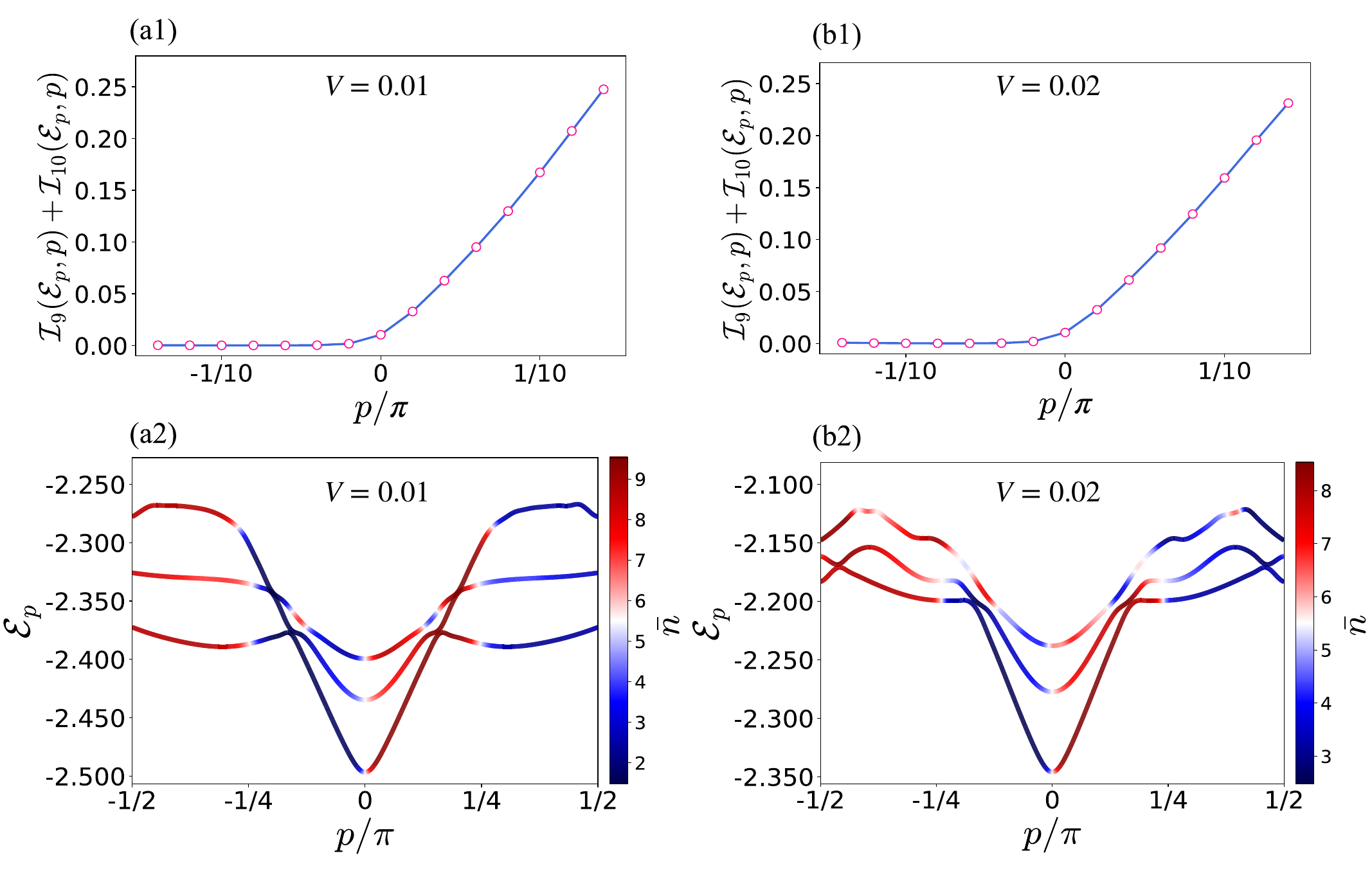}
\caption{The spectral weights and average positions for $L_y=10$, $n_\phi=1/4$, with $V=0.01$ in (a1)(a2) and $V=0.02$ in (b1)(b2). The bond dimensions are $D=1500$ in both cases. }
\label{fig:L_y_10_V_001_002}
\end{figure}

The key features of the spectral functions predicted by the $\chi$LL theory remain robust against a weak harmonic trapping potential on the strip with $L_y=10$, as shown in Fig.~\ref{fig:L_y_10_V_001_002}, where we plot the spectral weights and average positions for $V=0.01$ and $0.02$.

\begin{figure}[htb]
\centering
\includegraphics[width=0.9\linewidth]{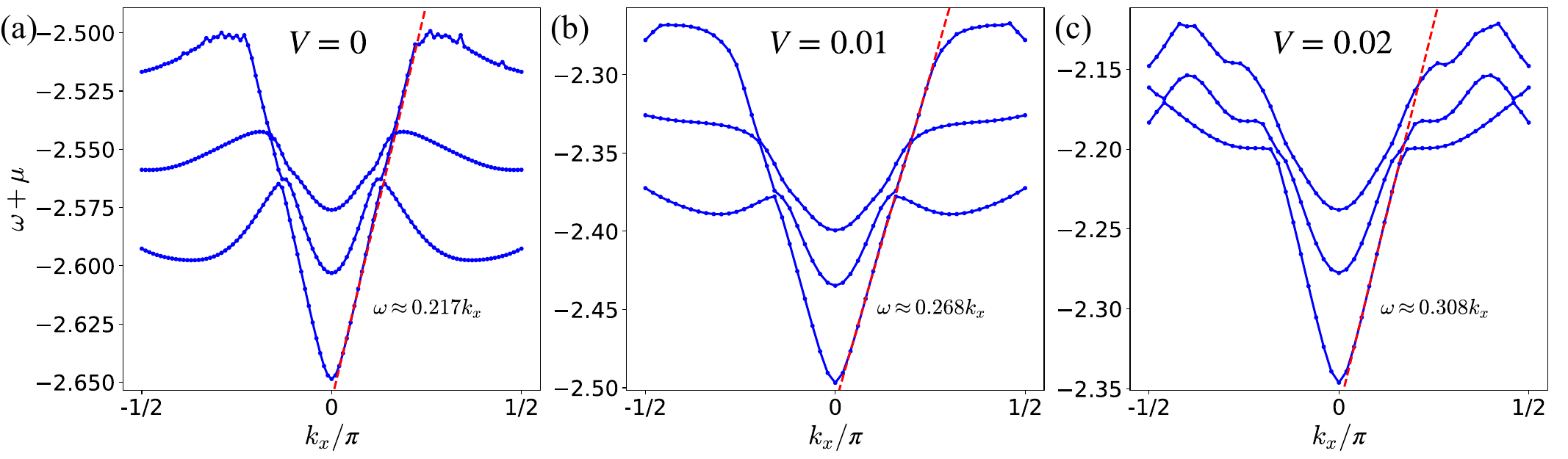}
\caption{The fitted velocity of charge-one chiral edge modes on an infinitely long strip with $L_y=10$, $n_{\phi}=1/4$, and harmonic trapping potential $V=0,0.01,0.02$ in (a,b,c), respectively.}
\label{fig:edge_velocity}
\end{figure}

The velocity of the lowest chiral edge branch is non-universal and depends on the confining potential. We find that this velocity will increase when we increase the harmonic trap $V$ as demonstrated in Fig.~\ref{fig:edge_velocity} for the charge-one chiral edge states on a strip with $L_y=10$.

\clearpage

\section{Chiral edge excitations for $L_y=12$}

To demonstrate the universality of our results and further establish the conditions under which a well-behaved spectrum of chiral excitations could be observed on discrete lattice systems, we perform calculations on an infinitely long strip of width $L_y=12$. We consider two distinct setups, both of which host a $\nu=1/2$ FCI in the bulk and has one boson per column, consistent with the main text: (1) magnetic flux $n_{\phi}=1/4$ in the presence of a finite harmonic trap $V$, and (2) magnetic flux $n_{\phi}=1/5$ with $V=0$.

\begin{figure} [h]
\includegraphics[width=0.45\linewidth]{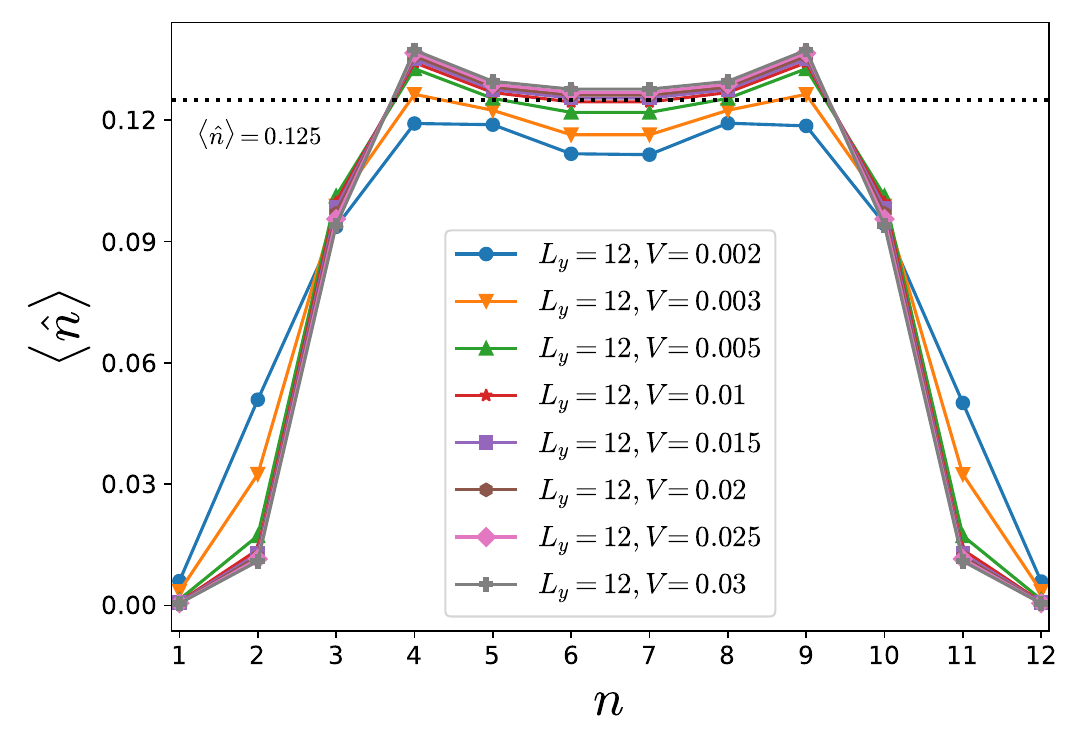}
\caption{Ground-state densities on an infinite strip with $L_y=12$, $n_{\phi}=1/4$, and trapping potential $V$ varied from $0.002$ to $0.03$. The bond dimension here is $D=500$ for illustration (the difference in densities for for $D=500$ and $d=2000$ is of the order $10^{-4}$). }
\label{fig:Ly_12_phi_1_4_N_vary_V}
\end{figure}

\begin{figure}[h]
\includegraphics[width=0.9\linewidth]{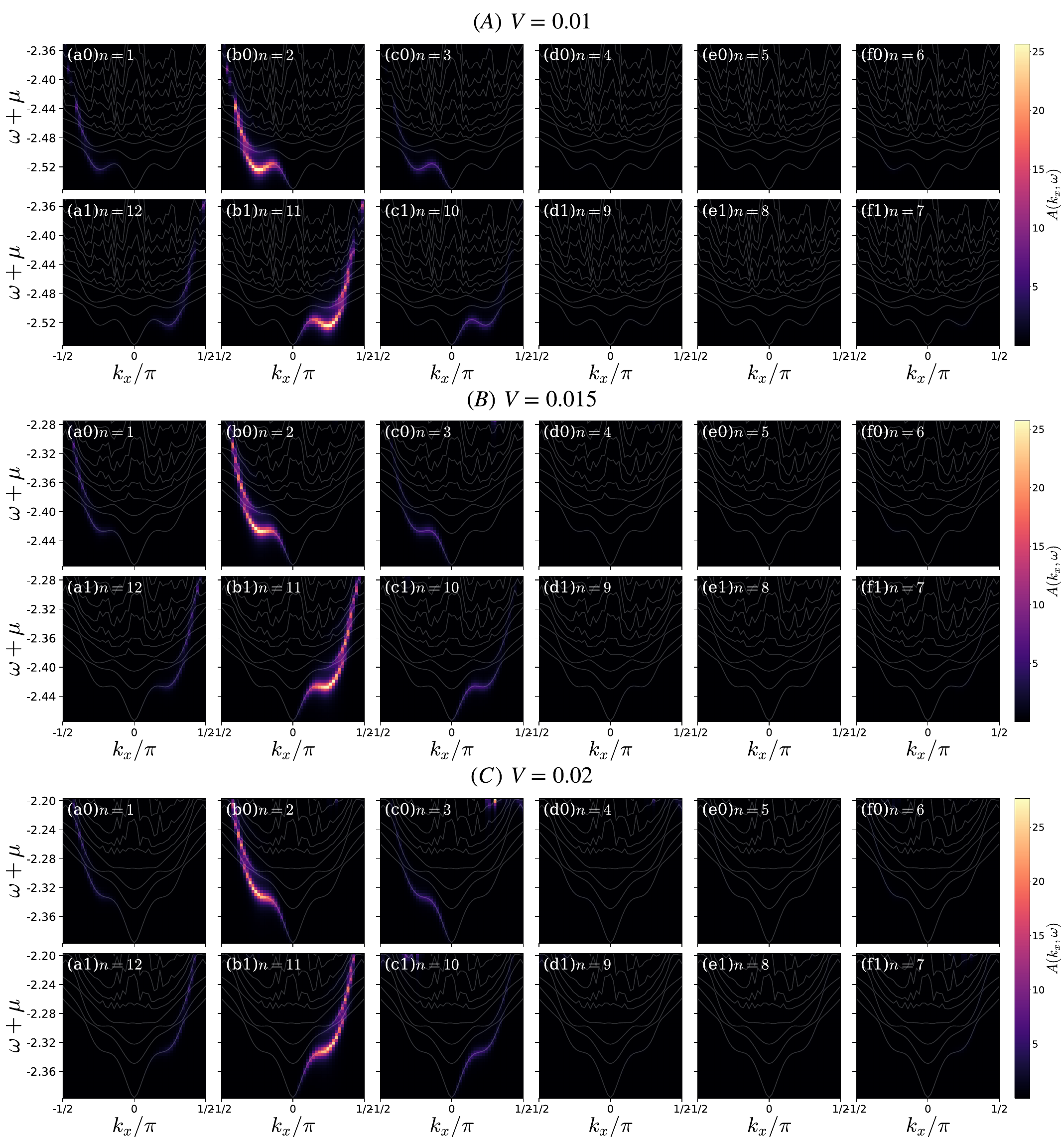}
\caption{The spectral functions of charge-one excitations on each row $n=1,...,12$ of an infinite long strip with width $L_y=12$, $n_\phi=1/4$, and trapping potential $V=0.01$ in (A), $V=0.015$ in (B), and $V=0.02$ in (C). The bond dimension is $D=2000$, the Lorentzian broadening factor is $\eta=0.005$, and $\mu$ is the chemical potential. The width of all the energy windows is chosen as $0.2$, with which it is clear that the gaps at $k_x\neq 0$ increase as $V$ increases.}
\label{fig:Ly_12_phi_1_4_spectral_functions_V_0_01_0_015_0_02}
\end{figure}

\begin{figure}[h]
\includegraphics[width=0.8\linewidth]{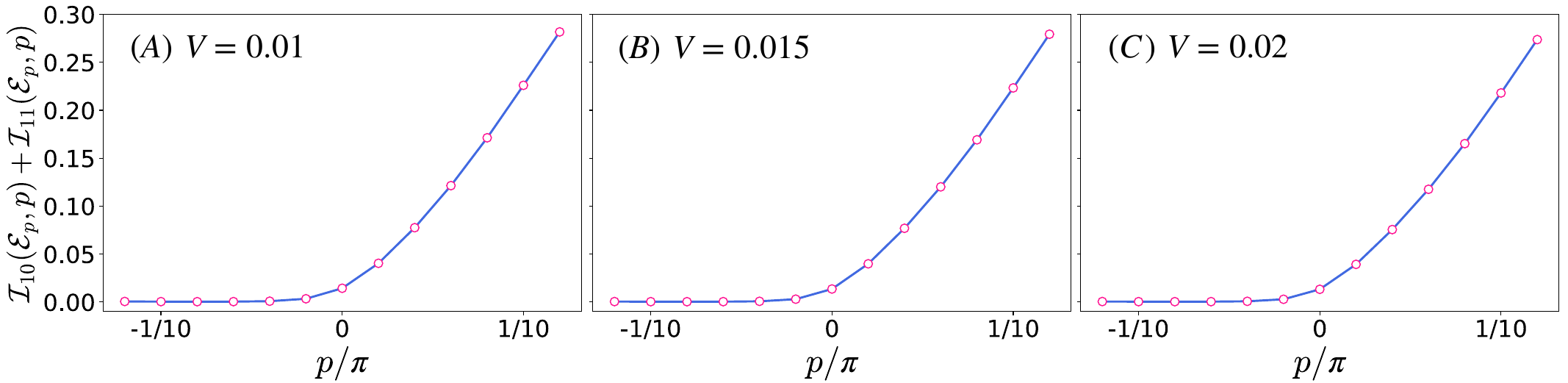}
\caption{Spectral weight $\mathcal{I}_{10}(\mathcal{E}_p,p)+\mathcal{I}_{11}(\mathcal{E}_p,p)$ of the lowest chiral excitaions near $p=0$ on an infinite long strip with $L_y=12$, $n_\phi=1/4$, and tapping potential $V=0.01$ in (A), $V=0.015$ in (B), $V=0.02$ in (C). 
}
\label{fig:spectral_weight_Ly12site10_plus11_V_0_015_phi_1_4}
\end{figure}

In the first case ($n_{\phi}=1/4$), following the discussion in the main text, a bulk particle density of $n_b=1/8$ is required to realize a $\nu=1/2$ FCI state. As demonstrated in Fig.~\ref{fig:Ly_12_phi_1_4_N_vary_V}, a finite harmonic trapping potential is essential to achieve this density. The resulting spectral functions in the charge-one sector for $V=0.01, 0.015,$ and $0.02$ are presented in Fig.~\ref{fig:Ly_12_phi_1_4_spectral_functions_V_0_01_0_015_0_02}. Notably, the lowest chiral excitations are consistent with the predictions of $\chi$LL theory; their corresponding spectral weights, $\mathcal{I}_{10}(\mathcal{E}_p,p)+\mathcal{I}_{11}(\mathcal{E}_p,p)$ near $p=0$, are shown in Fig.~\ref{fig:spectral_weight_Ly12site10_plus11_V_0_015_phi_1_4}. Furthermore, we observe a local gapped excitation minimum at $k_x\neq 0$ near the edge. However, this mode can be suppressed by increasing the trapping potential $V$, as evidenced by the comparison of the spectra across different values of $V$ in Fig.~\ref{fig:Ly_12_phi_1_4_spectral_functions_V_0_01_0_015_0_02}.

\begin{figure}[h]
\includegraphics[width=0.45\linewidth]{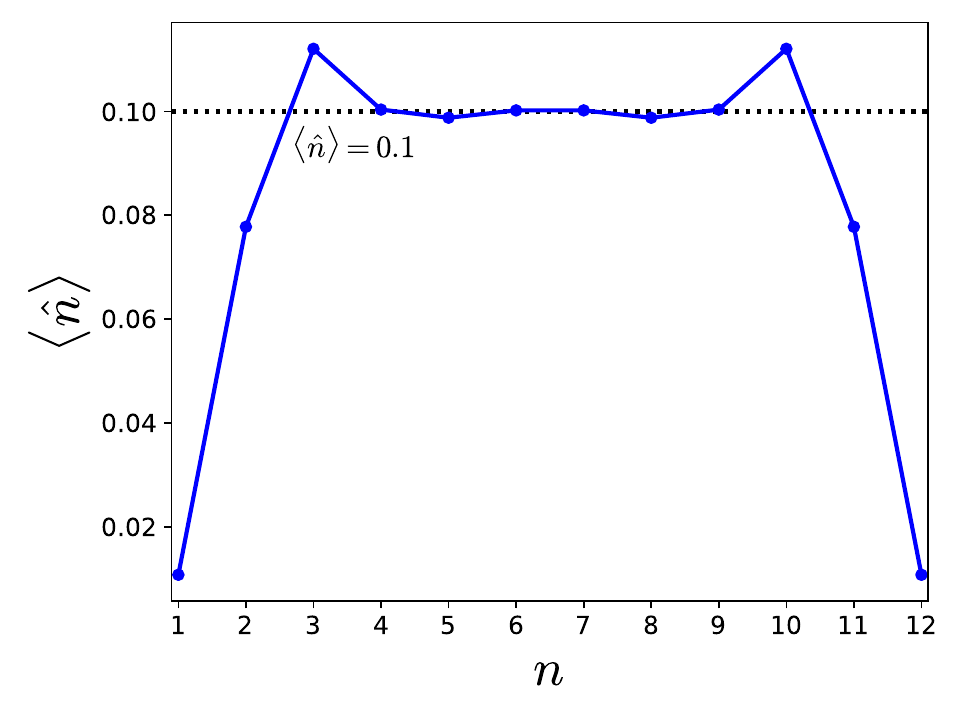}
\caption{Ground-state densities on an infinite long strip with $L_y=12$, $n_{\phi}=1/5$, and no trapping potential. The bond dimension here is $D=2000$.}
\label{fig:B_Hof_GS_N_density_Ly_12BD_2000Vtrap_0.0phi_1_5obc_A}
\end{figure}

\begin{figure}[h]
\includegraphics[width=0.9\linewidth]{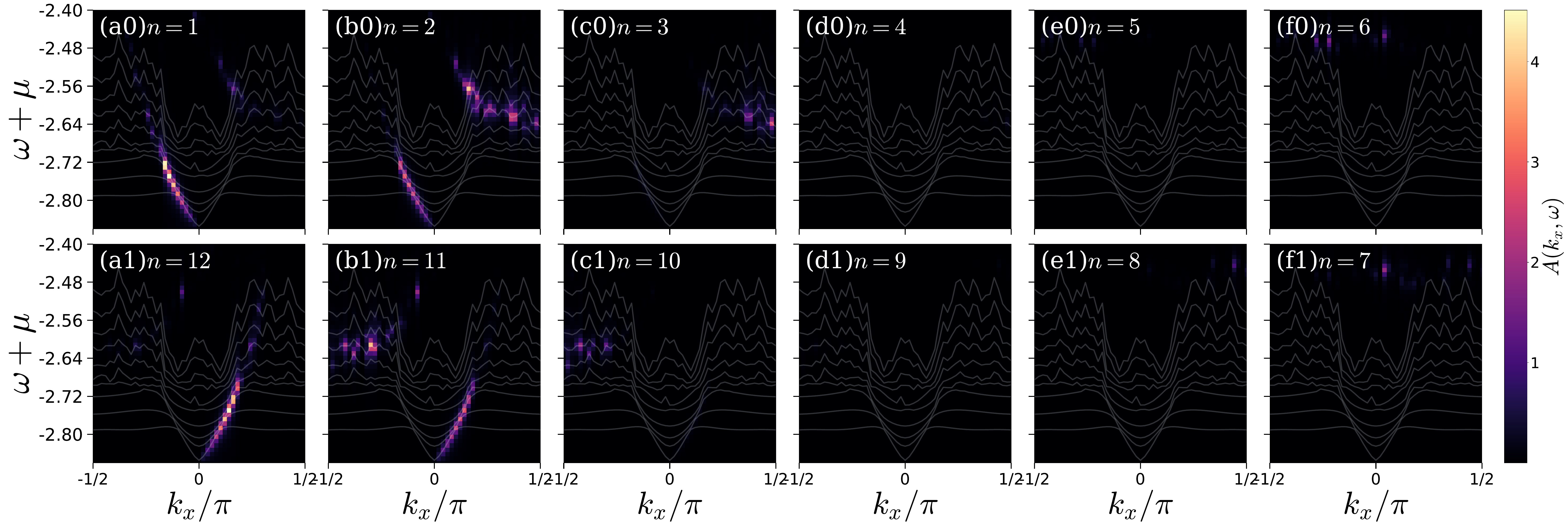}
\caption{The spectral functions of charge-one excitations on each row $n=1,...,12$ of an infinite long strip with width $L_y=12$, $n_\phi=1/5$, and no trapping potential. The bond dimension is $D=2000$, the Lorentzian broadening factor is $\eta=0.01$.}
\label{fig:Ly_12_Charge1_Vtrap0.0_phi_1_5_BD2000_etaw0.01_eta_k0magma_site_0_to_11}
\end{figure}

\begin{figure}[h]
\includegraphics[width=0.9\linewidth]{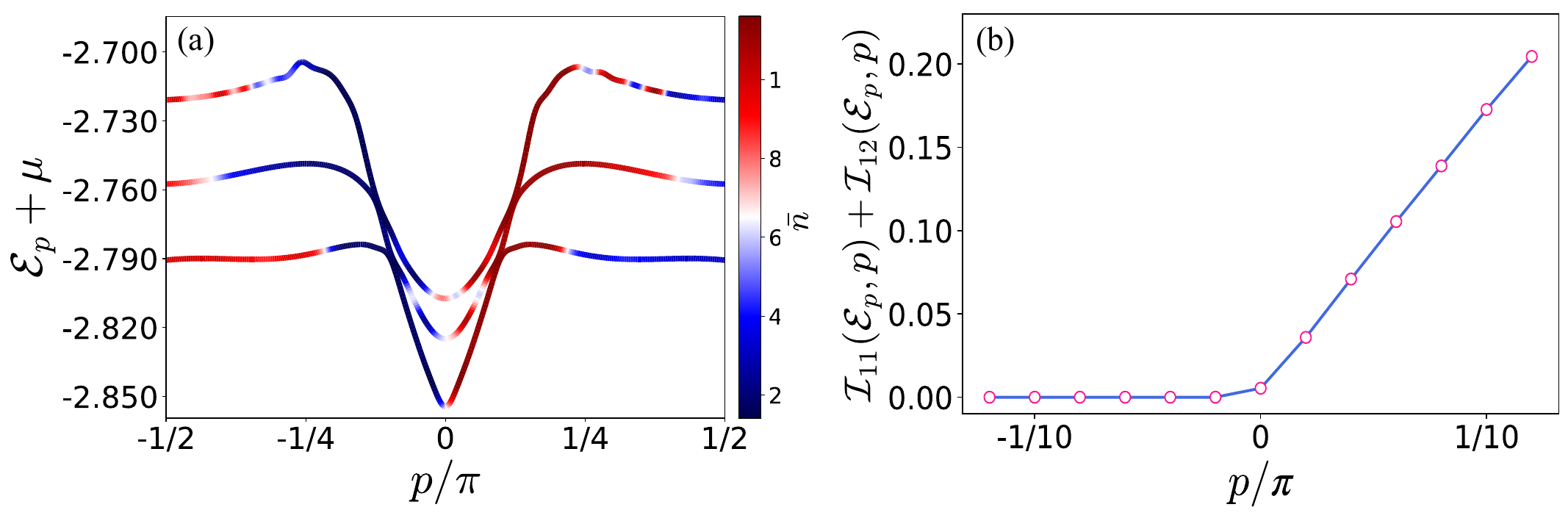}
\caption{Average positions and spectral weight and of charge-one excited states on an infinite long strip with $L_y=12$, $n_{\phi}=1/5$, and no trapping potential. (a) Average positions $\overline{n}$ of the three lowest energy levels of the excited states. (b) Spectral weight $\mathcal{I}_{11}(\mathcal{E}_p,p)+\mathcal{I}_{12}(\mathcal{E}_p,p)$ of the lowest chiral excitations near $p=0$. 
}
\label{fig:COM_and_Intensity_Ly_12_V0_0_charge1_phi_1_5}
\end{figure}

In the second case, we set the magnetic flux to $n_{\phi}=1/5$. As predicted in Ref.~\cite{he2017realizing}, a $\nu=1/2$ FCI state can be realized when the bulk particle density is $n_{b}=1/10$. By placing one boson in each column, we find that the intrinsic box-like potential provided by the edges of the strip is sufficient to achieve this density without an additional harmonic trap [Fig.~\ref{fig:B_Hof_GS_N_density_Ly_12BD_2000Vtrap_0.0phi_1_5obc_A}]. The corresponding charge-one spectral functions are presented in Fig.~\ref{fig:Ly_12_Charge1_Vtrap0.0_phi_1_5_BD2000_etaw0.01_eta_k0magma_site_0_to_11}, with the largest spectral weights localized at the edges ($n=1,2$ and $n=11,12$). The average positions and the spectral weight $\mathcal{I}_{11}(\mathcal{E}_p,p)+\mathcal{I}_{12}(\mathcal{E}_p,p)$ near $p=0$ are shown in Fig.~\ref{fig:COM_and_Intensity_Ly_12_V0_0_charge1_phi_1_5}. The behavior of these chiral edge modes remains consistent with $\chi$LL theory, demonstrating that our results apply generally to $\nu=1/2$ FCIs, provided the bulk state is correctly established.

\begin{figure}[h]
\includegraphics[width=1.0\linewidth]{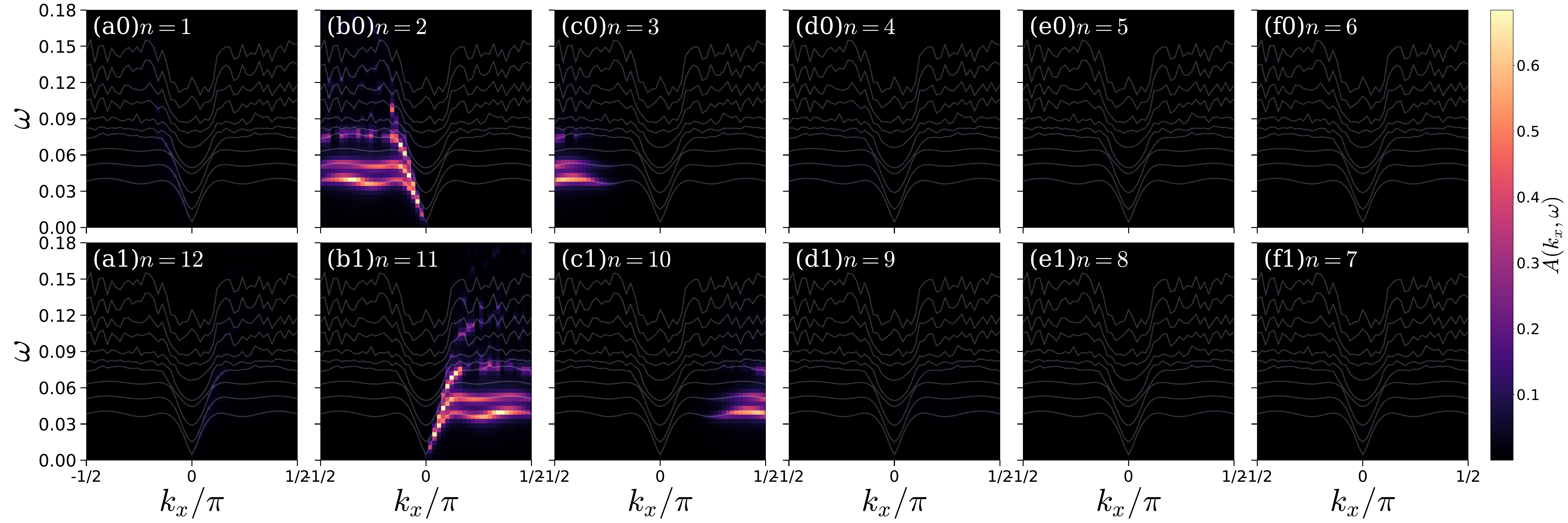}
\caption{The spectral functions of charge-zero excitations on each row $n=1,...,12$ of an infinite long strip with width $L_y=12$, $n_\phi=1/5$, and no trapping potential. The bond dimension is $D=2000$, and the Lorentzian broadening factor is $\eta=0.003$.}
\label{fig:Ly_12_Charge0_Vtrap0.0_phi_1_5_BD2000_etaw0.003_eta_k0magma_site_0_to_11}
\end{figure}

\begin{figure}[h]
\includegraphics[width=0.45\linewidth]{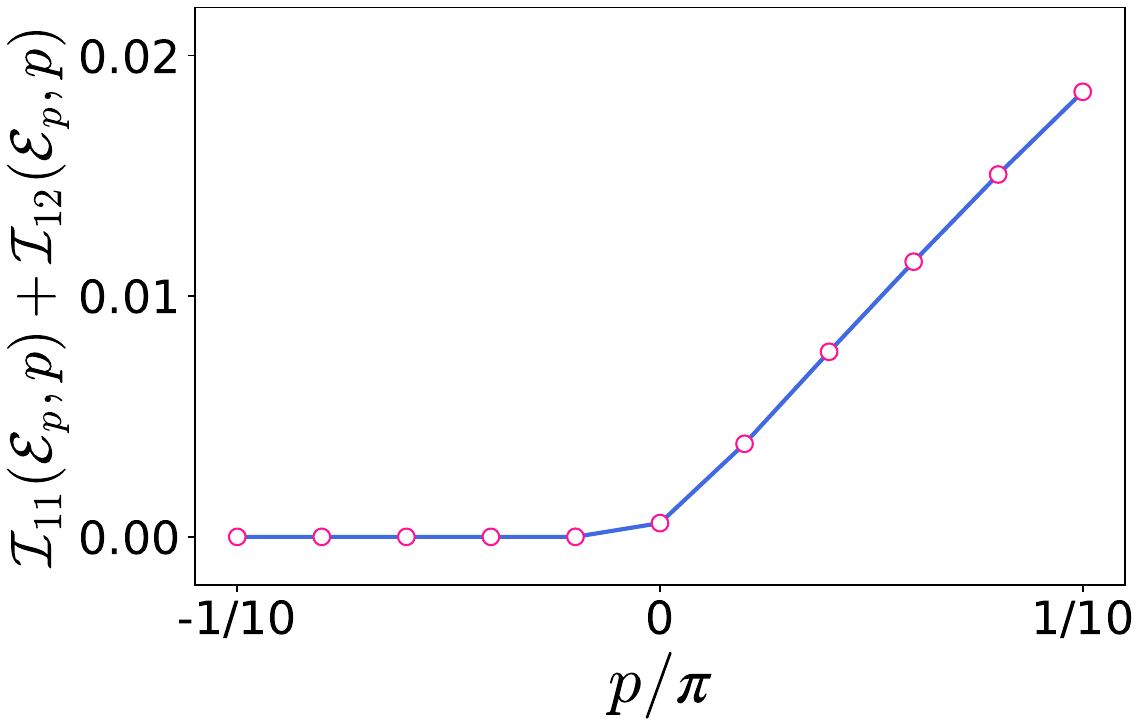}
\caption{Spectral weight $\mathcal{I}_{11}(\mathcal{E}_p,p)+\mathcal{I}_{12}(\mathcal{E}_p,p)$ of lowest charge-zero excited states near $p=0$ on an infinite long strip with $L_y=12$, $n_{\phi}=1/5$, and no trapping potential.}
\label{fig:spectral_weight_Ly12_V_0_charge0_phi_1_5_site11_plus12}
\end{figure}

In the second case, we also calculate the spectral functions in the charge-zero sector, as shown in Fig.~\ref{fig:Ly_12_Charge0_Vtrap0.0_phi_1_5_BD2000_etaw0.003_eta_k0magma_site_0_to_11}, where the maximum spectral weight is localized at rows $n=2$ and $n=11$. Comparing these results with Figs.~\ref{Ly10_Charge0_Vtrap00_Full} and \ref{Ly10_Charge0_Vtrap001_Full} for $L_y=10$, we observe that increasing the spatial separation between the edge modes by two rows reduces the energy gap at $k_x=0$ from the order of $10^{-2}$ to $10^{-3}$. Furthermore, the spectral weight $\mathcal{I}_{11}(\mathcal{E}_p,p)+\mathcal{I}_{12}(\mathcal{E}_p,p)$ of the lowest excited states near $p=0$ is consistent with $\chi$LL theory [Fig.~\ref{fig:spectral_weight_Ly12_V_0_charge0_phi_1_5_site11_plus12}]. These findings indicate that once the spatial separation is sufficiently large, the hybridization between the edges is suppressed, enabling the recovery of gapless chiral edge modes in both the charge-zero and charge-one sectors.

\clearpage

    \FloatBarrier

\bibliographystyle{apsrev4-2}
\bibliography{reference}